\documentclass[12pt,a4paper]{article}

\usepackage{a4}

\usepackage{amsfonts}
\usepackage{amssymb}
\usepackage{epsfig}
\usepackage{psfig}
\usepackage{psfrag}
\usepackage{dsfont}

\usepackage[sc,small]{caption2}
\setcaptionwidth{14cm}

\newcommand{\beqn}{\begin{eqnarray}}
\newcommand{\eeqn}{\end{eqnarray}}
\newcommand{\be}{\begin{equation}}
\newcommand{\ee}{\end{equation}}
\newcommand{\non}{\nonumber \\}

\newcommand{\vol}{{\cal V}}
\newcommand{\vols}{{\cal V}^{\rm (str)}}

\newcommand{\ca}{{\cal A}}
\newcommand{\cn}{{\cal N}}
\newcommand{\cm}{{\cal M}}
\newcommand{\cl}{{\cal L}}
\newcommand{\cf}{{\cal F}}

\newcommand{\co}{{\cal O}}
\newcommand{\tr}{{\rm tr}}
\newcommand{\bb}[2]{\!\left[^{#1}_{#2}\right]\!}

\newcommand{\tht}{\vartheta}

\newcommand{\thbw}[2]{\vartheta[{ #1 \atop #2} ]}
\newcommand{\thba}[2]{\vartheta[\!\!\begin{array}{c}{\scriptstyle#1}%
                        \\[-1.6mm]{\scriptstyle #2}\end{array}\!\!]}
\newcommand{\thbap}[2]{\vartheta''[\!\!\begin{array}{c}{\scriptstyle#1}%
                        \\[-1.6mm]{\scriptstyle #2}\end{array}\!\!]}
\newcommand{\bw}[2]{[{ #1 \atop #2} ]}
\newcommand{\ba}[2]{[\!\!\begin{array}{c}{\scriptstyle#1}
                        \\[-1.6mm]{\scriptstyle #2}\end{array}\!\!]}

\newcommand{\bfg}{\mbox{\boldmath$\gamma$}}
\newcommand{\bfe}{\mbox{\boldmath$\epsilon$}}

\begin{document}

\title{}
\author{}
\date{}
\thispagestyle{empty}

\begin{flushright}
\vspace{-3cm}
{\small MIT-CTP-3486 \\
        NSF-KITP-04-41 \\
        hep-th/0404087}
\end{flushright}
\vspace{1cm}

\begin{center}
{\Large\bf Loop Corrections to Volume Moduli and \\[.3cm]

Inflation in String Theory}
\end{center}

\vspace{1.5cm}

\begin{center}

{\bf Marcus Berg}$^{\dag}$
{\bf,\hspace{.2cm} Michael Haack}$^{\dag}$
{\bf\hspace{.1cm} and\hspace{.2cm} Boris K\"ors}$^{*}$
\vspace{0.5cm}

\end{center}

\hbox{
\parbox{7cm}{
\begin{center}
{\it
$^{\dag}$Kavli Institute for Theoretical Physics \\
University of California \\
Santa Barbara, \\
California 93106-4030, USA\\
}
\end{center}
}
\hspace{-.5cm}
\parbox{8cm}{
\begin{center}
{\it
$^*$Center for Theoretical Physics \\
Laboratory for Nuclear Science \\
and Department of Physics \\
Massachusetts Institute of Technology \\
Cambridge, Massachusetts 02139, USA \\
}
\end{center}
}
}

\vspace{1cm}

\begin{center}
{\bf Abstract} 
\end{center} 

The recent progress in embedding inflation in string
theory has made it clear that the problem of moduli stabilization 
cannot be ignored in this context. 
In many models a special role is played by the volume modulus,
which is modified in the presence of mobile branes. The 
challenge is to
stabilize this modified volume while keeping the inflaton mass 
small compared to the Hubble parameter. It is then 
crucial to know
not only how the volume modulus is modified, but also 
to have control over the dependence 
of the potential on the inflaton field. 
We address these questions
within a simple setting: toroidal $\cn=1$ type IIB orientifolds. 
We calculate corrections to the superpotential and show how the holomorphic 
dependence on the properly modified volume modulus arises.
The potential then explicitly 
involves the inflaton, leaving room for lowering the inflaton mass 
through moderate 
fine-tuning of flux quantum numbers. 

\clearpage


\section{Introduction}

Inflation has become one of the cornerstones of
our picture of the early universe and its evolution. 
The successes of inflation
include the explanation of the apparent homogeneity and isotropy 
of the universe at large scales,
and the prediction of 
a spectrum of density fluctuations in the cosmic 
microwave background
that agrees with observation. 
Recently there have been several interesting 
attempts to embed inflation in string theory 
by combining elements of string theory model building such as 
background fluxes and D-branes. 
In this paper, we will focus mainly on the model of Kachru, Kallosh, Linde, 
Maldacena, McAllister, and Trivedi (KKLMMT) 
\cite{Kachru:2003sx,Kachru:2003aw}, but our conclusions are equally 
relevant for e.g.\
D3/D7-brane inflation \cite{Dasgupta:2002ew,Hsu:2003cy,Koyama:2003yc}. 

The central idea of inflation in D-brane models 
is to realize inflaton fields by
open string moduli that parametrize the 
positions of branes.\footnote{Other more exotic 
candidates for the inflaton were proposed in \cite{DeWolfe:2004qx}.} 
The motion of the brane then roughly corresponds to the rolling of the 
inflaton. 
More precisely, these models typically use
 the standard framework of single-field 
slow-roll inflation, where the flatness of the effective potential 
for the inflaton 
is measured by the slow-roll parameters 
\beqn 
\epsilon = \frac{M_{\rm Pl}^2}{2} \left( \frac{V'(\varphi)}{V(\varphi)} \right)^2 \ll 1 \ , \quad 
\eta = M_{\rm Pl}^2 \left| \frac{V''(\varphi)}{V(\varphi)} \right| \ll 1 \ , 
\eeqn
primes denoting derivatives of $V(\varphi)$ with respect to the 
canonically normalized inflaton field $\varphi$.
Early models of brane inflation relied on the assumption that all 
other scalars, in particular the geometrical moduli for the background, can be 
ignored or frozen while the brane-position scalars evolve. 
It may be considered one of the main merits of 
\cite{Kachru:2003sx,Kachru:2003aw}
that this issue was addressed in a 
model that in principle allows stabilization of 
all geometrical moduli of the compactification 
space. However, 
the situation is complicated by the mixing of
the geometrical 
background moduli and the open string moduli 
in the effective action; it is not obvious that one can fix the former
while evolving the latter \cite{Kachru:2003sx}. 

In particular, let us consider
the volume modulus. 
Under the assumption that
a model with just a single
K\"ahler modulus can be found, 
it was argued in \cite{Kachru:2003aw} that the volume of the 
internal space can be
stabilized in type IIB orientifold compactifications with 3-form fluxes 
\cite{Giddings:2001yu}, if one includes
non-perturbative effects: either superpotential 
contributions from 
Euclidean D3-brane instantons \cite{Witten:1996bn}, or gaugino condensation 
on the world-volume of wrapped D7-branes 
(we will concentrate on the latter in the following). In either case, 
a superpotential is generated that depends holomorphically 
on the volume modulus and on the inflaton field, as dictated by supersymmetry. 
Now, 
in the presence of mobile D3-branes, the K\"ahler modulus includes 
not only the volume
but also  the D3-brane scalars,
as alluded to in the previous paragraph. 
It was argued in \cite{Kachru:2003sx} 
that it is this combination, as opposed 
to the actual ``geometrical'' volume, that 
is stabilized along the lines described in \cite{Kachru:2003aw}. 
This mixing produces a mass term for the D3-brane scalars, 
a combination of which can naturally serve as a
candidate for the inflaton field. 
This mass gives a contribution of order one to the slow-roll parameter 
$\eta$ and thus seems to spoil 
inflation in this class of models if no additional contribution 
to the mass arises, e.g.\ 
through quantum corrections.\footnote{This result was confirmed 
using a completely 
different method in \cite{Buchel:2003qj}. Also, 
in the following we will always use the term ``inflaton mass'' 
instead of ``$\eta$'', although the mass of the inflaton 
is strictly speaking only defined at the minimum 
of the potential. Finally, note that the parameter $\epsilon$ is
usually much smaller than $\eta$ in the KKLMMT model, at least if the
inflaton field is much smaller than the Planck mass 
\cite{Kachru:2003sx,Iizuka:2004ct}. We, therefore, concentrate 
on the inflaton mass problem in the following.}

Of course, the actual mass depends crucially on {\it what} 
combination of the volume modulus and
the D3-brane scalars is stabilized, and how the inflaton enters 
the superpotential. The key to understanding both issues is the 
gauge kinetic function that appears 
in the effective Lagrangian for the gauge fields on the D7-branes.
This is because the gauge kinetic function determines the non-perturbative 
superpotential, and holomorphy of the superpotential allows one to read off 
the correct K\"ahler modulus. The dependence of the 
superpotential on the inflaton field, which we determine,
leads to additional 
contributions to its mass and confirms the expectation of 
\cite{Kachru:2003sx} that it can be fine-tuned to 
small values. 

An important potential problem with the form of 
the K\"ahler modulus $\rho$ suggested in 
\cite{Kachru:2003sx}  
is that at first sight it seems to be in conflict with supersymmetry of the 
effective theory \cite{Hsu:2003cy,renatastalk}, because 
it appears to violate holomorphy of the gauge kinetic 
function.\footnote{ See 
\cite{Louis:1996ya} for a nice introduction to 
holomorphic couplings in string theory.} More concretely,
one might compute the Wilsonian coupling of the D7-brane gauge group 
by reduction of the Dirac-Born-Infeld (DBI) action of the 
D7-branes on the 4-cycle the branes 
are wrapped around, as we review in section \ref{inflation}.
The D7-brane gauge coupling that results from this reduction
does not seem to be the real part of a holomorphic function of the 
corrected K\"ahler modulus
$\rho$.\footnote{We would like to stress that this puzzle 
is not restricted 
to the present class of cosmological models; it is a general 
problem of 
the effective supergravity action that follows from string theory 
in the presence of D-branes. It seems to us that up to now,
the problem has simply been ignored.}  
To summarize, there are three questions we would like to address:
\pagebreak
\vspace{-2mm}
\begin{itemize}
\item 
How does the modified modulus $\rho$ depend on the D3-brane scalars?
\item
How does the gauge kinetic function become a holomorphic function 
of this modified modulus? (We call these two issues 
collectively the ``rho problem''.)
\item
How does the non-perturbative superpotential depend on the open string 
scalars, in particular on the inflaton candidate $\varphi$? 
\end{itemize}
It is the main purpose of this paper to shed some light on the 
solution to these problems. We propose that the
dependence of the gauge kinetic function 
on the D3-brane scalars due to
{\it open-string one-loop corrections} leads both to a solution of the 
rho problem and to additional dependence of the superpotential 
on the open-string scalars. These corrections 
arise from the M\"obius and annulus diagrams at Euler characteristic
zero.\footnote{The order of string perturbation theory,
which is given in terms of the Euler characteristic $\chi$ and the
dilaton $\Phi$  as $e^{-\chi \Phi}$,
should not be confused with the
open (loop) vs.\ closed (tree) channel interpretations; 
the annulus diagram always has $\chi=0$, but it can be
computed two different ways. See
\cite{Angelantonj:2002ct} for a comprehensive 
introduction to open strings.}
By comparison, the tree-level action
from dimensional reduction 
of the bulk supergravity and the D-brane DBI action 
come with powers $e^{-2\Phi}$ and $e^{-\Phi}$ of the string coupling.
Thus, the three questions above  
can be addressed simultaneously 
by calculating the gauge kinetic function of the 
D7-branes at the open-string one-loop level. 

The actual KKLMMT setup
is reviewed in the next section, 
but for the computations later in the paper we 
consider a simplified setting: toroidal $\cn=1$ type IIB 
orientifolds (in the following
we will drop ``toroidal'' and simply talk about type IIB 
orientifolds). 
In doing so, we are certainly 
not able to capture the details 
of a hypothetical analogous calculation in the KKLMMT background, but it 
enables us to understand the basic qualitative picture 
in a controlled way.
Also, we actually perform the calculation in a T-dual picture, 
where six T-dualities are performed to turn the D3- and D7-branes into 
D9- and D5-branes, respectively. \label{tdual}
Working in this T-dual picture makes it easier 
to compare our results to the existing literature 
on open string loop corrections in orientifolds, where
the equivalent D9/D5-brane language is usually preferred.  
In this language, the D3-brane scalars 
are mapped to continuous Wilson line moduli. 
This means that for our purpose of investigating
the rho problem, we want to compute
the dependence of the D5-brane gauge coupling on the D9-brane 
Wilson lines. In particular, we want the dependence due to
closed string exchange between D-branes (and O-planes), or
equivalently, due to open string one-loop dia\-grams. Such 
corrections are usually 
referred to as one-loop {\it threshold corrections}
 to the gauge coupling constants 
(see e.g. \cite{Kaplunovsky:1987rp} for 
earlier work in this context, 
including the heterotic string).\footnote{The 
role of threshold corrections for moduli stabilization through 
non-perturbative superpotentials due to gaugino condensation has recently 
been discussed also in
\cite{Gukov:2003cy}.} 
In fact, without much additional effort, we can 
ask the slightly more general question 
of how both the D5-brane and D9-brane 
gauge couplings depend on both the D5-brane and D9-brane Wilson line 
moduli. Hence, our proposal for an answer to 
the three questions above appears as a special case of a more
general result.

Our conclusions are that the D3-brane scalars and thus the inflaton 
field $\varphi$ indeed 
appear in the modified $\rho$ modulus in a form that
was qualitatively anticipated in \cite{Kachru:2003sx}. 
The one-loop correction to the gauge kinetic function  is 
found to be of exactly the 
right form to reinstate its holomorphy, when expressed in terms of 
the modified modulus $\rho$. 
Hence the rho problem is solved in this (simplified) setting, and the 
inflaton mass problem of \cite{Kachru:2003sx} 
is manifest. 
Happily, the inflaton mass problem may be alleviated
by certain additional corrections to the 
gauge kinetic function, and thus to the non-perturbative superpotential,
which depend on the inflaton.
This can help lowering the inflaton mass. 
We expect there to be quantitative modification 
of our orientifold results in the KKLMMT 
and D3/D7-brane inflationary models, 
but qualitatively, we expect our conclusions to remain the same.

The paper is organized as follows.
In the next section, we review the two models of inflation 
where we want to apply our results: the KKLMMT model and D3/D7-brane
inflation. We then proceed 
by describing our method of calculation, the background-field
method in type IIB orientifolds
\cite{Bachas:bh,Bachas:1996zt,AnBaDu99}, 
in section
\ref{background}, and we 
compute the one-loop corrections to the D9- and D5-brane 
gauge couplings in the 
$\mathbb{T}^2\times\mathbb{T}^4/\mathbb{Z}_2$ orientifold 
\cite{Bianchi:1990yu,Gimon:1996rq,Gimon:1996ay},
and their dependence on the Wilson lines along the
$\mathbb{T}^2$, in section \ref{seconeloop}.
This model actually has unbroken ${\cal N} = 2$ supersymmetry
in four dimensions, but the computation 
can easily be generalized to ${\cal N} = 1$ orientifolds
on $\mathbb{T}^6/\mathbb{Z}_N$ with even $N$  (see e.g. 
\cite{Angelantonj:1996uy,Aldazabal:1998mr}), or to 
$\mathbb{T}^6/(\mathbb{Z}_N \times \mathbb{Z}_M)$ models
\cite{Berkooz:1996dw,Zwart:1997aj}. We 
 carry out this generalization
in section \ref{none}, focusing on the examples of $\mathbb{Z}'_6$
and $\mathbb{Z}_2 \times \mathbb{Z}_2$. 
Finally, in section \ref{inter}
we interpret our results in the 
context of string-theoretic models for inflation.
Some of the relevant formulas and more technical details are 
collected in the appendices. 

{\bf Reading Guide:}
For the reader who is interested in results
and not details, we propose reading the review section
\ref{inflation},
and then jumping straight to section \ref{inter}, where the 
implications for inflation in string theory are discussed. 
It requires some of the notation introduced in section 
\ref{technicalstuff}, but it does not rely on understanding the
calculations of section \ref{technicalstuff} in any detail.
This reader might
also want to have a look at our two ``side remarks'' 
about the prepotential and the 
special coordinates in the $\cn=2$ case, 
pages \pageref{remark1}-\pageref{remark2}.


\section{The KKLMMT model for D-brane inflation} 
\label{inflation}

\begin{figure}[h]
\begin{center}
  \resizebox{12cm}{!}{\psfig{figure=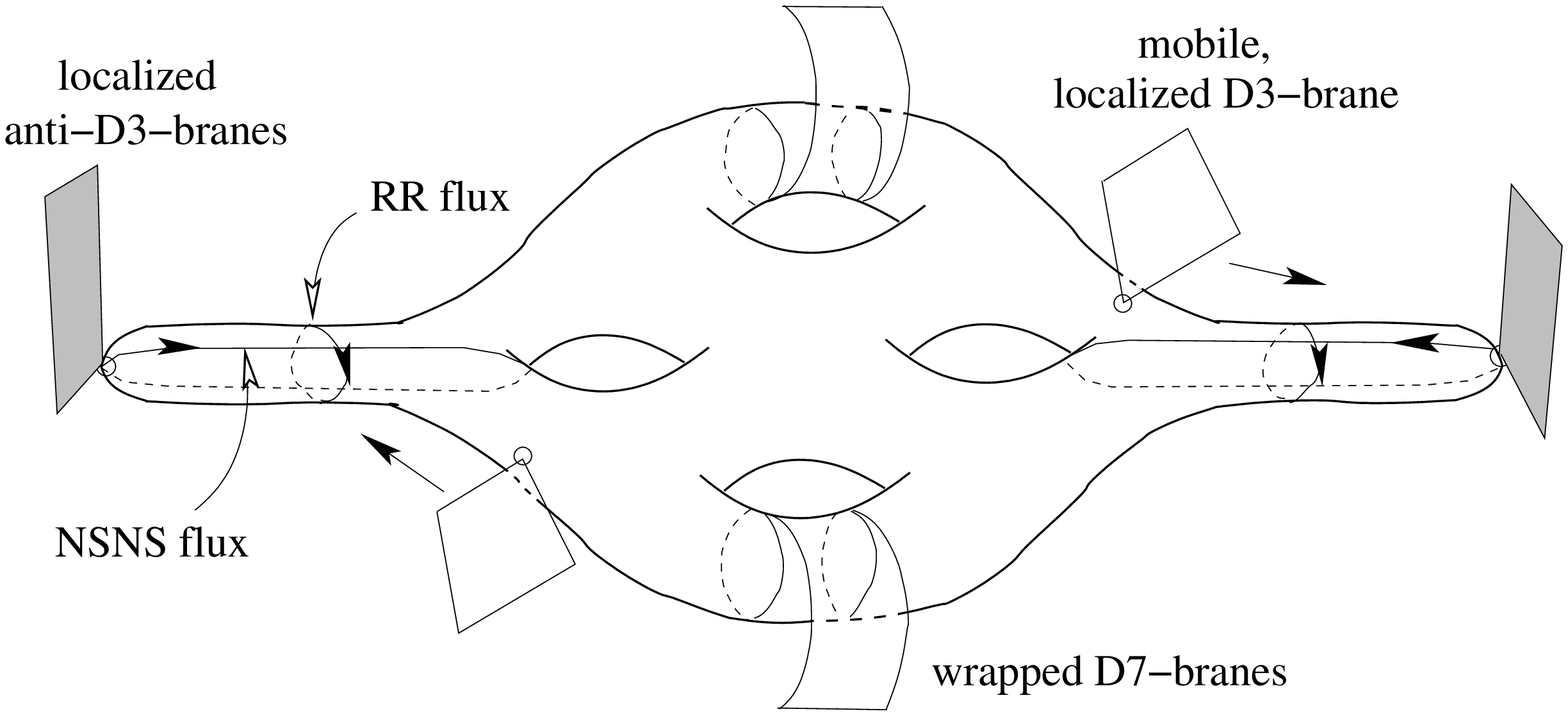,width=12cm}}
\caption{The KKLMMT model} \label{fruehkoelsch}
\end{center}
\end{figure}

This section is a brief review of the basic ingredients 
that go into the string-theoretic models of inflation 
that our results can be applied to. The one we shall be concentrating 
on was introduced 
in \cite{Kachru:2003sx} and is often referred to as the KKLMMT 
model.\footnote{We would like to 
emphasize that our results are important also for any other 
effective theory in which closed-string and 
open-string moduli appear simultaneously. 
Another example, D3/D7-brane inflation, will be 
mentioned later on.} 
It is based on a type IIB compactification on a Calabi-Yau manifold 
that has a discrete symmetry, 
which is quotiented out together with the world-sheet parity $\Omega$,
producing an orientifold (an example with $\mathbb{Z}_2$ symmetry is sketched 
in figure \ref{fruehkoelsch}). 
This is analogous to the
T-dualized $\Omega$-projection applied in \cite{Kachru:2002he}
and allows turning on imaginary self-dual 3-form fluxes despite 
the fact that the fluctuations of the corresponding potentials are 
projected out of the spectrum.\footnote{The effective action of these 
orientifold models has for instance been discussed 
in \cite{Frey:2002hf,D'Auria:2003jk,Berg:2003ri,Camara:2003ku,Grimm:2004uq}. 
Further generalizations 
with non-abelian gauge groups and chiral matter
have also been proposed in \cite{Blumenhagen:2003vr}.} 

The Calabi-Yau manifold has  
deformed conifold singularities, with a deformation parameter
that is fixed by the values of the fluxes 
\cite{Giddings:2001yu}. These fluxes  stabilize not only the 
deformation parameter but also {\it all} the other complex structure 
moduli and the complexified dilaton.\footnote{See \cite{Giryavets:2003vd}
for explicit examples of this stabilization of complex structure moduli 
in Calabi-Yau compactifications with fluxes.} However, the K\"ahler moduli, 
such as the overall volume modulus, remain unfixed by
this flux stabilization, as is manifest in the no-scale 
structure of the resulting 
effective potential. Although the no-scale structure is broken by 
$\alpha'$-corrections \cite{Becker:2002nn}, the 
known $\alpha'$-corrections are not sufficient to argue for 
stabilization of the K\"ahler moduli. In order to fix also 
them, Kachru, Kallosh, Linde and Trivedi 
(KKLT) \cite{Kachru:2003aw} resorted to non-perturbative effects like
Euclidean D3-brane instantons \cite{Witten:1996bn} or gaugino
condensation on wrapped D7-branes in order to generate 
an additional contribution to the superpotential that explicitly 
depends on the K\"ahler moduli. In this way all closed string moduli 
are stabilized, albeit in an AdS minimum. 

Momentarily we will recall how to 
lift this AdS minimum to a dS minimum, but
let us first elaborate a bit more on the non-perturbative superpotential, 
focusing on the version using gaugino condensation. In this case 
it takes the form 
\be \label{supergeneral}
W_{\rm nonpert} \;\sim\; e^{- \alpha f} \; ,
\ee
where $f$ is the D7-brane gauge kinetic function. To leading order 
in string perturbation theory, and 
ignoring the open string moduli $\phi^i$ for the moment, we have
$f=-i\rho$ and hence
\be \label{supergaugino}
W_{\rm nonpert} \;\sim\; C e^{i \alpha \rho}\ , \qquad (\phi^i=0)\ ,
\ee
where the imaginary part of $\rho$ is the volume 
of the 4-cycle that the D7-branes are wrapped around,
measured in the Einstein-frame metric. The constants $C$ and $\alpha$ 
depend on e.g.\ the beta function of the D7-brane gauge theory.
To derive (\ref{supergaugino}), let us assume that 
there is only one K\"ahler modulus $\rho$. Then
the volume of the wrapped 4-cycle is given by the 2/3 power
of the overall six-dimensional volume, 
and the reduction of the DBI action leads to 
\be \label{reducedbi}
\cl_{\rm DBI} ~\sim~ -\frac14 \vol^{2/3}\; \tr \; F_{\mu \nu} F^{\mu \nu} \ ,
\label{Twothreeterm}
\ee
among other terms. From (\ref{Twothreeterm})
one can read off the real part of the gauge kinetic function:
\be \label{realpart}
{\rm Re}\; f \;=\; \vol^{2/3} \;=\; -\frac{i}{2} (\rho - \bar \rho)\ , 
\ee
which can be taken as a defining equation for the imaginary 
part of $\rho$.\footnote{Of course, one has to make sure that this definition 
leads to a viable K\"ahler coordinate on the moduli space. This is
clear from the appendix of \cite{Giddings:2001yu}.}
As $f$ has to be holomorphic in the field variables, it follows that 
$f(\rho)=-i \rho$, which leads to (\ref{supergaugino}). 

This non-perturbative superpotential stabilizes
the volume, but in an AdS minimum.
In order to lift the negative cosmological constant to a positive value, 
several possibilities have been proposed.\footnote{Note that 
non-perturbative 
effects release us from the shackles of the no-go 
theorems \cite{deWit:1986xg} that prohibit 
compactifications with fluxes and/or branes to four 
dimensions with a positive cosmological 
constant.} 
The original KKLT approach \cite{Kachru:2003aw} was to add anti-D3-branes 
at the tip of the deformed conifolds, 
as in figure \ref{fruehkoelsch}. 
This breaks supersymmetry explicitly, so the authors of
\cite{Burgess:2003ic} suggested replacing the effect of the 
anti-D3-branes by a D-term potential due to the introduction
of a background for the 
gauge fields on the world-volume of the D7-branes. 
In this scenario, supersymmetry is only broken spontaneously, 
and $\cn =1$ supersymmetric Lagrangians can be used 
straightforwardly.\footnote{A 
different approach has recently been put forward in 
\cite{Saltman:2004sn}, where the potential energy 
is positive because one expands around a relative dS minimum as 
opposed to an absolute AdS minimum.} 
Here we focus on the original model 
involving anti-D3-branes, but like in 
\cite{Kachru:2003sx,Kachru:2003aw}, we mostly 
ignore their effects except for their contribution to the vacuum
energy.\footnote{Consequences of 
soft supersymmetry breaking 
in effective actions derived from D-brane models (in orientifolds) 
have recently 
been discussed in \cite{Camara:2003ku,Kors:2003wf}.}

It was the idea of KKLMMT \cite{Kachru:2003sx} to study 
brane-anti-brane inflation 
\cite{Dvali:1998pa}\footnote{For a recent review of D-brane cosmology and more references see 
\cite{Quevedo:2002xw}.} 
in the previously described KKLT background,
by adding mobile D3-branes to the D7- and 
anti-D3-branes already present in the KKLT model. This approach solves
one of the generic problems that brane-anti-brane inflation
had struggled with; in a flat geometry, the attractive potential 
between a brane and an anti-brane is too steep to 
allow for slow-roll inflation. Placing the anti-branes at the tip
of the curved-geometry throat as in fig.\ \ref{fruehkoelsch} 
reduces the attractive 
force between the mobile D3- and the anti-D3-branes
by gravitational redshift (due to the warp factor in the metric), 
and the potential can become flat enough to allow for 
slow-roll inflation, at least in principle. 

In practice, the story  
is more complicated due to the ``rho problem'' 
outlined in the introduction, and
this was realized in \cite{Kachru:2003sx}. To decide whether slow-roll
inflation is possible or not,
it is not sufficient to consider only the attractive force between 
the branes and anti-branes,
but one has to take into account the other contributions to the 
potential as well. The potential generated by fluxes and gaugino condensation 
leads to a stabilization of the geometric moduli, but in the presence of 
mobile D3-branes, the K\"ahler modulus $\rho$ that is fixed is 
not the geometric volume, 
rather it is a combination of the volume and the D3-brane scalars. 
As the inflaton field $\varphi$ is supposed 
to be represented by D3-brane scalars,
expanding the potential around the minimum shows that in the KKLMMT model the 
inflaton has a mass that (after canonically normalizing the field) 
is of the order of the Hubble parameter:
\be \label{inflatonmass}
m^2_\varphi = 2 H^2 = \frac23 V_{\rm dS} \; ,
\ee 
with $V_{\rm dS}$ the vacuum energy density at the de Sitter minimum.
This mass is too large to allow for 
slow-roll inflation. Still there is hope; 
this $m^2_{\varphi}$ was derived by neglecting any explicit dependence of the 
superpotential on the open string scalars, and thus on the 
inflaton. It was already indicated in \cite{Kachru:2003sx}
that such a dependence could contribute to the inflaton mass, 
allowing the value (\ref{inflatonmass}) to be lowered. 
It is clear from (\ref{supergeneral}) that the open string scalars 
can enter the superpotential directly, 
i.e.\ $W_{\rm nonpert}=W_{\rm nonpert}(\rho,\phi)$, 
if the gauge kinetic function 
receives corrections depending on them. 
In other words, apart from the dependence
through $f(\rho(\phi))$ that we already argued for,
$W$ could also depend on $\phi$ through additional explicit
dependence $f=f(\rho(\phi),\phi)$.
We will see that such corrections typically do arise at 
the open-string one-loop level.

One of the important lessons here is that the question of 
the inflaton mass cannot be discussed separately 
from the issue of volume stabilization, and that the precise 
form in which the open string scalars 
enter into the definition of $\rho$ can have a large impact 
on the physical outcome.  
This precise form can be determined by direct computation,
as we will discuss later, but let us now briefly review
general arguments why
such a dependence is expected at all, following \cite{Kachru:2003sx}.  
It was conjectured in 
\cite{DeWolfe:2002nn} that the K\"ahler potential for the
volume modulus $\rho$ in the presence of D3-brane scalars
$\phi^i, \ i=1,2,3$, should be modified to 
\be \label{kmod}
K(\rho, \bar \rho, \phi, \bar \phi) 
\; = \; -3 \ln(-i(\rho - \bar \rho) + 
k(\phi,\bar \phi))\ ,
\label{DG}
\ee
where $k(\phi,\bar \phi)$ is the (so far unknown) K\"ahler potential 
of the metric on the 
Calabi-Yau manifold. This leads to a 
kinetic term for the 3-brane scalars of the form
\be \label{d3kin}
\cl \; \sim \; \frac{k_{i \bar \jmath}\, \partial_\mu \phi^i \partial^\mu 
\bar \phi^{\bar \jmath}}{-i(\rho - \bar \rho) + k(\phi,\bar \phi)} + \ldots \ ,
\ee
where $k_{i \bar \jmath}$ is the derivative
of $k(\phi,\bar \phi)$ with respect to $\phi^i$ and 
$\bar{\phi}^{\bar \jmath}$, and
the dots include further contributions to the kinetic terms 
of the 3-brane scalars involving single derivatives of 
$k(\phi,\bar \phi)$. Comparing this to the kinetic term 
that one would infer from considering the 
DBI action of a D3-brane transverse to the Calabi-Yau, i.e.\
\be
\cl \; \sim \; \frac{k_{i \bar \jmath} \, \partial_\mu \phi^i \partial^\mu 
\bar \phi^{\bar \jmath}}{\vol^{2/3}} + \ldots\ , 
\ee
where $\vol$ is the volume of the Calabi-Yau as measured in the 
Einstein frame, suggests the identification
\be
-i(\rho - \bar \rho) 
\; = \; \vol^{2/3} - k(\phi,\bar \phi)\ ,
\label{identif}
\ee
so that the imaginary part of the K\"ahler modulus is indeed
a mixture of the geometric volume and a function of the 
3-brane scalars. Thus we see that the form (\ref{identif}) follows 
from the conjectured form (\ref{DG}).\footnote{Another argument 
comes from the analogy to the heterotic string. In 
compactifications on either a Calabi-Yau manifold or a torus,
the K\"ahler moduli are corrected in the presence of 
Wilson line moduli, the heterotic analogs of 
open-string scalars \cite{Witten:1985xb,LopesCardoso:1994is}.}

The identification (\ref{identif}) 
leads to an intriguing puzzle \cite{Hsu:2003cy,renatastalk}. 
Comparing with (\ref{reducedbi}) shows that the gauge coupling of the
D7-brane gauge group as derived from a reduction of the DBI action
is not the real part of a holomorphic function in the corrected $\rho$.
More precisely, in (\ref{reducedbi}) a term of the type 
$k(\phi,\bar \phi)\, \tr (F_{\mu \nu} F^{\mu \nu})$ is missing to complete 
the imaginary part of the modified $\rho$.  
The restoration of 
the holomorphy of the gauge kinetic 
functions is the ``rho problem'' described in the
introduction. As we will argue in the following and as already anticipated 
in the introduction, this missing term should arise at 
open string one-loop level. The computation we will present  
confirms the conjectured form of the 
corrected $\rho$, making it compatible with supersymmetry, 
and thus solves the puzzle (at least in our simplified setting,
but as mentioned, we expect this to be true more generally). 
Moreover, additional terms 
depending on the open string scalars arise at this order.

The reasons why the corrections 
must arise at open-string one-loop level are easy to state.
First, the missing term does not involve the 
ten-dimensional dilaton, whereas the  term
from reduction of the DBI 
action has a factor $e^{-\Phi_{10}}$, i.e.\
it is open string tree-level.\footnote{Note that 
this $e^{-\Phi_{10}}$ is implicit in (\ref{reducedbi}) and 
(\ref{realpart}) but appears when we
ex\-press the volume in the string frame.} 
Therefore, the missing term comes with a power of the
dilaton appropriate for string diagrams of Euler characteristic zero. 
Second, for several coincident D-branes the scalars $\phi^i$ 
would carry a representation of the 
world-volume gauge group, and non-abelian versions of the missing term,
e.g.\ $\tr (k(\phi,\bar \phi))\, \tr(F_{\mu \nu} F^{\mu \nu})$,
would involve at least two traces over gauge indices.
This requires an open-string diagram with two boundaries,
but there is no such diagram at tree-level.

We conclude this review section with a few remarks
on D3/D7-inflation;
the rho problem arises also in that context. 
There, the idea is to consider a system of D3- and D7-branes
that have four non-compact directions in common. In the absence of 
world-volume fluxes, this system is supersymmetric and the 
distance between the D3- and D7-branes along the two directions transverse
to the D7-branes is a massless modulus. Turning on 
a non-self-dual magnetic background flux for the gauge fields on the 
D7-branes breaks supersymmetry, and leads to an attraction between the
D3- and D7-branes \cite{Dasgupta:2002ew}. In other words, the 
scalar parametrizing the distance feels a potential,
and this potential turns out to be flat enough to allow for slow-roll 
inflation with the distance scalar as the inflaton field. The 
rho problem then arises in this 
string-theoretic model of inflation,
just like in the KKLMMT model.


\section{One-loop corrections to the volume modulus} 
\label{technicalstuff}

In this section we are going to compute the dependence 
of D3- and D7-brane gauge kinetic functions 
on the open string scalars by determining the complete one-loop 
correction, which also involves terms depending on the 
background complex 
structure moduli. As explained in the previous section, this produces an 
explicit expression for how 
the non-perturbative superpotential  
due to gaugino condensation depends on the open string scalars,
and addresses the rho problem and the inflaton mass problem.

\begin{figure}[h]
\begin{center}
  \resizebox{12cm}{!}{\psfig{figure=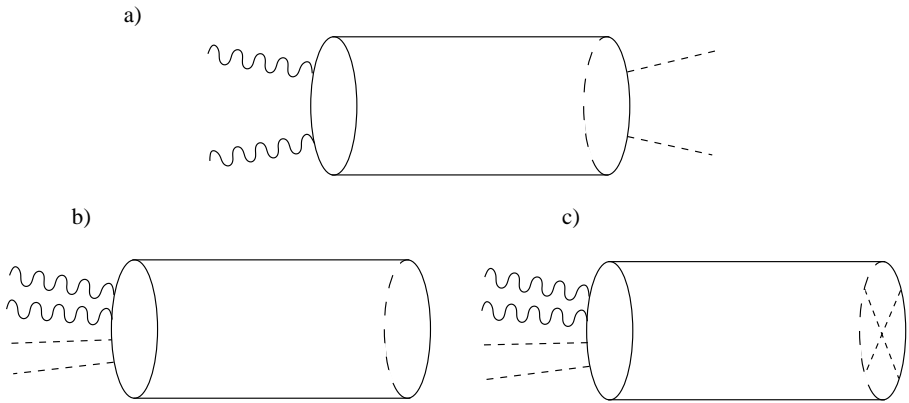,width=12cm}}
\caption{D-brane gauge kinetic term corrections.
a) Non-planar annulus amplitude, b) Planar annulus amplitude, 
c) M\"obius amplitude.} \label{amplitudes}
\end{center}
\end{figure}

One could in principle obtain the desired renormalization
from string amplitudes with vertex operator insertions,
depicted in figure \ref{amplitudes}.  
The wiggly lines are vector insertions, and
the dashed lines are (open string) scalar insertions.
For the KKLMMT model, diagram $a)$ with one end on 
D5-branes and the other on D9-branes would be sufficient
to determine the renormalization of the D5-brane gauge kinetic function
and its dependence on the D9-brane Wilson lines 
to quadratic order (recall from page \pageref{tdual} that
we work in the D9/D5-picture).
For the other dependences we are interested in, 
computation of diagrams $b)$ and $c)$
would also be necessary. \\ 

Here, however, we will use a convenient shortcut to gauge 
coupling renormalization: the background field method.


\subsection{The background field method} 
\label{background}

The background field method was introduced in \cite{Bachas:bh} and 
used there and in \cite{Bachas:1996zt,AnBaDu99} to calculate 
threshold corrections to gauge coupling constants in type IIB 
orientifolds. The basic idea is to study how the one-loop vacuum 
energy is deformed by
the presence of a constant background gauge field strength, where
the background field is turned on in a 
$U(1)$ subgroup of the gauge group factor of interest.
The deformed vacuum energy is straightforward to calculate,
since the background field only modifies the boundary 
conditions; the worldsheet conformal field theory is still free.
By expanding the deformed vacuum energy for small background
fields, one can extract the zero-momentum limits
of the string amplitudes in fig.\ \ref{amplitudes}.
In models with both 
D5- and D9-branes, one can consider a background 
in either the D5- or D9-brane gauge group, or both. 
Here we will consider the most 
general case, in which we turn on constant background gauge field strengths 
in both gauge groups at the same time, denoted by ${\cal F}_i$ and 
${\cal F}_a$,
respectively. The indices $i$ and $a$ 
enumerate gauge group factors in the D9- and D5-brane gauge groups.
In components, the gauge field background reads 
\beqn 
A_\mu = {\cal F}_{\mu\nu} x^\nu \ , \qquad {\rm only\; \, e.g.}\ 
\; {\cal F}_{23}\not=0\ . 
\eeqn
The expressions for the relevant one-loop diagrams are 
then expanded to quadratic order in 
the field strengths ${\cal F}_i$ or ${\cal F}_a$ around 
${\cal F}_i={\cal F}_a=0$, which yields
the correction to the gauge coupling.  
In principle one could limit oneself to turn on only one type of
background field and invoke T-duality to infer the corresponding terms for the 
other gauge groups. However, since we want to consider the effect of 
both types of Wilson lines (denoted by $\vec a_i$ and $\vec a_a$) 
on both types of gauge couplings $g_i$ and $g_a$, we 
turn on both types of gauge fields concurrently.\footnote{More precisely, 
for non-abelian gauge groups, it is in fact sufficient 
to turn on only one of the gauge fields and both types of Wilson lines, 
but for abelian gauge groups
it is not; one would miss cross-term corrections such as eq.\ (\ref{cross}).}

Let us concretize these introductory words in three schematic formulas.
The one-loop vacuum energy in type I theory receives four
contributions, from the torus, Klein bottle, M\"obius strip and 
annulus diagrams,\footnote{Note that 
we have moved the factor of $1/2$ 
that appears explicitly in $\Lambda_{\rm 1-loop}({\cal F},\vec a)$
in much of the literature, e.g.\ in \cite{AnBaDu99},
to our definition of the amplitudes, eq.\ (\ref{loopch}).}
\be \label{oneloop}
\Lambda_{\rm 1-loop}({\cal F},\vec a)  = 
{\cal T} + {\cal K} + {\cal M}({\cal F},\vec a) 
+ {\cal A}({\cal F},\vec a)\ ,
\ee
where we made explicit the fact
that only diagrams with boundaries can have insertions
of background gauge 
fields and Wilson lines. If one expands the vacuum energy
to second order in the back\-ground field, 
the coefficient of the 
quadratic term  
directly gives the one-loop threshold correction to the gauge 
group for which the background was turned on. Omitting indices 
(that enumerate the gauge group factors) for backgrounds and Wilson lines, 
the expansion schematically looks like
\be \label{expand}
\Lambda_{\rm 1-loop}({\cal F},\vec a) = \Lambda^{(0)} + \frac12 
\left(\frac{{\cal F}}{2 \pi}\right)^2
\Lambda^{(2)}(\vec a) + \ldots \ ,
\ee
where $\Lambda^{(0)}$ is the one-loop induced cosmological constant,
and the one-loop corrected gauge coupling 
can be identified as
\be \label{couplingcorrect}
\left.\frac{4\pi^2}{g^2}\right|_{\rm 1-loop} = 
\left.\frac{4\pi^2}{g^2}\right|_{\rm tree} 
+ \frac{1}{\sqrt{-g_4}}\; \Lambda^{(2)}(\vec a)\ .
\ee
Here ``tree'' signifies
open string tree-level, which in principle means disk diagram, although
in practice the gauge kinetic term is of course easier to obtain
by dimensional reduction of the DBI action. 
Unfortunately, the type I literature is littered with pitfalls 
when it comes to the precise meaning of ``tree-level'',
so this would be an appropriate place to elaborate
on this issue.


\subsection{Tree-level effective action of type I on $\mathbb{T}^2\times$ K3} 
\label{tree}

Before we present the relevant loop calculations 
in section \ref{seconeloop}, let us collect 
some results on the 
known ``tree-level'' supergravity action for the model considered in
that section: the orientifold $\mathbb{T}^2\times\mathbb{T}^4/\mathbb{Z}_2$.

This orientifold is a special case of a type I compactification on 
$\mathbb{T}^2\times$ K3, with the K3 at a particular 
orbifold point.
We are interested in computing 
corrections to the gauge 
kinetic terms of the 
Yang-Mills theories supported by D9- and D5-branes.
The reader may rightly wonder why we begin by considering an $\cn =2$
orientifold when we are interested in $\cn=1$
models; the detailed reason is explained at the beginning of section
\ref{none}, but the basic idea is that the {\it relevant part} of the
full result in $\cn=1$ models is very similar to the result in this
$\cn=2$ model, and the latter is simpler. 

We have already bemoaned the fact that some of the literature on 
Kaluza-Klein (KK) reduction of type I is not precise in the usage 
of the term ``tree-level''. More concretely, it
secretly incorporates various 
terms that really only arise at 
string one-loop level, and one may wonder whether 
including some terms but excluding others is
consistent from the point of view of string perturbation 
theory --- hence our quotation marks on ``tree-level''.
This potentially confusing situation 
comes about because part of the relevant literature concerns
heterotic-type I duality, and certain terms are loop corrections on
the type I side but tree-level on the heterotic side.
Here we use type I terminology exclusively.
 
Some general aspects of compactifications of type I strings on 
$\mathbb{T}^2\times$ K3 have been discussed
in \cite{Antoniadis:1996vw}, and we review the relevant results here.
However, since the explicit factors  will turn out  to be
important for our conclusions later on, we redo some of their 
analysis and adapt it to our conventions. 

The closed string spectrum contains hypermultiplets and vector
multiplets. The hypermultiplets, which will not be of great concern
here, consist of the geometric moduli 
of the K3, moduli from an 
expansion of the antisymmetric 
tensors into the harmonic forms of the K3, and the six-dimensional 
dilaton. More important for our purposes are the vector 
multiplets. There are $3+N_9+N_5$ of them, where $N_9$ and $N_5$ 
denote the number of vector fields from the open string sector of the 
9- and 5-branes, respectively. The three additional vector multiplets arise in 
the closed string sector. There are four KK vectors from
the metric and the antisymmetric 2-form due to the 
presence of the two 1-forms of the torus. One of them, the 
graviphoton, resides in the supergravity multiplet and the 
other three are contained in the three closed string vector multiplets. 
Their scalar components are given as follows. Let 
\be \label{torusmetric}
G = \frac{\sqrt{G}}{{\rm Im} (U)} \left( 
  \begin{array}{cc}
  1 & {\rm Re}(U) \\
  {\rm Re}(U) & |U|^2 
  \end{array}
\right) 
\ee
be the metric of the torus in string 
frame.\footnote{We
use the same letter $G$ for both the torus metric and its determinant, 
but the determinant 
always occurs in the form $\sqrt{G}$, so no confusion should arise.} 
Then $U=(G_{45} + i \sqrt{G})/G_{44}$ 
is its complex structure modulus, which belongs to one of the 
closed string vector multiplets. In the absence of Wilson line moduli 
the other two scalars are given by 
\be \label{ssprime}
S = \frac{1}{2 \pi \sqrt{2}} \left(b + i e^{-\Phi_{10}}  
\vols_{\rm K3} \sqrt{G} \alpha'^{-3} \right)\ , \quad 
S\, ' = \frac{1}{2 \pi \sqrt{2}} \left(B_{45} + i e^{-\Phi_{10}} 
\sqrt{G} \alpha'^{-1} \right)\ ,
\ee
where $b$ is the scalar dual to $B_{\mu\nu}$, $\Phi_{10}$ is the 
ten-dimensional dilaton and $\vols_{\rm K3}$ denotes 
the volume of the K3-manifold measured in the string frame metric. 
Moreover, we keep $\alpha'$ explicit in this section 
to have better control over numerical factors.
The three scalars $U,S,S\, '$ span the moduli space $(SU(1,1)/U(1))^3$ with 
prepotential ${\cal F}^{(0)} = SS\, 'U$. 

We included an extra factor $1/(2 \pi \sqrt{2})$ in 
(\ref{ssprime}) as compared to the definition of \cite{Antoniadis:1996vw}
because we want the relations $g^{-2}_{(9)} = {\rm Im}(S)$ and 
$g^{-2}_{(5)} = {\rm Im}(S\, ')$ to hold. 
Let us check this explicitly for the case of $S\, '$,
by reducing the 5-brane DBI action on the 
torus. To this end, we start with the 
standard expression for the DBI action 
\be \label{dbi5}
T_5 \int \!\! d^6\xi\ e^{-\Phi_{10}} \sqrt{{\rm det}
(-g_{6}+ 2\pi\alpha' F_{(5)})} ~=~ 
- \frac14 T_5 (2 \pi \alpha')^2 \int \!\! d^6 \xi \sqrt{-g_6} \, 
e^{-\Phi_{10}}\, {\rm tr} \, F_{(5)}^2+ \ \cdots \ ,
\ee
where $T_5 = (1/\sqrt{2}) 2 \pi (4 \pi^2 \alpha')^{-3}$. The 
factor of $1/\sqrt{2}$ in $T_5$ arises in type I theory and 
is absent in type IIB, see e.g.\ \cite{Bachas:1998rg}.  
Reducing (\ref{dbi5}) on a torus of volume $(2 \pi)^2 \sqrt{G}$ 
leads to a gauge coupling 
\be \label{gaugec}
g^{-2}_{(5)} = \frac{1}{2 \pi \sqrt{2}} \, e^{-\Phi_{10}} 
\sqrt{G} \alpha'^{-1} = {\rm Im}(S\, ') \ ,
\ee
where we choose the normalization such that
$-\frac14 g^{-2} \int d^4 x \sqrt{-g_4} \; {\rm tr} \, F^2$ 
is the kinetic term for vector fields in four dimensions.

Including  
Wilson line moduli of the open string vector fields, i.e.\ 
components of the (higher-dimensional) vectors along the 
$\mathbb{T}^2$, leads to a modification of the
expression (\ref{ssprime}) for the scalars 
$S$ and $S\, '$,
cf.\ \cite{LopesCardoso:1994is,Antoniadis:1996vw}. 
Explicitly, we parametrize the Wilson line moduli by a 2-vector 
$\vec a = (a_4,a_5)$ with an 
index $i$ or $a$ for the stack of D9- or D5-branes respectively,
where $a_4$ and $a_5$ are related to the internal components 
$V_4$ and $V_5$
of the corresponding higher-dimensional vectors
as $(a_{4},a_5) = \sqrt{\alpha'} (V_{4},V_5)$. 
To be precise, the $\vec a$ are defined as components with respect 
to the basis of the dual lattice 
$(\vec e^{\, 4},\vec e^{\, 5})$. 
The original compactification lattice is then
$(\vec e_4,\vec e_5)$ with conventions  such that  
\beqn \label{lattice}
(a_I \vec e^{\, I})\cdot( a_J \vec e^{\, J}) = G^{IJ} a_I a_J \ , \quad 
\vec e^{\, I} \cdot \vec e_J = \delta^I_J \; , \qquad I,J\in \{4,5\} \; ,
\eeqn 
where $G^{IJ}$ is the inverse of the metric (\ref{torusmetric}). 
Note that \cite{Antoniadis:1996vw}
considers the case when the gauge group is broken to the abelian
subgroup, so for notational simplicity
we restrict to that case in this section, but the string computations
in later sections will be performed also for the non-abelian case. 
Adopting to
our conventions, the modified scalars are given by\footnote{In 
\cite{Antoniadis:1996vw} it was shown that there 
appears a further correction to ${\rm Im}(S)$, given by 
${\rm Im}(S) \rightarrow {\rm Im}(S) + 
\sqrt{G}\, \delta/(2 \, {\rm Im}\, (S\, '))$, 
where $\delta$ is the correction to the Einstein-Hilbert 
term arising at open string one-loop level. This 
redefinition is even higher order in an 
expansion in $e^{\Phi_{10}}$ and we will ignore it in the following. 
It is, however, important to establish the duality to the 
heterotic string \cite{Antoniadis:1996vw}.}
\be \label{ss'}
S = S|_{A=0} + \frac{1}{8 \pi} \sum_a a^a_4 A_a \ , \quad 
S\, ' = S\, '|_{A=0} + \frac{1}{8 \pi} \sum_i a^i_4 A_i\ . 
\ee
Here we used the complex Wilson line modulus
\be \label{newa}
A = U a_4 - a_5\ ,
\ee
which is the complex
combination that makes the metric of the four-dimensional
scalar manifold manifestly K\"ahler.
The modifications (\ref{ss'}) 
of the scalars $S$ and $S\, '$ are not supposed
to be obvious; one way to see that they must be modified is to
consider the reduction of the kinetic term of the 3-form RR field
strength in six dimensions. This term includes a Chern-Simons
correction in the presence of open string fields.

In fact, we can fix the relative factor 
$1/8\pi$ between the leading term and the 
one-loop correction in (\ref{ss'})  by inspection
of  precisely this Chern-Simons-corrected kinetic term.
It can be reduced on the torus according to 
\be
dB - \frac{\kappa_{10}^2}{g_{10}^2} \, \omega_3 \quad 
\stackrel{\mathbb{T}^2}{\longrightarrow} \quad 
\partial_\mu B_{45} - \frac{1}{2\sqrt{2}} \sum_i
\Big((\partial_\mu a^i_4) a_5^i - (\partial_\mu a^i_5) a_4^i\Big) \ ,
\ee
where $\omega_3$ is the standard Yang-Mills Chern-Simons 3-form.  
We used the type I relation $g_{10}^2 \kappa_{10}^{-1} = 
2 (2 \pi)^{7/2} \alpha'$ and 
$\kappa_{10}^2 = \frac12 (2 \pi)^7 \alpha'^4$, 
cf.\ \cite{Polchinski:rr}. This leads to 
$\kappa_{10}^2/g_{10}^2 =
\alpha' / (2 \sqrt{2})$ and we absorbed 
$\alpha'$ in the definition of 
the fields $\vec a_i$ as explained above (\ref{lattice}). 
Thus the relative factor between ${\rm Re}(S\, ')$ and 
$\sum_i a_4^i \! \stackrel{\leftrightarrow}{\partial}\! a^i_5$
contains an additional $1/\sqrt{2}$ as compared to 
(3.6) of \cite{Antoniadis:1996vw}. Taking into account the
overall factor introduced in (\ref{ssprime}), we arrive 
at the $1/8\pi$ factor in the modification of  $S\, '$ given in
(\ref{ss'}). A similar argument should hold for
the modification of $S$. 

The full moduli space of the vector multiplets 
was identified in \cite{Angelantonj:2003zx} to 
be a space called 
$L(0,N_5,N_9)$ in \cite{deWit:1991nm}, which is
homogeneous but not symmetric. 
The corresponding K\"ahler potential is determined
by a holomorphic prepotential that was derived in 
\cite{Antoniadis:1996vw}. An explicit KK-reduction 
of the ten-dimensional type I action (including the 9-brane vector 
fields but not those from the 5-branes) leads to 
${\cal F}^{(0)}=S (S\, ' U - \frac{1}{8\pi} \sum_i A_i^2)$, where we
adjusted the formula of \cite{Antoniadis:1996vw} to 
our conventions. To derive the 
form of the prepotential including the 5-brane vectors, one 
first has to compactify to six dimensions and add in 
the kinetic term for the 5-brane vectors (\ref{dbi5}) 
and the Chern-Simons term that is needed to cancel 
anomalies \cite{Berkooz:1996iz}
\be
\cl_{\rm CS}\ d{\rm vol} ~\sim~ - B \wedge F_{(5)} \wedge F_{(5)} \ , 
\ee
where $d{\rm vol}$ is the six-dimensional volume form. 
Taking these terms into account in a further 
reduction to four dimensions leads to what is called 
the ``tree-level'' prepotential in  
\cite{Antoniadis:1996vw}, i.e.\footnote{To be more precise, 
one has to take into account additional counterterms in
the derivation, cf.\ 
\cite{Antoniadis:1996vw,D'Auria:2004cu}.}
\be \label{treeprepot}
\cf^{(0)}=S S\, ' U - \frac{1}{8\pi} S \sum_i A_i^2 - \frac{1}{8\pi} 
S\, ' \sum_a A_a^2\ .
\ee
We will come back to this prepotential in section \ref{ac}.

The important bottom line of this section is that the 
geometric closed string moduli are corrected 
in the presence of open strings and D-branes, cf.\ (\ref{ss'}).
As discussed in the 
previous section, we want to show
that the gauge kinetic functions are holomorphic in 
the {\it corrected} closed string moduli fields $S,S\, '$, or more precisely,
in their $\cn =1$ analogs. We will see
that, in the $\cn=2$ case, the correction terms present 
in (\ref{ss'}) arise as open string one-loop contributions to the 
tree-level result (\ref{gaugec}).


\subsection{One-loop threshold corrections in ${\cal N}=2$}
\label{seconeloop}

Let us now specialize to the orbifold limit of 
K3 by considering the $\mathbb{T}^2\times\mathbb{T}^4/\mathbb{Z}_2$
orientifold. (The 
parts of the following 
computation that can be considered standard
are collected in appendix \ref{appampl}.)
The orientifold group 
is generated by $\{\Omega,\Theta\}$, where
$\Omega$ is world-sheet parity and $\Theta$ is a
reflection along the $\mathbb{T}^4$. 
Threshold corrections to gauge couplings in this model 
were studied in \cite{Bachas:1996zt}, 
for the case where the
background field and Wilson lines are turned on only on
D9-branes. 
It is true that this situation is T-dual to the case with 
background field and Wilson lines on D5-branes, but 
as we already pointed out, 
mixing between D9-brane and D5-brane gauge groups (that may occur for 
$U(1)$ group factors) cannot be obtained by T-dualizing
the results in \cite{Bachas:1996zt},
and the same holds for any dependence of the D9-brane coupling on the 
D5-brane Wilson lines and vice versa; this dependence is
crucial for the  application we are interested in. 

We first summarize a few important features of the
$\mathbb{T}^2\times\mathbb{T}^4/\mathbb{Z}_2$ orientifold.
Tadpole conditions (cancellation of RR charge)
imply the presence of 32 units of D9- and 32 units of D5-brane charge, 
with maximal gauge group 
$U(16)_{{\rm D}9}\times U(16)_{{\rm D}5}$. 
The one-loop vacuum energy (\ref{oneloop})  becomes
\beqn \label{oneloop2}
\Lambda_{\rm 1-loop} &=& 
{\cal T} + {\cal K} + 
{\cal M}_{9} + {\cal M}_{5} + 
{\cal A}_{99} + {\cal A}_{55} + {\cal A}_{95} + {\cal A}_{59}\ . 
\nonumber 
\eeqn
We use labels $i$ and $a$ for stacks of D9$_i$- and 
D5$_a$-branes, and assume that the maximal gauge group is broken to a subgroup 
\beqn \label{gauge}
G = \bigotimes_i U(N_{i}) \times \bigotimes_a U(N_{a}) \ , \quad 
\sum_i N_i = \sum_a N_a = 16 \ , 
\eeqn
through the presence of Wilson lines along $\mathbb{T}^2$, 
denoted by $\vec a_i$ or $\vec a_a$. 
For example, one may want to consider 
breaking to the abelian subgroup $N_i=N_a=1$;
we will consider both  abelian and non-abelian groups. 
The overall $U(1)$ factors in the two $U(16)$'s are actually
anomalous, and will 
decouple from the low energy 
theory \cite{Berkooz:1996iz}. Given a
configuration of branes, a 
background gauge field strength 
${\cal F}_i$ or ${\cal F}_a$ can be turned on on any individual stack 
of branes. 
Each stack is represented in the CFT by a boundary state
\beqn 
| {\rm D9}_i , IJ \rangle = | {\rm D9}_i ( {\cal F}_i , \vec a_i ) , 
IJ \rangle \ , \quad  
| {\rm D5}_a , IJ \rangle = | {\rm D5}_a ( {\cal F}_a , \vec a_a ) , 
IJ \rangle \ , 
\eeqn
with Chan-Paton (CP) indices $I,J$. The explicit form of 
the boundary states is standard (for a review see \cite{Gaberdiel:2000jr}), 
but will not be needed here. 
The elements of the orientifold group act on the boundary 
states e.g.\ by 
\beqn 
\Omega | {\rm D9}_i , IJ \rangle &=& 
(\gamma_{\Omega i})_{IK} | \Omega \cdot {\rm D9}_i , LK \rangle (\gamma^{-1}_{\Omega i})_{LJ} \ ,  
\non
\Theta | {\rm D9}_i , IJ \rangle &=& 
(\gamma_{\Theta i})_{IK} | \Theta \cdot {\rm D9}_i , KL \rangle (\gamma^{-1}_{\Theta i})_{LJ} \ ,  
\eeqn
where $\Omega \cdot {\rm D9}_i$ and 
$\Theta \cdot {\rm D9}_i$ schematically 
represent the action on the string world-sheet fields. 
The unitary matrices $\gamma$ summarize the action on the gauge bundle. 
In this notation it is evident that 
a $32\times 32$ matrix $\gamma$ labeled by $i$ or $a$ only acts on the 
respective stack, with all other entries vanishing, and 
\beqn 
\gamma_{\Omega 9} = \sum_i \gamma_{\Omega i}\ , \quad 
\gamma_{\Theta 9} = \sum_i \gamma_{\Theta i}\ , 
\eeqn
and so on.  
The solution of the tadpole constraints fixes the only non-vanishing blocks 
to the form given in (\ref{gamma1}), (\ref{gammaomega}) 
and (\ref{gamma2}) in the appendix.\footnote{We actually 
use the 
solution presented e.g.\ in \cite{AnBaDu99}, 
not that of the original literature \cite{Gimon:1996rq}. The former has the 
advantage that $\gamma_{\Theta i}$ is diagonal.}
In this $32\times 32$ matrix formulation, the original
32 D-branes of either type are pairwise 
related under $\Omega$, breaking $U(32)$ to $SO(32)$ as in type I, and  
further subjected to the $\Theta$-projection. This breaks the gauge group to  
$U(16)$, 
 without further rank reduction. Therefore, 
the $32+32$ Wilson lines in the Cartan subalgebra
are really $16+16$ independent pairs. T-dualizing to D3- and D7-branes 
localized on the 2-torus, 
this means the $32+32$ branes can be moved pairwise out of the 
fixed locus of the T-dual $\Omega$-projection $\Omega R (-1)^{F_L}$, 
$R$ being a reflection of all six internal coordinates, 
and $F_L$ the left-moving space-time 
fermion operator. Apart from breaking the gauge group,
the Wilson lines have the effect of introducing 
shifts $\vec a_i$ and $\vec a_a$ in the spectrum of KK states. 


\subsubsection{Couplings of non-abelian gauge groups}

We will first determine threshold corrections to 
non-abelian gauge group factors, i.e.\ to 
 $SU(N)$ groups, postponing the discussion of 
$U(1)$ factors to the next subsection.

To describe the embedding of the Wilson line in the gauge group 
we introduce charge matrices of the form
\beqn \label{wi}
W_i = {\rm diag}( \underbrace{0,\, ...\, ,0}_{p_i\ {\rm entries}}, 
{\bf 1}_{N_i},-{\bf 1}_{N_i},0,\, ...,0) 
\eeqn 
with non-vanishing entries in the block of the $i$-th factor of the gauge 
group, and similarly for $W_a$.  
The two factors of the gauge group (\ref{gauge}) are just by definition 
the subgroups of the two
$U(16)$ that commute with all $W_i$ or $W_a$. 
For example, the $W_i$ belong to $U(1)_i$, the 
overall factor in $U(N_i)=U(1)_i\times SU(N_i)$. 
The Wilson lines take their values in
these $U(1)$ factors. 

To specify the background gauge fields in some direction of 
the $SU(N_i)$ subgroup, we can choose the matrices 
\beqn \label{qi}
Q_i = \frac12 {\rm diag}(  \underbrace{0,\, ...\, ,0}_{p_i}, 
                   \underbrace{1,-1,0,\, ...,0}_{N_i}, 
                   \underbrace{-1,1,0,\, ...,0}_{N_i}, 
                   \underbrace{0,\, ...,0}_{32-p_i-2N_i})  
\eeqn 
with just four non-vanishing entries. The first $1,-1$ pair
occurs at the position of the  
first non-vanishing 
block ${\bf 1}_{N_i}$ in $W_i$, the second pair
at the position of the block $-{\bf 1}_{N_i}$ 
in $W_i$. 
Together, these matrices specify the background in $32\times 32$ 
matrix notation, and 
the matrix valued gauge field strengths and Wilson lines are 
$Q_i {\cal F}_i$ and $W_i \vec a_i$, etc. 
To keep the expressions for the annulus diagrams compact, 
we introduce the following notation for the background fields:
\beqn \label{fa}
{\bf F}_{ij} &=& (Q_i{\cal F}_i \otimes {\bf 1}_{32}) \oplus 
 ({\bf 1}_{32} \otimes (-Q_j{\cal F}_j)) \ , \non
{\bf \vec A}_{ij} &=& (W_i\vec a_i \otimes {\bf 1}_{32}) \oplus ({\bf 1}_{32} \otimes (-W_j \vec a_j)) \ , 
\eeqn
and similarly for 
$({\bf F}_{ab},{\bf \vec A}_{ab})$, $({\bf F}_{ai},{\bf \vec A}_{ai})$. 
The background is now tensor-valued, with one factor for 
each end of the  open string in question. 
The matrices $\gamma$ are then also tensor-valued:
$\bfg_i = \gamma_i \otimes {\bf 1}_{32}$ or 
$\bfg_j = {\bf 1}_{32} \otimes \gamma_j$ etc., depending on whether
the matrix $\gamma$ acts on the left or the right end of the string. 
The trace on CP indices is defined as the product of the traces 
on both ends, e.g.\ 
\beqn \label{example}
{\rm tr} \left( \bfg_{\Theta i}\bfg_{\Theta j} {\bf F}_{ij}^2 \right) &=& 
{\rm tr} \Big( ( \gamma_{\Theta i} \otimes {\bf 1}_{32}) ({\bf 1}_{32} \otimes 
\gamma_{\Theta j} ) 
 \Big[( (Q_i{\cal F}_i)^2 \otimes {\bf 1}_{32} ) 
\non
&&  \hspace{1cm}
\oplus 2 ( (Q_i{\cal F}_i) \otimes (-Q_j{\cal F}_j) ) \oplus 
 ( {\bf 1}_{32} \otimes  (-Q_j{\cal F}_j)^2 ) 
\Big] \Big)
\non
&=& {\cal F}_i^2 {\rm tr}(\gamma_{\Theta i}Q_i^2){\rm tr}(\gamma_{\Theta j}) 
- 2 {\cal F}_i {\cal F}_j {\rm tr}(\gamma_{\Theta i}Q_i){\rm tr}
 (\gamma_{\Theta j}Q_j) \non
& & \mbox{} + {\cal F}_j^2 {\rm tr}(\gamma_{\Theta i}){\rm tr}(\gamma_{\Theta j}Q_j^2) \non
&=& 0 
 \ ,
\eeqn
where for the last
equality we used (\ref{qi}) and (\ref{gamma2}). 
More generally, one has
\beqn \label{traces}
&&
{\rm tr}(\gamma_{\Theta i}Q^{2n-1}_i W_i^n)= {\rm tr}(\gamma_{\Theta i}
Q^{2n}_i W_i^{2n})= 
{\rm tr}(\gamma_{i}Q^{2n-1}_i W_i^n )= {\rm tr}(\gamma_{i}Q^{2n}_i 
W_i^{2n-1})= 0 \ , 
\non
&&  
{\rm tr}(\gamma_{\Theta i} Q^{2n}_iW_i^{2n-1}) = \frac{4i}{2^{2n}}\ , \quad 
{\rm tr}(\gamma_i Q^{2n}_iW_i^{2n}) = \frac{4}{2^{2n}} \ , 
\non
&&
{\rm tr}(\gamma_{\Theta i} W_i^{2n-1}) = 2i N_i\ , \quad 
{\rm tr}(\gamma_i W_i^{2n}) = 2N_i \ .
\eeqn
For the M\"obius strip there is only one boundary, so we
write
\be
{\bf F}_i = Q_i{\cal F}_i\ , \quad
{\bf \vec A}_i = W_i\vec a_i\ .
\ee
without tensor products.
To exclude the two overall $U(1)$ factors in the two $U(16)$ gauge groups, 
as mentioned above, we impose the additional conditions 
\beqn \label{anou1}
\sum_i N_i \vec a_i = \sum_a N_a \vec a_a = 0
\eeqn 
on the Wilson line moduli. 
Expanding the amplitudes to leading quadratic order in the background 
fields, eq.\ (\ref{total}) in the appendix  
gives the total one-loop correction 
\beqn \label{correction}
\tilde{\cal M}_{i} &=& 
-\pi^{-2}  \sqrt{G} \, 
 {\rm tr}\left( \gamma_{\Omega\Theta i}^{-1} \gamma_{\Omega\Theta i}^{\rm T} 
{\bf F}_{i}^2   
 \thba{\vec{0}}{\vec{0}} (2 {\bf \vec A}_{i}, 8ilG )  \right)
 \ , 
\non 
\tilde{\cal M}_{a} &=& 
-\pi^{-2} \sqrt{G} \, {\rm tr}\left( \gamma_{\Omega a}^{-1} \gamma_{\Omega a}^{\rm T} {\bf F}_{a}^2 
 \thba{\vec{0}}{\vec{0}} (2 {\bf \vec A}_{a}, 8ilG )  \right)
\ , 
\non
\tilde{\cal A}_{ij} &=& 
 (16 \pi^2)^{-1} \sqrt{G}\,  {\rm tr} \left( 
\bfg_{\Theta i}\bfg_{\Theta j}^{-1}  
{\bf F}_{ij}^2  \thba{\vec{0}}{\vec{0}} ({\bf \vec A}_{ij}, 2ilG )
 \right) 
 \ , 
\non
\tilde{\cal A}_{ab} &=& 
 (16 \pi^2)^{-1} \sqrt{G}\,  {\rm tr} \left( 
\bfg_{\Theta a}\bfg_{\Theta b}^{-1} 
 {\bf F}_{ab}^2 \thba{\vec{0}}{\vec{0}} ({\bf \vec A}_{ab}, 2ilG ) \right) 
 \ , 
\non
\tilde{\cal A}_{ia} + \tilde{\cal A}_{ai} &=&  
 (32 \pi^2)^{-1} \sqrt{G}\, {\rm tr}\left( ( \bfg_{i}\bfg_{a}^{-1}  
 + \bfg_{\Theta i}\bfg_{\Theta a}^{-1} )  
 {\bf F}_{ia}^2   \thba{\vec{0}}{\vec{0}} ({\bf \vec A}_{ia}, 2ilG ) \right) 
 \ . 
\eeqn
where we set $\alpha'=1/2$.
Here, $G$ is the metric on the torus (\ref{torusmetric}).
The sums over string oscillators 
have collapsed to numbers,
due to ${\cal N}=2$ supersymmetry \cite{Bachas:1996zt}
and only KK states originating in the torus 
reduction from $D=6$ to $D=4$ contribute.
These contributions appear in the form 
of Wilson-line-shifted KK momentum sums,
that we have written as (genus-two) Jacobi theta functions.
Some useful properties of theta functions are collected in appendix
\ref{thetas}.

One can proceed to directly evaluate the traces in the
amplitudes (\ref{correction}) 
with the help of (\ref{traces}) as in the example
(\ref{example}). 
This evaluation is straightforward but fairly tedious, and we will not
repeat it here.
Instead, we will condense the trace evaluation to a
simple prescription that is hopefully more transparent.
Let us first note that all traces 
 in (\ref{correction}) are effectively over 
$2N\times2N$ matrices for some $N \in \{N_i, N_j, N_a, N_b\}$. 
Moreover, these $2N\times2N$ matrices are 
diagonal, and the first $N$ elements on the diagonal are 
either the same as or the negative of the next $N$ elements. Thus one can 
express each trace in terms of traces over $N\times N$ matrices that
consist of the first $N$ elements of the corresponding 
$2N\times2N$ matrix. In the following we denote such a trace 
with tr$_{N}$, and use the same letter for the matrix. 
With this prescription, we arrive at the following form of the 
one-loop amplitudes:
\beqn \label{correction2}
\tilde{\cal M}_{i} &=& 
-\pi^{-2}  \sqrt{G} \, 
 {\cal F}_i^2 \, {\rm tr}_{N_i}\!\!
 \left( \gamma_{\Omega\Theta i}^{-1} \gamma_{\Omega\Theta i}^{\rm T} Q_{i}^2   
 \right) \left[ \tht(2 \vec a_{i}) + \tht(-2 \vec a_{i})  
 \right]
\non 
&=& - (2 \pi^2)^{-1} \sqrt{G} {\cal F}_i^2 
 \left[ \tht(2 \vec a_{i}) + \tht(-2 \vec a_{i})  
 \right]\ , 
\non [3mm]
\tilde{\cal M}_{a} &=& 
-\pi^{-2}  \sqrt{G} \, 
 {\cal F}_a^2 \, {\rm tr}_{N_a}\!\!
 \left( \gamma_{\Omega a}^{-1} \gamma_{\Omega a}^{\rm T} Q_{a}^2   
 \right) \left[ \tht(2 \vec a_{a}) + \tht(-2 \vec a_{a})  
 \right]
\non 
&=& - (2 \pi^2)^{-1} \sqrt{G} {\cal F}_a^2 
 \left[ \tht(2 \vec a_{a}) + \tht(-2 \vec a_{a})  
 \right]\ , 
\non [3mm]
\tilde{\cal A}_{ij} &=& 
(16 \pi^2)^{-1} \sqrt{G}\, 
 \left( {\cal F}_i^2 {\rm tr}_{N_i}\!\! \left( \gamma_{\Theta i} Q_i^2 \right) 
 {\rm tr}_{N_j}\!\! \left( \gamma_{\Theta j}^{-1} \right) + 
 {\cal F}_j^2 {\rm tr}_{N_i}\!\! \left( \gamma_{\Theta i} \right) 
 {\rm tr}_{N_j}\!\! \left( \gamma_{\Theta j}^{-1} Q_j^2 \right) 
 \right) 
\non
&&
 \hspace{1cm} \left[ \tht(\vec a_{i} - \vec a_{j}) 
 + \tht(-\vec a_{i} + \vec a_{j}) 
 - \tht(\vec a_{i} + \vec a_{j}) - \tht(-\vec a_{i} - \vec a_{j})  
 \right] 
\non
&=& 
(32 \pi^2)^{-1} \sqrt{G}\, ({\cal F}_i^2 N_j + {\cal F}_j^2 N_i) \non
&& 
 \hspace{1cm} \left[ \tht(\vec a_{i} - \vec a_{j}) 
 + \tht(-\vec a_{i} + \vec a_{j}) 
 - \tht(\vec a_{i} + \vec a_{j}) - \tht(-\vec a_{i} - \vec a_{j})  
 \right] \ ,
\non [3mm]
\tilde{\cal A}_{ab} &=& 
(16 \pi^2)^{-1} \sqrt{G}\, 
 \left( {\cal F}_a^2 {\rm tr}_{N_a}\!\! \left( \gamma_{\Theta a} Q_a^2 \right) 
 {\rm tr}_{N_b}\!\! \left( \gamma_{\Theta b}^{-1} \right) + 
 {\cal F}_b^2 {\rm tr}_{N_a}\!\! \left( \gamma_{\Theta a} \right) 
 {\rm tr}_{N_b}\!\! \left( \gamma_{\Theta b}^{-1} Q_b^2 \right) 
 \right) 
\non
&&
 \hspace{1cm} \left[ \tht(\vec a_{a} - \vec a_{b}) 
 + \tht(-\vec a_{a} + \vec a_{b}) 
 - \tht(\vec a_{a} + \vec a_{b}) - \tht(-\vec a_{a} - \vec a_{b})  
 \right] 
\non
&=& 
(32 \pi^2)^{-1} \sqrt{G}\, ({\cal F}_a^2 N_b + {\cal F}_b^2 N_a) \non
&& 
 \hspace{1cm} \left[ \tht(\vec a_{a} - \vec a_{b}) 
 + \tht(-\vec a_{a} + \vec a_{b}) 
 - \tht(\vec a_{a} + \vec a_{b}) - \tht(-\vec a_{a} - \vec a_{b})  
 \right] \ ,
\non [3mm]
\tilde{\cal A}_{ia} + \tilde{\cal A}_{ai} &=&  
(32 \pi^2)^{-1} \sqrt{G}\, 
 \left( {\cal F}_i^2 {\rm tr}_{N_i}\!\! \left( \gamma_{i} Q_i^2 \right) 
 {\rm tr}_{N_a}\!\! \left( \gamma_{a}^{-1} \right) + 
 {\cal F}_a^2 {\rm tr}_{N_i}\!\! \left( \gamma_{i} \right) 
 {\rm tr}_{N_a}\!\! \left( \gamma_{a}^{-1} Q_a^2 \right) 
 \right) 
\non
&& 
 \hspace{1cm} \left[ \tht(\vec a_{i} - \vec a_{a}) 
 + \tht(-\vec a_{i} + \vec a_{a}) 
 + \tht(\vec a_{i} + \vec a_{a}) + \tht(-\vec a_{i} - \vec a_{a})  
 \right] \non
&&
\mbox{} \hspace{-.5cm} + (32 \pi^2)^{-1} \sqrt{G}\, 
 \left( {\cal F}_i^2 {\rm tr}_{N_i}\!\! \left( \gamma_{\Theta i} Q_i^2 \right) 
 {\rm tr}_{N_a}\!\! \left( \gamma_{\Theta a}^{-1} \right) + 
 {\cal F}_a^2 {\rm tr}_{N_i}\!\! \left( \gamma_{\Theta i} \right) 
 {\rm tr}_{N_a}\!\! \left( \gamma_{\Theta a}^{-1} Q_a^2 \right) 
 \right) 
\non
&& 
 \hspace{1cm} \left[ \tht(\vec a_{i} - \vec a_{a}) 
 + \tht(-\vec a_{i} + \vec a_{a}) 
 - \tht(\vec a_{i} + \vec a_{a}) - \tht(-\vec a_{i} - \vec a_{a})  
 \right] \non
&=& (32 \pi^2)^{-1} \sqrt{G}\, \left( {\cal F}_i^2 N_a + {\cal F}_a^2 N_i \right) 
 \left[ \tht(\vec a_{i} - \vec a_{a}) 
 + \tht(-\vec a_{i} + \vec a_{a}) \right]\ .
\eeqn
For notational simplicity, we abbreviated $\thba{\vec{0}}{\vec{0}}$ 
as $\tht$ and 
left the second argument of the theta functions implicit. 
Moreover, we directly 
omitted terms that vanish in the non-abelian case due 
to the appearance of a single factor of $Q_i$ or $Q_a$ in the trace. 

To finish the computation, we want to
integrate these expressions over the world-sheet modulus $l$
(see eq.\ (\ref{directch})).
One has to take extra care of massless 
fields propagating in the loop due to coincident D-branes. 
In particular, for each of the theta functions in (\ref{correction2})
we must distinguish between zero and nonzero argument. 
When an argument is zero, massless modes 
appear, and we have to introduce an 
explicit IR cutoff\footnote{Here we mean IR in the open string channel, 
i.e.\ cutting off large values of $t$, c.f.\ (\ref{loopch}), 
or equivalently small values of $l \sim 1/t$.} 
to regulate the integral over  $l$ 
for the massless states (i.e.\ the 
zero modes $\vec n^{\rm T}=(0,0)$ in the theta function
(\ref{thetamatrix})). 
If an argument is nonzero, 
the Wilson lines act as an effective IR cutoff for that 
integral and all 
states are massive; then
the only contributions from massless states come from the ${\cal A}_{ii}$ 
and ${\cal A}_{aa}$
amplitudes. As far as UV divergences are concerned, we know that they 
all 
cancel in the end, but it is still useful
to introduce a UV cutoff in addition to the IR cutoff,
to  check that this indeed happens.

Now for the explicit integration, 
beginning with the case of vanishing first
argument of the theta function.
Using \cite{AnBaDu99} we have
for annulus amplitudes that
\beqn \label{ghil1}
{\cal A}_{\rm KK}(\Lambda^2, \vec 0) &=& \int_0^{2\Lambda^2} dl\ 
\thba{\vec{0}}{\vec{0}}(\vec 0, 2ilG) e^{-\pi \chi /l} = 
\frac{1}{2\sqrt{G}} \int_{1/\Lambda^2}^\infty \frac{dt}{t} 
\thba{\vec 0}{\vec 0} (0,itG^{-1}) e^{-2 \pi \chi t} \non
&=&  
\frac{1}{2\sqrt{G}} \Big( \Lambda^2 \sqrt{G} 
-\ln (8 \pi^3 \chi \sqrt{G} U_2 |\eta(U)|^4) \Big)\ ,
\eeqn
where $8 \pi^3 \chi$ corresponds to $\mu^2$ in \cite{AnBaDu99}. 
This integral is truly divergent for $\chi\rightarrow 0$;
this is the usual field-theory IR divergence due to massless modes.

For non-vanishing first argument, we argued above that there is
no need to introduce the IR cutoff $\chi$;
the Wilson lines act as IR regulators. 
Let us check that this works, by keeping $\chi$ for now.
Using \cite{Ghilencea:2002ff} we have
\beqn \label{ghil2}
{\cal A}_{\rm KK}(\Lambda^2, \vec a) &=& \int_0^{2\Lambda^2} dl\ 
\thba{\vec{0}}{\vec{0}}(\vec a, 2ilG)e^{-\pi\chi/l} = 
\frac{1}{2\sqrt{G}} \int_{1/\Lambda^2}^\infty \frac{dt}{t} 
\thba{\vec a}{\vec 0} (0,itG^{-1})e^{-2\pi\chi t } \non
&=&  
\frac{1}{2\sqrt{G}}
 \Bigg[- \ln \left(2 \pi \chi 
+ \frac{|A|^2}{\sqrt{G} U_2} \right)+ \Lambda^2 \sqrt{G} - 
\ln \frac{\sqrt{G} U_2}{|A|^2}  \\
&& \qquad\qquad\qquad\qquad\qquad\qquad\qquad
-\ln \left| \frac{\tht_1(A, U)}{\eta(U)} \right|^2 
+ 2\pi U_2 a_4^2 \Bigg] \nonumber \; ,
\eeqn
where we used the complex Wilson line $A$ introduced in (\ref{newa}),
and $U_2={\rm Im}\, U$.
The first logarithm in (\ref{ghil2}) is
the contribution of states with $\vec n^{\rm T} = (0,0)$, which
was the source of the
IR divergence in the previous case.
It is clear that as long as $|A|\neq 0$, we can set $\chi=0$ in 
this first term, and then it cancels against the third term. 
Thus, provided $|A|\neq 0$,
we can remove the IR regulator $\chi=0$ as promised 
and find the  answer
for nonvanishing first argument of the theta function:
\beqn
{\cal A}_{\rm KK}(\Lambda^2, \vec a) 
&=& \frac{1}{2\sqrt{G}} \Big( \Lambda^2 \sqrt{G} 
-\ln \left| \frac{\tht_1(A, U)}{\eta(U)} \right|^2 
+ 2\pi U_2 a_4^2 \Big) \; .
\eeqn

Before continuing, let us make
a quick consistency check. In the form (\ref{ghil2}),
that includes the IR cutoff $\chi$, we could have taken $|A|\rightarrow 0$
which should yield (\ref{ghil1}). 
To see this, it suffices to know
that the other terms in (\ref{ghil2}) are actually finite in the 
limit  $|A|\rightarrow 0$
as can be seen from the expansion (cf. (\ref{firstder}) together
with eq.\ (\ref{thirdder}))
\beqn \label{thetexp}
\ln \left| \frac{\tht_1(A, U)}{\eta(U)} \right|^2 
&=& 
\ln (2\pi)^2 + \ln \left| \eta(U) \right|^4 + \ln \left| A \right|^2 
\\[-2mm]
&& \qquad\qquad - \frac{\pi^2}{3} {\rm Re} \Big( E_2(U) A^2\Big) 
 + {\cal O}(A^4) \ , 
\nonumber 
\eeqn
where $E_2(U)$ is the holomorphic second Eisenstein series
(\ref{Eisen2}). 
The logarithmic term $\ln|A|^2$ cancels the third term in 
(\ref{ghil2}), the
contribution of massive modes is manifestly IR finite,
and taking
$|A|\rightarrow 0$ we recover (\ref{ghil1}).

To use the same formula for the M\"obius strip amplitude, one 
simply needs to 
keep track of the different modular 
transformation, $t=1/(8l)$ instead of $t=1/(2l)$
that we use for the annulus, 
and one finds 
\beqn
{\cal M}_{\rm KK}(\Lambda^2, \vec a) &=& \int_0^{2\Lambda^2} dl\ 
\thba{\vec{0}}{\vec{0}} (2\vec a, 8ilG) \non
&=&
\frac{1}{8\sqrt{G}} \int_{1/(4\Lambda^2)}^\infty \frac{dt}{t} 
\thba{2 \vec a}{\,\, \vec 0} (0, itG^{-1}) 
= \frac14 {\cal A}_{\rm KK}(4\Lambda^2, 2 \vec a) \ . 
\eeqn
These expressions can now be used when integrating
(\ref{correction2}).
We only give the result for the 
correction to the D5-brane gauge couplings; the correction for 
D9-branes can be obtained from this via the replacement 
$(a,b,i)\rightarrow (i,j,a)$. It is
\beqn \label{correct}
\delta \left( \frac{4\pi^2 }{g_a^{2}} \right)
&=& -2 \sqrt{G} 
{\cal A}_{\rm KK}(4 \Lambda^2,2\vec a_a) \non
&& 
\mbox{} + \sqrt{G}
\sum_{b\neq a} N_b \left({\cal A}_{\rm KK}(\Lambda^2,\vec a_a -\vec a_b)
-{\cal A}_{\rm KK}(\Lambda^2,\vec a_a +\vec a_b)
\right) \non
&& 
\mbox{} + \sqrt{G} N_a
\left({\cal A}_{\rm KK}(\Lambda^2,\vec 0)
-{\cal A}_{\rm KK}(\Lambda^2,2 \vec a_a)
\right) \non
&& 
\mbox{} + \frac12 \sqrt{G} 
\sum_{i} N_i {\cal A}_{\rm KK}(\Lambda^2,\vec a_a - \vec a_i)\ .
\eeqn
One important check of this result is
that it is UV finite, due to tadpole cancellation. It is 
easy to carry over the actual check from 
\cite{Bachas:1996zt}, using (\ref{ghil1}) and (\ref{ghil2}).
In these two equations it is
obvious that the $\Lambda$-dependent terms are independent of the 
Wilson lines, so they drop out of the contribution to 
(\ref{correct}) from the 55-annulus amplitudes ${\cal A}_{ab}$ (given 
in the second and third row). Moreover, using $\sum_i N_i = 16$,
it is also evident that the UV-divergent terms cancel between 
the M\"obius and 95-annulus amplitudes, leaving a UV-finite 
result in which the cutoff $\Lambda$ can be taken to infinity.
Using (\ref{ghil1}) and (\ref{ghil2})
and assuming for concreteness that 
all $\vec a_a$ and $\vec a_i$ are distinct and nonzero, 
the result (\ref{correct})
can finally be expressed as
\beqn \label{correct2}
\delta \left( \frac{4\pi^2 }{g_a^{2}} \right)
&=& - \frac12  N_a \ln (8 \pi^3 \chi \sqrt{G} U_2)
+(6-3N_a) \ln|\eta(U)| 
+ \frac12 \pi U_2 \sum_i N_i (a_i)_4^2 
\non
&& \mbox{} + (2 + N_a) 
\ln|\tht_1(2A_a,U)| - \frac12 \sum_{i} N_i \ln|\tht_1(A_a - A_i,U)| \non
&& \mbox{} +  
\sum_{b\neq a} N_b 
\ln\left|\frac{\tht_1(A_a+A_b,U)}{\tht_1(A_a-A_b,U)}\right| \ .
\eeqn
This formula is one of the main results of this paper. 
Note that we have used the extra condition (\ref{anou1}) that 
excludes Wilson lines in the anomalous $U(1)$ factors. 
The first term is the contribution of massless fields. 
All the other terms --- except the 
third one --- are the real part of a holomorphic function in the variables
$A$ and $U$. We will come back to the interpretation of the third term in 
section \ref{inter}. 


\subsubsection{Couplings of abelian gauge groups}
\label{ac}

Before we go on to generalize the result to the ${\cal N}=1$ case 
and draw our conclusions for inflationary models in string theory, let 
us first, for completeness, also discuss the corrections to the
gauge couplings of (non-anomalous) $U(1)$ group factors, although it is 
not the case relevant for the discussion of the KKLMMT-model. Readers 
who are more interested in the application of our result to that 
model can therefore skip this subsection.

To deal with the $U(1)$ case, we  choose the generators 
specifying the background fields to lie in the $U(1)$ factors. For 
each of the $U(N)$ factors in (\ref{gauge}) we have one $U(1)$, whose
background can be characterized by replacing the $Q_i$ of (\ref{qi}) with matrices equal to 
the $W_i$ of (\ref{wi}), i.e.\ take 
\be
Q_i = W_i\ , \quad Q_a = W_a\ , 
\ee
and the $W_i$ and $W_a$ defined as before. 
As the overall two $U(1)$ inside $U(16)_{\rm D9} \times U(16)_{\rm D5}$ are 
anomalous \cite{Berkooz:1996iz}, we also have to impose 
\be \label{nonanom} 
\sum_i N_i {\cal F}_i= 0\ , \quad \sum_a N_a {\cal F}_a= 0
\ee
for the field strengths in this case. 
Repeating the steps of the non-abelian case leading to (\ref{correction2}), 
we end up with 
\beqn \label{correctionu1}
\tilde{\cal M}_{i} &=& 
 -\pi^{-2} \sqrt{G} N_i {\cal F}_i^2 
 \left[ \tht(2 \vec a_{i}) + \tht(-2 \vec a_{i})  
 \right]\ , \\ [3mm]
\tilde{\cal M}_{a} &=& 
 -\pi^{-2} \sqrt{G} N_a {\cal F}_a^2 
 \left[ \tht(2 \vec a_{a}) + \tht(-2 \vec a_{a})  
 \right]\ , 
\non [3mm]
\tilde{\cal A}_{ij} &=& 
(16 \pi^2)^{-1} \sqrt{G}\, N_i N_j ({\cal F}_i^2 + {\cal F}_j^2) \non
&& 
 \hspace{1cm} \left[ \tht(\vec a_{i} - \vec a_{j}) 
 + \tht(-\vec a_{i} + \vec a_{j}) 
 - \tht(\vec a_{i} + \vec a_{j}) - \tht(-\vec a_{i} - \vec a_{j})  
 \right] \non
&& \hspace{-.5cm} 
 \mbox{} + (16 \pi^2)^{-1} \sqrt{G}\, N_i N_j {\cal F}_i {\cal F}_j \non
&& 
\hspace{1cm} \left[ \tht(\vec a_{i} - \vec a_{j}) 
 + \tht(-\vec a_{i} + \vec a_{j}) 
 + \tht(\vec a_{i} + \vec a_{j}) + \tht(-\vec a_{i} - \vec a_{j})  
 \right] \ ,
\non [3mm]
\tilde{\cal A}_{ab} &=& 
(16 \pi^2)^{-1} \sqrt{G}\, N_a N_b ({\cal F}_a^2 + {\cal F}_b^2) \non
&& 
 \hspace{1cm} \left[ \tht(\vec a_{a} - \vec a_{b}) 
 + \tht(-\vec a_{a} + \vec a_{b}) 
 - \tht(\vec a_{a} + \vec a_{b}) - \tht(-\vec a_{a} - \vec a_{b})  
 \right] \non
&& \hspace{-.5cm} 
 \mbox{} + (16 \pi^2)^{-1} \sqrt{G}\, N_a N_b {\cal F}_a {\cal F}_b \non
&& 
\hspace{1cm} \left[ \tht(\vec a_{a} - \vec a_{b}) 
 + \tht(-\vec a_{a} + \vec a_{b}) 
 + \tht(\vec a_{a} + \vec a_{b}) + \tht(-\vec a_{a} - \vec a_{b})  
 \right] \ ,
\non [3mm]
\tilde{\cal A}_{ia} + \tilde{\cal A}_{ai} &=&  
(16 \pi^2)^{-1} \sqrt{G}\, N_i N_a \left( {\cal F}_i^2 
+ {\cal F}_i {\cal F}_a + {\cal F}_a^2 \right) 
 \left[ \tht(\vec a_{i} - \vec a_{a}) 
 + \tht(-\vec a_{i} + \vec a_{a}) \right]\ . \nonumber 
\eeqn
The main difference to the non-abelian case appears in the 
annulus diagrams. Now there are also off-diagonal terms present,
mixing different gauge groups, and in the given basis the gauge 
kinetic terms will 
no longer be a simple sum of terms for the stacks labeled by
 $i$ and $a$, but 
of the bi-linear form 
\beqn 
-\frac14 \sqrt{-g_4} \left( 
\sum_{ij} g^{-2}_{ij} {\cal F}_i {\cal F}_j + \sum_{ia} g^{-2}_{ia} {\cal F}_i {\cal F}_a + 
\sum_{ab} g^{-2}_{ab} {\cal F}_a {\cal F}_b \right) \ . 
\eeqn 
In the non-abelian case the cross-terms are not allowed by gauge
invariance, which is encoded in the fact that 
$\tr_{N_i}(Q_i)=0$ for the non-abelian $Q_i$ of (\ref{qi}), 
whereas $\tr_{N_i}(Q_i)=N_i$ for the abelian ones, cf.\ (\ref{wi}), which 
we use presently. To remove the extra factors of $N_i$ from the 
gauge couplings, we 
could redefine the gauge fields by $\sqrt{N_i}$, but here we
leave the expressions as they are. 

For the terms proportional to ${\cal F}_i^2$ and  
${\cal F}_a^2$ the cancellation of the UV-divergence 
proceeds in the same way as in the non-abelian case. On the other hand,
for the cancellation in the mixed terms, 
it is important that we decoupled the anomalous $U(1)$ by 
imposing (\ref{nonanom}). The same condition also implies that 
the mixed terms would be absent for vanishing Wilson lines. 

We then derive, for abelian gauge groups,
\beqn \label{correctabelian}
\delta \left( \frac{4\pi^2}{g_{ab}^{2}}\right) &=& \delta_{ab} \Bigg[
-\frac32 N_a^2 \ln (8 \pi^3 \chi \sqrt{G} U_2)
+(12-9N_a) N_a \ln|\eta(U)| \non
&& \hspace{1cm} + \pi U_2 N_a \sum_i N_i (a_i)_4^2 
+ (4 + N_a) N_a
\ln|\tht_1(2A_a,U)| \non
&& \hspace{1cm} - N_a \sum_{i} N_i \ln|\tht_1(A_a-A_i,U)| + 2 N_a 
\sum_{c\neq a} N_c 
\ln\left|\frac{\tht_1(A_a+A_c,U)}{\tht_1(A_a-A_c,U)}\right| \Bigg] 
\non
&& 
\mbox{} - (1 - \delta_{ab}) N_a N_b
\ln|\tht_1(A_a-A_b,U)\tht_1(A_a+A_b,U)| \ , \non [3mm]
\delta \left( \frac{4\pi^2}{g_{ai}^{2}}\right) &=& - N_a N_i \Big[
2 \pi U_2 (a_a)_4 (a_i)_4 + \ln|\tht_1(A_a-A_i,U)|\Big]\ .
\label{cross}
\eeqn
There are no summations over repeated indices here; rather, 
all summations have been written explicitly. Moreover, we 
have used (\ref{anou1}) and (\ref{nonanom}) in the derivation. 
Again, the correction to $g_{ij}^{-2}$ 
can be recovered from $\delta g_{ab}^{-2}$ by replacing 
$(a,b,c,i)\rightarrow (i,j,k,a)$ and we assume 
also here that all $\vec a_i$ and $\vec a_a$ are non-zero and 
distinct. As before, there is a contribution from massless modes, 
one term depending on the open string
scalars that is not the real part of a holomorphic function 
in $U$ and $A$ and various others that are.

\label{remark1}
Let us close this section with two side remarks. 
We stress again  that the result for the couplings 
of the 9-brane gauge group and the 5-brane group are exactly 
the same, related just by exchanging indices $a \leftrightarrow i$. 
This means that, in the case where the gauge group has been broken to the 
abelian subgroup, 
string theory seems to choose a different symplectic section as 
the one used in the supergravity literature, see e.g.\ 
\cite{Angelantonj:2003zx,Hsu:2004hi}, where the gauge groups 
are treated asymmetrically.\footnote{Note that the coordinates 
used in \cite{Angelantonj:2003zx} are related to ours 
in the following way: $s \leftrightarrow S\, ', t \leftrightarrow U,
u \leftrightarrow S, x^k \leftrightarrow A_a\,\,  
({\rm resp.}\,\, A'$ in (\ref{gferrara})), 
$y^r \leftrightarrow A_i\,\,  
({\rm resp.}\,\, A$ in (\ref{gferrara})).} 
As we only derived the gauge couplings 
for the open string vectors and not for the closed string KK vectors, 
we cannot say precisely what symplectic section is  used by
 string theory. Let us discuss this point in a little more detail.
Consider the ``tree-level'' prepotential of 
\cite{Antoniadis:1996vw} given in (\ref{treeprepot}), when 
absorbing the $1/(4\pi)$ in the definition of the Wilson line moduli. 
If one calculates the gauge couplings according to\footnote{We
refer again to \cite{Louis:1996ya} for notations and conventions on $\cn =2$ 
gauge couplings. Note, though, that we include an additional factor 
of $2$ in the definition of the couplings as compared to 
\cite{Louis:1996ya}, in order to get the same 
normalization in the relations 
$g^{-2}_{99} = {\rm Im}(S)$, etc., as in section \ref{tree}, cf. 
(\ref{gferrara}) below.}
\be \label{gc}
g^{-2}_{\Lambda \Sigma} = \frac{i}{2} (\cn_{\Lambda \Sigma} - 
\bar \cn_{\Lambda \Sigma}) \ , \quad \cn_{\Lambda \Sigma} = 
\bar F_{\Lambda \Sigma} + 2 i \, 
\frac{{\rm Im}(F_{\Lambda \Xi})\, {\rm Im}
(F_{\Sigma \Upsilon})X^\Xi X^\Upsilon}{{\rm Im}(F_{\Xi \Upsilon})X^\Xi 
X^\Upsilon}\ ,
\ee
and performs the symplectic transformation \`a la \cite{Angelantonj:2003zx}
one derives the following coupling constants for the open string gauge 
groups (for notational simplicity we consider only one 9-brane vector $A$ and
one 5-brane vector $A'$ here):
\beqn \label{gferrara}
g^{-2}_{\rm 99} & = & S_2\ , \non
g^{-2}_{\rm 55} & = & S\, '_2 + 
2 \,\frac{S\, '_2 ( 2 U_2 S_2 - A'_2\,\!\!^2) (A'_1\,\!\!^2 + A'_2\,\!\!^2) 
+ (A'_2\,\!\!^2 - A'_1\,\!\!^2) S_2 A_2^2}{(2U_2 S_2 - A'_2\,\!\!^2)^2}\ 
, \non
g^{-2}_{\rm 95} & = & 2 \,
\frac{A_2 A'_2 S_2}{2U_2 S_2 - A'_2\,\!\!^2}\ . 
\eeqn
Obviously, the 55 and 99 gauge couplings are rather different and
therefore correspond to a different symplectic section as 
chosen by the string in our background field 
calculation. In the absence of charged fields,
the two choices of symplectic section lead to an equivalent
set of equations of motion and Bianchi identities, see e.g.\ 
\cite{Ceresole:1995jg,deWit:1995zg}. In the closed string sector 
such charged fields would arise e.g.\ through gauging the theory by 
turning on background fluxes 
\cite{Tripathy:2002qw,Angelantonj:2003zx}.

Finally, let us make a remark about the ``tree-level''
approximation using the prepotential (\ref{treeprepot}). The term 
``tree-level'' is used in analogy to the perturbative 
heterotic string, where the last term of (\ref{treeprepot}) is absent. 
In the heterotic theory $S\, '$ is usually called $T$ and is a K\"ahler 
modulus that is independent of the ten-dimensional dilaton, 
i.e.\ the string coupling constant $g_s$. 
Thus both terms in the heterotic analog of (\ref{treeprepot}) have the same 
dependence on the dilaton and their sum does, in fact, correspond 
to a consistent tree-level truncation, cf.\  
\cite{Ceresole:1995jg,deWit:1995zg}. 
Here, however, both $S_2$ and $S\, '_2$
depend on the dilaton, 
as is obvious from (\ref{ssprime}), and therefore
the first and the second two terms of (\ref{treeprepot}) have a 
different dilaton dependence. This fact, that there are 
two independent gauge couplings $S_2$ and $S\, '_2$ which should  
both be large in perturbation theory, raises the question 
in which sense it is possible in open string perturbation theory to 
truncate the prepotential to the ``tree-level'' terms of 
(\ref{treeprepot}). 
To see the problem more explicitly, consider for example the 55 gauge 
coupling in (\ref{gferrara}). Expanding the second term 
in the couplings $S_2$ and $S\, '_2$ gives to leading order
\be \label{gzero}
\frac{S\, '_2 (A'_1\,\!\!^2 + A'_2\,\!\!^2)}{S_2 U_2}\ .
\ee
%
This term is of order $\co (g_s^0)$. A term of the same order would,
however, be generated e.g.\ by a ``one-loop''-correction to the 
prepotential of the form\footnote{The argument does not depend on the 
particular form chosen here. Any 
$\delta \cf \sim f_1(U) f_2(A) A'\,\!^n$, with 
$n \geq 2$ and $f_1$, $f_2$ some arbitrary functions, would 
lead to the same conclusion. Comparing this with (\ref{correctabelian}),
it is obvious that such terms indeed do appear at one-loop level.} 
\be
\delta \cf ~\sim~ U A'\,\!^2\ .
\ee
We call this ``one-loop'' because it is neither multiplied by $S$ 
nor $S\, '$. Taking such a term into account and performing the same 
symplectic transformation as in \cite{Angelantonj:2003zx} leads to 
an additional contribution to the 55 gauge coupling of order 
$\co (g_s^0)$ and proportional to 
$U_2$.
Thus it is doubtful whether a truncation to the prepotential 
(\ref{treeprepot}) and the form (\ref{gferrara}) of the 
gauge couplings would be consistent 
in string perturbation theory, since it does not appear to correspond to 
a systematic expansion of the effective Lagrangian (in particular, of the
gauge couplings) in powers of 
the string coupling. 
For example, only some terms of the order $\co (g_s^0)$ would be
included, others left out. 

We do not claim that previous literature on the subject 
is wrong; we merely wish to emphasize
that ``tree-level'' should not be taken 
too literally.

\label{remark2}


\subsection{Generalization to ${\cal N}=1$}
\label{none}

In this section we want to generalize our results 
to the case of interest with only  
${\cal N}=1$ supersymmetry. In terms of toroidal orientifold 
models, we will
use a background $\mathbb{T}^6/\mathbb{Z}_N$ or 
$\mathbb{T}^6/(\mathbb{Z}_N \times \mathbb{Z}_M)$. In order to 
be able to employ the results of \cite{AnBaDu99}, we  
first concentrate on the $\mathbb{Z}_6'$ orientifold. Another 
reason for choosing this model is that the discussion of 
Wilson lines is rather similar to the one 
in the $\mathbb{Z}_6$ orientifold given in \cite{Cvetic:2000st}.
It turns out that the one-loop corrections to the 
gauge kinetic function (and thus to the non-perturbative 
superpotential) do not contain any terms quadratic in the 
Wilson line moduli and thus cannot help to reduce the
inflaton mass in a KKLMMT like scenario. To show that this is 
not a generic problem, we also consider the 
$\mathbb{Z}_2 \times \mathbb{Z}_2$ model of \cite{Berkooz:1996dw}. 
We do not go into the details as much as in the $\mathbb{Z}_6'$ case 
but our results show that the one-loop corrections in this model 
are capable to lower the inflaton mass by fine-tuning.


\subsubsection{The $\mathbb{Z}_6'$ model}

This orientifold is defined in terms of the eigenvalues 
$\exp(2\pi iv)$, $v=(1,-3,2)/6$, of 
the generator $\Theta$, acting on the three complex coordinates of a 
$\mathbb{T}^6=\mathbb{T}^2_1\times \mathbb{T}^2_2\times\mathbb{T}^2_3$. 
The first and third of the 
three 2-tori are assumed to allow a crystallographic 
$\mathbb{Z}_6$ and $\mathbb{Z}_3$ operation, respectively. 
Since $\Theta^3$ is just 
identical to the geometric operation of the generator of 
the $\mathbb{Z}_2$ we considered in 
the previous section on $\mathbb{T}^4/\mathbb{Z}_2\times \mathbb{T}^2$, 
the $\mathbb{Z}_6'$ model 
includes 32 D5-branes extended along the third 2-torus 
$\mathbb{T}^2_3$, in addition to the 
32 space-time filling D9-branes. The moduli space of 
the untwisted moduli is given by 
three copies of $SU(1,1)/U(1)$ for each of the three 
generic K\"ahler moduli of the 
three 2-tori, and one extra copy each for the dilaton and the only complex 
structure modulus of the model, the 
complex structure of the second 2-torus \cite{Ibanez:1992hc}. 
The complex structure of the 
third 2-torus $\mathbb{T}^2_3$, that we denote by $U$, 
will actually appear 
in the one-loop correction to the gauge couplings discussed below, 
but this is not a modulus, since it 
is fixed to a rational value by compatibility
with the orbifold action. 

For application to the KKLMMT model, we are mainly 
interested in the dependence of the 5-brane gauge couplings on the 
Wilson line moduli of the 9-branes along their 
common world volume directions, which is the 
third torus. This dependence is completely 
contained in the 95 annulus. Moreover, the only  
sectors that can depend on 
Wilson lines along the third torus are those in which this 
torus is left invariant, i.e.\ those with 
insertions of the identity or $\Theta^3$.
The relevant amplitudes 
are given in (\ref{keq03}) of appendix \ref{z6}, where 
we also give all the other amplitudes for completeness. 
The important point that allows 
to reduce much of the calculation to the ${\cal N}=2$ case of the 
previous section is the 
fact that the amplitudes in the sectors with insertions $\Theta^k,\ k=0,3$, 
are formally identical to those arising in the 
case of $\mathbb{T}^2\times\mathbb{T}^4/\mathbb{Z}_2$. 
Due to the fact that the element $\Theta^3$ of the orbifold group is 
exactly the same as the $\mathbb{Z}_2$ generator in the $\cn = 2$ case
discussed above, the result formally exactly carries over to the case
at hand, up to an overall factor that we will determine. 

T-duality along all six internal directions again maps 9- and 
5-branes to 3- and 7-branes, localized on the third torus. 
It is clear that these now have to be moved in 
sets of six at least: the orbifold generator 
$\Theta$ identifies three of them and the T-dual world sheet parity 
$\Omega R(-1)^{F_L}$ acts geometrically 
as a reflection on the 2-torus, and thus identifies these 
three with another set of three images. 
When analyzing the allowed Wilson lines in the next subsection, we 
will use this geometric intuition of moving sets of six branes. 

Thus, the most important difference 
to the $\cn=2$ case is that the gauge group 
and the allowed Wilson lines are 
different. The latter have to be compatible with the operation 
of the orbifold generator on 
the third torus, while in the previous section, the Wilson lines 
were turned on in a 2-torus that 
was invariant under the orbifold action. 
Thus, here in the $\cn =1$ case we have to go into some detail
to solve this compatibility condition.
As usual, the action of $\Theta^k$ on the CP labels is encoded 
in a $32\times 32$ matrix $\gamma_{\Theta^k}$. 
Without Wilson lines, the tadpole cancellation conditions for 
twisted tadpoles are \cite{Aldazabal:1998mr}
\beqn
\tr (\gamma_{\Theta^k 9}) &=& \tr (\gamma_{\Theta^k 5}) ~=~ 0 \ , 
\quad k=1,3,5 , \non
\tr (\gamma_{\Theta^2 9}) &=& \tr (\gamma_{\Theta^2 5}) ~=~ -8 \ , \non
\tr (\gamma_{\Theta^4 9}) &=& \tr (\gamma_{\Theta^4 5}) ~=~ 8 
\eeqn
and the solution with the maximal gauge group is given by
\beqn \label{maximal}
\gamma_{\Theta 9} = \gamma_{\Theta 5} = 
{\rm diag}( \beta {\bf 1}_4,\beta^5 {\bf 1}_4, \beta^9  {\bf 1}_8, 
\bar \beta {\bf 1}_4,\bar \beta^5 {\bf 1}_4, \bar \beta^9  {\bf 1}_8) 
\eeqn
and $\gamma_{\Theta^k 9} = \gamma_{\Theta 9}^k$, 
$\gamma_{\Theta^k 5} = \gamma_{\Theta 5}^k$, where 
we used $\beta=e^{i\pi/6}$. 
This choice of matrices $\gamma$ implies the gauge group 
\beqn 
\left( U(4)^2 \times U(8) \right)_{{\rm D}9} \times 
\left( U(4)^2 \times U(8) \right)_{{\rm D}5} \ . 
\eeqn
We will see in the next subsection how this gauge group can be 
broken by turning on continuous Wilson lines. 


\subsubsection{Wilson lines in the $\mathbb{Z}_6'$ orientifold}

\begin{figure}[h]
 \begin{center}
\psfrag{a1}[bc][bc][.7][0]{$\vec a_1$}
\psfrag{a1o}[bc][bc][.7][0]{$\vec a_1^\Theta$}
\psfrag{a1o2}[bc][bc][.7][0]{$\vec a_1^{\Theta^2}$}
\psfrag{a2}[bc][bc][.7][0]{$\vec a_2$}
\psfrag{a2o}[bc][bc][.7][0]{$\vec a_2^\Theta$}
\psfrag{a2o2}[bc][bc][.7][0]{$\vec a_2^{\Theta^2}$}
\psfrag{ma1}[bc][bc][.7][0]{$-\vec a_1$}
\psfrag{ma1o}[bc][bc][.7][0]{$-\vec a_1^\Theta$}
\psfrag{ma1o2}[bc][bc][.7][0]{$-\vec a_1^{\Theta^2}$}
\psfrag{ma2}[bc][bc][.7][0]{$-\vec a_2$}
\psfrag{ma2o}[bc][bc][.7][0]{$-\vec a_2^\Theta$}
\psfrag{ma2o2}[bc][bc][.7][0]{$-\vec a_2^{\Theta^2}$}
\psfrag{lp}[bc][bc][.7][0]{\hspace{-1cm} dual lattice points}
\psfrag{fp}[bc][bc][.7][0]{fixed points}
   \resizebox{11cm}{!}{\psfig{figure=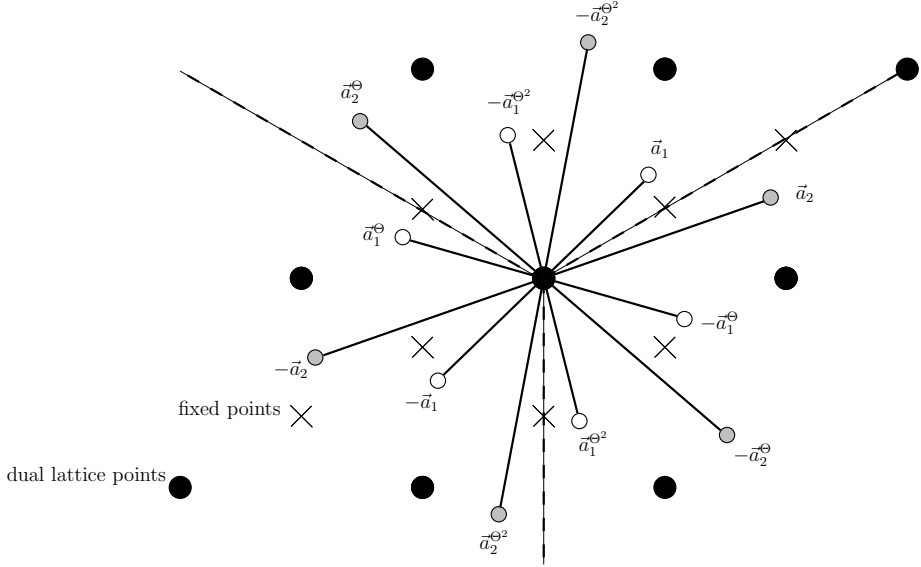}}
   \vspace{0mm}
   \caption{$\mathbb{Z}_3$-symmetric Wilson lines in $\mathbb{Z}_6'$} 
\label{z3w}
 \end{center}
\end{figure}

The classification of Wilson lines in the $\mathbb{Z}_6'$ orientifold 
has not been 
considered in the literature so far. However, 
our discussion will be 
very similar to that for
the $\mathbb{Z}_6$ orientifold \cite{Cvetic:2000st}, 
and we will be able to make use of the results for 
$\mathbb{Z}_3$ \cite{Cvetic:2000aq} as well. 
To introduce Wilson lines it is convenient to reorder the blocks in 
$\gamma_{\Theta 9}$ in the following way 
(by abuse of notation we still use $\gamma_{\Theta 9}$ after 
the reordering): 
\beqn 
\gamma_{\Theta 9} =  
{\rm diag}( \beta {\bf 1}_{4-n_9},\beta^5 {\bf 1}_{4-n_9}, \beta^9  
{\bf 1}_{8-n_9}, 
\bar \beta {\bf 1}_{4-n_9},\bar \beta^5 {\bf 1}_{4-n_9}, \bar \beta^9  
{\bf 1}_{8-n_9}, \gamma^{[6n_9]}_{\Theta 9}) 
\eeqn
with
\beqn 
\gamma^{[6n_9]}_{\Theta 9} = 
{\rm diag}( \beta ,\beta^5 , \beta^9 , 
\bar \beta ,\bar \beta^5 , \bar \beta^9 ) \otimes {\bf 1}_{n_9} = 
\gamma^{[6]}_{\Theta 9} \otimes {\bf 1}_{n_9} 
\eeqn
and similarly for the 5-branes, using $n_5$ and 
$\gamma^{[6]}_{\Theta 5}$. 
(We use bracketed superscripts to denote the size of the matrix). 
Obviously, $n_5,n_9 \leq 4$ has to hold. 
The most general ansatz for the Wilson lines that leaves 
(at least) the gauge group
\beqn  
\left( U(4-n_9)^2 \times U(8-n_9) \times U(n_9) \right)_{{\rm D}9} 
\times \left( U(4-n_5)^2 \times U(8-n_5)\times U(n_5) \right)_{{\rm D}5}
\eeqn
intact is 
\beqn 
\gamma_{W9} = {\rm diag}({\bf 1}_{32-6n_9}, \gamma_{W9}^{[6]} 
\otimes {\bf 1}_{n_9}) \ , \quad 
\gamma_{W5} = {\rm diag}({\bf 1}_{32-6n_5}, \gamma_{W5}^{[6]} 
\otimes {\bf 1}_{n_5}) \ . 
\eeqn
In order for the matrices $\gamma_W$ to describe Wilson lines along the 
third 2-torus, they have to satisfy three conditions.
Tadpole cancellation has to remain fulfilled, they need to be 
compatible with the orientifold operation on the third 2-torus, 
and finally they have to be unitary. 
The tadpole constraints are satisfied if \cite{Cvetic:2000st,Cvetic:2000aq} 
\be \label{tadz6}
\tr ((\gamma^{[6]}_{\Theta 9})^k(\gamma_{W9}^{[6]})^p)  = 
\tr ((\gamma^{[6]}_{\Theta 5})^k(\gamma_{W5}^{[6]})^p) ~=~ 0 \ , 
\quad k=1,2,4,5\ , \quad p=0,1,2 \ . 
\ee
%
Note that there is no condition for $k=3$ because $\Theta^3$ 
acts as the identity on the third torus. 
Further, consistency conditions for the Wilson line to be 
compatible with the action of $\Theta$ on the third
2-torus have to be satisfied, 
\beqn \label{consistz6}
&&
(\gamma^{[6]}_{\Theta 9}\gamma^{[6]}_{W9})^6=-{\bf 1}_6\ , \quad 
(\gamma^{[6]}_{\Theta 5}\gamma^{[6]}_{W5})^6=-{\bf 1}_6\ , 
\non 
&& 
((\gamma^{[6]}_{\Theta 9})^2\gamma^{[6]}_{W9})^3=-{\bf 1}_6\ , \quad 
((\gamma^{[6]}_{\Theta 5})^2\gamma^{[6]}_{W5})^3=-{\bf 1}_6\ . 
\eeqn
The T-dual geometrical interpretation goes as follows: 
Originally, with the maximal gauge group given above, 
all 32+32 D7- and D3-branes 
are located at the origin. One can then move six in a 
$\mathbb{Z}_3\times \mathbb{Z}_2$ invariant fashion, 
two sets of three being identified under the T-dual world 
sheet parity, and the elements of each set of three are
identified under $\Theta$, leaving just a single independent 
brane. Moving $n_9$ coinciding sets of 6 
D3-branes then leaves a $U(n_9)$ on the mobile stack, 
while reducing the rank of the total gauge group by 
$3n_9-n_9=2n_9$. The tadpole consistency requires that 
one takes one brane each from the three sets that made up 
$(U(4)\times U(4)\times U(8))_{{\rm D}9}$, which 
explains the breaking pattern. 
Guided by the geometrical intuition that the Wilson line that 
corresponds to the 
T-dual separation of D3-branes from the origin  
should reflect the fact that there are two triplets of branes, which 
are separately identified under $\Theta$, but not mixed, we now 
make an ansatz, where the Wilson line is block diagonal in $3\times 3$ blocks, 
i.e.\ we choose 
\beqn 
\gamma_{W9}^{[6]} = {\rm diag}( \gamma^{[3]}_{W9} , \bar\gamma^{[3]}_{W9} ) \ . 
\eeqn
%
Here $\bar\gamma^{[3]}_{W9}$ is the complex conjugate of 
$\gamma^{[3]}_{W9}$.
For the blocks, we adopt the form of the most general 
Wilson line consistent 
with a $\mathbb{Z}_3$ twist \cite{Cvetic:2000aq}\footnote{Despite such claims in 
the literature, this does not seem to imply that the Wilson lines in the 
$\mathbb{Z}_6'$ model
on the third torus, where $\Theta$ 
is of order 3, are fully classified by the solution for $\mathbb{Z}_3$.}  
\beqn 
\gamma^{[3]}_{W9} = b_1 {\bf 1}_3 + b_2 \zeta + b_3 \zeta^2 \ , 
\eeqn
having defined the permutation matrices of three elements via 
\beqn 
\zeta = \left( \begin{array}{ccc} 
0 & 1 & 0 \\ 0 & 0 & 1 \\ 1 & 0 & 0 
\end{array} \right) 
\ , \quad 
\zeta^2 = \zeta^{\rm T} \ . 
\eeqn
The three coefficients $b_i$ are a priori free complex parameters. 
This choice automatically satisfies the tadpole constraints 
(\ref{tadz6}). Evaluating the 
consistency conditions (\ref{consistz6}), one finds 
\beqn 
(\gamma^{[6]}_{\Theta 9}\gamma^{[6]}_{W9})^6 &=& -{\bf 1}_6 
\left( b_1^3+b_2^3+b_3^3-3b_1b_2b_3 \right)^2 \ , \non
((\gamma^{[6]}_{\Theta 9})^2\gamma^{[6]}_{W9})^3 &=& -{\bf 1}_6
\left( b_1^3+b_2^3+b_3^3-3b_1b_2b_3 \right) \ . 
\eeqn
%
Upon diagonalizing the matrix $\zeta$ via the unitary transformation 
\beqn 
P_3 =\frac{1}{\sqrt{3}} \left( \begin{array}{ccc}
1 & 1 & 1 \\ 1 & \alpha & \alpha^2 \\ 1 & \alpha^2 & \alpha 
\end{array}\right) \ , 
\eeqn 
$\alpha=e^{2\pi i/3}$, one has 
\beqn 
P_3 \gamma^{[3]}_{W9} P_3^\dag = {\rm diag}( b_1 + b_2 + b_3, b_1 
+ b_2 \alpha^2 + b_3 \alpha , 
                                             b_1 + b_2 \alpha + b_3 \alpha^2) \ . 
\eeqn
In the diagonal form unitarity is most easily imposed, and 
implies that the three 
diagonal elements are just phases. The extra condition 
$b_1^3+b_2^3+b_3^3-3b_1b_2b_3=1$ means that the 
determinant has to be one as well, so that we can finally write 
\beqn 
P_3 \gamma^{[3]}_{W9} P_3^\dag = {\rm diag}( e^{i\varphi_1} ,
e^{i\varphi_2},e^{-i(\varphi_1+\varphi_2)} ) \ . 
\eeqn
This provides an explicit parametrization in terms of two periodic 
variables, which are related to the T-dual 
positions of the branes on the 2-torus. In order to implement the Wilson line 
in the open string KK spectrum as shifts of momenta, we define $\vec a$ 
through   
\beqn 
\varphi_1 = \vec e\, \vec a \ , \quad 
\varphi_2 = \vec e\, \vec a^\Theta\ , \quad 
-\varphi_1-\varphi_2 = \vec e\, \vec a^{\Theta^2}\ , \quad 
\eeqn
where $\vec a$ is the Wilson line on the first D9-brane, i.e.\ 
the T-dual of the D3-brane position, 
and $\vec a^\Theta$ and $\vec a^{\Theta^2}$ are its images under 
the orbifold generator, 
acting on the third 2-torus. 
Explicitly, the action is $(a_4,a_5)^\Theta=(-a_5,a_5-a_4)$. 
Moreover, $\vec{e}$ can be chosen to be 
one of the basic lattice vectors $(\vec e_4,\vec e_5)$. 
Thus the complete Wilson line on the 9-branes is given by
\beqn \label{w6}
P_{32} \gamma_{W9} P_{32}^\dag & = & 
{\rm diag}( {\bf 1}_{32-6n_9} , \gamma^{[6n_9]}_{W9})\ , \non
\gamma^{[6n_9]}_{W9} &=& {\rm diag}
(e^{i \vec e\, \vec a},e^{i \vec e\, \vec a^\Theta},
e^{i \vec e\, \vec a^{\Theta^2}}, e^{-i \vec e\, \vec a},
e^{-i \vec e\, \vec a^\Theta},
e^{-i \vec e\, \vec a^{\Theta^2}}) \otimes {\bf 1}_{n_9}
\eeqn
with $P_{32} = {\bf 1}_{32-6n_9} \oplus ( P_3 \otimes {\bf 1}_{2n_9} )$. 
In the T-dual picture this describes $6n_9$ mobile D3-branes at 
positions given through $\pm \vec a^{\Theta^n}, n=0,1,2$, 
and supporting a mobile $U(n_9)$ gauge group.\footnote{Note that 
the lattice and the dual lattice are exchanged via T-duality.} This is 
depicted in figure \ref{z3w}. The points 
labeled by $\vec a$ have coordinates $a_I \vec e^{\, I}$, the 
$\vec e^{\, I}$ being the basis of the dual lattice, as in (\ref{lattice}). 

In this basis, where the Wilson line is diagonal, the operation of 
the orbifold no longer is, except for
\beqn 
&& 
P_6(\gamma^{[6]}_{\Theta 9})^3P_6^\dag =(\gamma^{[6]}_{\Theta 9})^3=
P_6(\gamma^{[6]}_{\Theta 5})^3P_6^\dag=(\gamma^{[6]}_{\Theta 5})^3=
{\rm diag}(i {\bf 1}_{3},-i {\bf 1}_{3}) \ , 
\eeqn
where we have defined $P_6=P_3\otimes {\bf 1}_2$, acting 
block-diagonally.
This is, however, all we need to evaluate the amplitudes 
in the $k=0,3$ sectors explicitly. This matrix is identical to the 
matrix representation (\ref{gamma2}) of the $\mathbb{Z}_2$ 
generator in the ${\cal N}=2$ model of the previous chapter. 
In the basis where the Wilson line is diagonal, we expect 
the orbifold generator to act in a way on the CP labels 
that matches with our geometrical intuition. Indeed, one finds that 
\beqn 
P_6 \gamma_{\Theta 9}^{[6]}  P_6^\dag = {\rm diag}(\beta \zeta, \bar\beta \zeta^2) \ .  
\eeqn
Thus, $\gamma_{\Theta 9}$ really just permutes the three CP 
labels of each of the two sets separately, as expected. 

To determine the matrix representation 
$\lambda={\rm diag}({\bf 0}_{32-6n_9},\lambda^{[6n_9]})$ 
of the surviving, say, 9-brane gauge fields in the mobile 
$U(n_9)$, one has to regard the projections 
\beqn 
\lambda^{[6n_9]} = \gamma^{[6n_9]}_{W9} \lambda^{[6n_9]} 
(\gamma^{[6n_9]}_{W9})^{-1} \ , \quad 
\lambda^{[6n_9]} = \gamma^{[6n_9]}_{\Theta 9} \lambda^{[6n_9]} 
(\gamma^{[6n_9]}_{\Theta 9})^{-1}\ . 
\eeqn
This leads to gauge fields represented by CP matrices 
\beqn \label{q6} 
\lambda^{[6n_9]} = {\rm diag} ({\bf 1}_3 \otimes Q_{n_9}, 
{\bf 1}_3 \otimes (-Q_{n_9})) \ , 
\eeqn
where $Q_{n_9}$ is an arbitrary $n_9\times n_9$ matrix in the 
adjoint of $U(n_9)$. 
A field strength in the Cartan subalgebra would now be given by e.g.\ 
\beqn \label{q6na} 
Q_{n_9}= \frac{1}{2\sqrt{3}} {\rm diag}(1,-1,0,\, ...\, ,0)\ ,
\eeqn
where we included the factor of $1/\sqrt{3}$ in order to 
normalize $\tr (\lambda^{[6n_9]})^2=1$.
All the above works analogously for the D5-branes, 
respectively their T-dual D7-branes. 


\subsubsection{Results for $\mathbb{Z}_6'$}

In this subsection we would like to use the above insights 
into the breaking of the gauge group to determine the 
dependence of the 5-brane gauge couplings on the 9-brane scalars, 
i.e.\ we consider the case with vanishing Wilson lines on the 
5-branes so that their gauge group is the unbroken 
$(U(4)\times U(4)\times U(8))_{{\rm D}5}$. Moreover, being 
interested in the 5-brane gauge couplings, we only consider 
a background for the 5-brane gauge fields, specified by a 
matrix similar to (\ref{qi}), where the position of the 
non-vanishing entries depends on the gauge group factor 
according to $Q_{U(4)_1}=\frac12 {\rm diag} (1,-1,0^{14},-1,1,0^{14})$,
$Q_{U(4)_2}=\frac12 {\rm diag} (0^4,1,-1,0^{14},-1,1,0^{10})$ and
$Q_{U(8)}=\frac12 {\rm diag} (0^8,1,-1,0^{14},-1,1,0^6)$, where
the ordering is chosen to be consistent with (\ref{maximal}).
Furthermore, we only give the dependence on the 9-brane scalars.
The full gauge couplings could be extracted from the 
formulas given in \cite{AnBaDu99} without much more difficulty. 
An important point to mention is that, 
as opposed to the $\cn =2$ case, it is no longer possible to go to 
an abelian limit (the Coulomb branch) by just turning on Wilson lines 
of the specified type, since a remnant non-abelian 
$U(4)_{{\rm D}9}\times U(4)_{{\rm D}5}$ cannot be broken this way. 
In the T-dual version this implies that one cannot move all the branes 
away from the origin. 

Recall that
the two amplitudes of interest for us
are formally given by the 95 annulus 
of the $\cn=2$ case discussed above, up to an overall factor 
of $1/3$. 
The main difference of the two cases is the fact that 
the Wilson lines in the $\cn=1$ model have to be consistent with the
orbifold operation, which amounts to moving the T-dual D3-branes 
in groups of multiples of six. In other words, for any Wilson line
$\vec{a}$ one has to turn on $\vec a^\Theta$ and $\vec a^{\Theta^2}$
at the same time. In addition, as in the $\cn =2$ case, also the 
negative of these values appear due to the world sheet parity projection.

Thus we can just copy the result from the last line of 
(\ref{correction2}) including
the different normalization factor to get
\be \label{95annulus}
\tilde{\cal A}_{ia}^{(0)} + \tilde{\cal A}_{ai}^{(0)} 
+ \tilde{\cal A}_{ia}^{(3)} + \tilde{\cal A}_{ai}^{(3)} 
= \frac13 (32 \pi^2)^{-1} \sqrt{G}\, {\cal F}_a^2 n_i
\sum_{m=0}^2  \left[ \tht(- \vec a^{\Theta^m}_{i}) 
 + \tht(\vec a^{\Theta^m}_{i}) \right]\ ,
\ee
where we allowed for various mobile stacks now, using $n_i$ 
instead of $n_9$ as before. As we already mentioned, the 
total sum of all $n_i$ is limited to four now. Moreover, 
in (\ref{95annulus}) we only wrote down the amplitudes for 
9-branes with a non-vanishing Wilson line.
Other D9-branes will only contribute universal terms, independent of 
the open string scalars. From (\ref{95annulus}) 
we can read off the dependence of the 5-brane gauge couplings on the
9-brane scalars according to 
\be \label{g5n1}
\delta \left( \frac{4\pi^2 }{g_a^{2}} \right) =
\frac16 \pi U_2 \sum_i n_i \sum_{m=0}^2 (a^{\Theta^m}_i)_4^2
- \frac16 \sum_{i} n_i \sum_{m=0}^2 \ln|\tht_1(A^{\Theta^m}_i,U)|
+ \ldots \ ,
\ee
where the dots stand for correction terms that are independent 
of the Wilson line moduli. We now note that the Wilson line moduli 
on the images under $\Theta$ are related 
by multiplication with a phase, i.e.\ $A^{\Theta}=e^{2\pi i/3}A$, 
which can be verified using the action of $\Theta$ on $\vec{a}$
given above eq.\ (\ref{w6}).\footnote{This mapping under $\Theta$ 
ensures that in the $\cn =1$ analog of the abelian 
gauge coupling $g_{ai}^{-2}$ 
(\ref{cross}), the first term on the right hand side drops out when 
summing over orbits of $\Theta$.
This is necessary to guarantee that the gauge coupling is the real part 
of a holomorphic function.} 
This implies that $A^{2m} + (A^{\Theta})^{2m} + (A^{2 \Theta})^{2m} = 0$ for 
all integers $m$ that are not multiples of $3$. 
Thus, using the fact that $\tht_1$ is an odd function in $A$, we see that
for small $|A|$
\be \label{g5n1b}
\delta \left( \frac{4\pi^2 }{g_a^{2}} \right) 
= \frac13 \pi U_2 \sum_i n_i \Big((a_i)_4^2 + (a_i)_5^2-(a_i)_4(a_i)_5\Big)
- \frac12 \sum_{i} n_i \ln|A_i| + \co (A^6) + \ldots \ .
\ee
The terms quadratic in the $A_i$ have canceled out. 
Since the $A_i$ 
are the candidate fields for the inflaton in the T-dual setting 
with D3-branes, this implies that the above 
gauge kinetic function would produce 
no extra contribution to the inflaton mass. 
This is not generic in $\cn=1$ orientifolds, but an 
accidental consequence of the global $\mathbb{Z}_3$ symmetry of the
Wilson lines, as we shall demonstrate in the following section. 


\subsubsection{The $\mathbb{Z}_2 \times \mathbb{Z}_2$ model}
\label{z2z2}

We would now like to discuss another Calabi-Yau orientifold model
with $\cn =1$ supersymmetry, the type IIB 
$\mathbb{T}^6/(\mathbb{Z}_2\times\mathbb{Z}_2)$ orientifold 
\cite{Berkooz:1996dw}. We have not worked
out all the details, but we intend to stress its
qualitative features here. In particular, we shall point out why the
one-loop correction in the $\mathbb{Z}_2\times\mathbb{Z}_2$ model 
relevant for the correction of the inflaton mass does not cancel
out, as it did in the previous section for $\mathbb{Z}_6'$.\footnote{We 
hope to give a more complete analysis
of this example in a forthcoming publication \cite{goldandglory}.}

This orientifold is defined by the action of the
three orbifold group elements $\Theta_p$, $p=1,2,3$, on
$\mathbb{T}^6=\mathbb{T}_1^2\times\mathbb{T}_2^2\times\mathbb{T}_3^2$,
where each $\Theta_p$ is a reflection along two 2-tori,
leaving the torus $\mathbb{T}_p^2$ invariant. The orbifold
symmetry does not impose any further requirements on the
background tori, such that all three complex structure moduli
remain in the spectrum, together with the generic three
complexified K\"ahler parameters. Thus, the untwisted moduli space
consists of a total of six copies of $SU(1,1)/U(1)$ plus one for
the dilaton \cite{Ibanez:1992hc}. When the 3-form fluxes are turned on, the complex
structure and the string coupling are assumed to get fixed. 
But compared to the previous $\mathbb{Z}_6'$
example, the modulus $U$ that appears in the relevant one-loop
corrections to the gauge coupling is not fixed universally by the
orbifold symmetry, and can, without fluxes, take any value. In analogy
with the $\mathbb{Z}_2$ K3-orientifold there are four types of
untwisted tadpole divergences in the Klein bottle, canceled by 32
D9-branes plus three sets of 32 D5$_p$-branes each, wrapped
around $\mathbb{T}_p^2$ respectively. Together they support
the maximal gauge symmetry $Sp(8)_{\rm D9}\times Sp(8)_{{\rm
D}5_1}\times Sp(8)_{{\rm D}5_2}\times Sp(8)_{{\rm D}5_3}$.\footnote{This
phenomenologically less interesting gauge group was actually one of
the main reasons to concentrate on the $\mathbb{Z}_6'$ orientifold
in the first place. It has, however, also been argued that the gauge
group may be changed to a group of unitary factors in the
presence of discrete torsion \cite{Aldazabal:1998mr}. The
$\mathbb{Z}_2\times\mathbb{Z}_2$ orientifold was also the starting
point of constructing supersymmetric intersecting brane models in
\cite{Cvetic:2001nr}, 
which do possess unitary gauge symmetries plus chiral matter.
Hence, it may turn out that this model allows for better
phenomenology than the standard solution with symplectic gauge
groups suggests. \label{nichsoschlimm}}

Again, we are now only interested in that part of the one-loop
amplitude that depends on the Wilson lines on the D9-branes and on
the gauge field background on one of the three stacks of
D5$_p$-brane. The latter is T-dual to the stack of D7-branes
that undergoes gaugino condensation, while the other D5-branes are
ignored for the moment. It is evident that again, the only relevant
amplitudes are
\beqn
\tilde\ca_{95_p}^{(0)} + \tilde\ca_{5_p 9}^{(0)} +
\tilde\ca_{95_p}^{(\Theta_p)} +\tilde\ca_{5_p
9}^{(\Theta_p)} \ ,
\eeqn
where the upper index $(\Theta_p)$ stands for the insertion of
$\Theta_p$ in the trace, and $(0)$ for the identity as
before. Formally, i.e.\ up to the concrete charge matrices to be
used in evaluating the traces, this amplitude is again identical
to the last equation of (\ref{correction2}) 
up to a different overall normalization, which this time is 
$1/4$ compared to $1/2$ (or $1/6$ in the $\mathbb{Z}_6'$ case). 
Now using the solution of \cite{Berkooz:1996dw}
for the operation of the orientifold group elements on the CP
indices, one can pick matrices $\gamma_{\Theta_p 9}$ and 
$\gamma_{\Theta_p 5_p}$ with
eigenvalues $\pm i$, that after diagonalizing become again
identical to $\gamma_{\Theta 9}$ and $\gamma_{\Theta 5}$ of the $\mathbb{Z}_2$
K3-orientifold as given in (\ref{gamma2}) with
$N=16$.\footnote{See equation (4.7) and the table above (4.15) in 
\cite{Berkooz:1996dw}. 
Pick $p=1$ 
which implies $\gamma_{\Theta_p 9}=\gamma_{\Theta_p 5_p}=-M_1$ 
with the claimed property $M_1^2=-{\bf 1}_{32}$.} Given this, we 
now would have to determine the consistent forms of Wilson lines on the
$p$-th 2-torus, defined by matrices $\gamma_{W9p}$, along
the same lines as for $\mathbb{Z}_6'$, and find the patterns of
gauge symmetry breaking.\footnote{In \cite{Berkooz:1996dw} 
it was already argued that
moving D5-branes out of the fixed points of the orbifold group
would still lead to symplectic gauge groups on the various
stacks.} However, for
the time being, we will not go through the procedure explicitly,
leaving it to future work \cite{goldandglory}, and 
simply follow geometric intuition. Thus, we just 
use the analog of (\ref{95annulus}), but now summing over the images of the
elements of $\mathbb{Z}_2\times\mathbb{Z}_2$ instead of
$\mathbb{Z}_6'$.

The two orbifold elements $\Theta_q$, $q\not=p$, only
act by reflection on $\mathbb{T}_p^2$, and so on $\vec a_{i}$. 
Thus the final form of the
relevant one-loop correction reads
\beqn \label{95annulusZ22}
\tilde\ca_{95_p}^{(0)} + \tilde\ca_{5_p 9}^{(0)} +
\tilde\ca_{95_p}^{(p)} +\tilde\ca_{5_p
9}^{(p)}
=
\frac12 (32\pi^2)^{-1} \sqrt{G}\, {\cal F}_{a}^2 n_i
\sum_{q\not=p} \sum_{m=0}^1 \left[ \tht(- \vec
a^{\Theta_q^m}_{i})
 + \tht(\vec a^{\Theta_q^m}_{i}) \right]\ ,
\eeqn
where now $\vec a_i^{\Theta_q}=-\vec a_i$. Since the
theta function is even, all contributions add up. In
particular, the outcome is identical to our  result 
(\ref{correction2}) for the
$\cn=2$ model, up to an overall
numerical factor. The expression for the correction to the gauge
coupling on the stack of D5-branes labeled by $a$ then reads
\beqn \label{g5n1z2}
\delta \left( \frac{4\pi^2 }{g_a^{2}} \right)
&=&
\pi U_2 \sum_i n_i  (a_i)_4^2 - \sum_{i} n_i \ln|\tht_1(A_i,U)| +
\ldots
\eeqn
Unlike in the $\mathbb{Z}_6'$ model, 
the terms quadratic in $A_i$ do not cancel out.
This leads to a new contribution to the inflaton mass,
as we will discuss in the next section.


\section{Interpretation}
\label{inter}

In this section we would like to interpret our results of section 
\ref{technicalstuff} and draw some conclusions for inflationary 
models in string theory. For readers who decided to skip section 
\ref{technicalstuff}, a few key facts from that section will be repeated 
here. We state our results for the $\cn=2$ and $\cn=1$ cases in turn. 
For $\cn=2$ we find a clean solution to the rho problem 
and for $\cn=1$ we describe the implications of our results 
for the inflaton mass problem. Let us also remind the reader that, as 
already stated in the introduction, the results have to be understood 
as giving a qualitative picture. Our toy models are not close enough 
to the actual KKLMMT model to allow for reliable quantitative predictions
(e.g.\ our calculation of the one-loop corrections to the gauge kinetic 
function of the 7-branes in section 
\ref{technicalstuff} neglected the warp factor and the fluxes; it would be 
very nice, but with present techniques very difficult, to perform our 
calculation in a more realistic setting). 
We will come back to this issue at the very end of this section. 

\subsection{$\cn=2$}
\label{intern2}

Let us start by reviewing the $\cn=2$ model that we discussed in 
sections \ref{tree} and \ref{seconeloop}. Readers who have gone 
through these sections can skip the following paragraph.

The model under consideration 
is the type IIB $\mathbb{T}^4/\mathbb{Z}_2 \times \mathbb{T}^2$ orientifold
\cite{Bianchi:1990yu,Gimon:1996rq,Gimon:1996ay}, i.e.\ we consider 
an orbifold limit of 
K3$\times \mathbb{T}^2$.
It contains 32 D9-branes and 32 D5-branes, wrapped around the 
torus $\mathbb{T}^2$. This leads to a gauge group $SU(16)_{{\rm D}9} \times 
SU(16)_{{\rm D}5}$ if all 5-branes are at the origin of the $\mathbb{T}^4$ 
and there are no Wilson lines on the 9-branes. 
The closed string spectrum contains hypermultiplets and vector
multiplets but for our purposes we can restrict to vector multiplets 
only. In addition to those from the open string sector,
there are three vector multiplets from the closed 
string sector, whose complex scalars are given by the complex structure 
modulus of the torus, $U$, and the two scalars $S$ and $S\, '$, given in 
(\ref{ssprime}) for the case of vanishing Wilson line moduli and 
in (\ref{ss'}) for the case with 
corrections due to non-vanishing Wilson line 
moduli \cite{LopesCardoso:1994is,Antoniadis:1996vw}. 
The scalars in the vector multiplets 
of the open string sector, on the other hand, are given by the Wilson line 
moduli on the 5- and 9-branes along the torus 
and are defined according to (\ref{newa}). 
Turning on Wilson lines breaks the gauge group to a product of unitary 
groups (\ref{gauge}), where the 
overall $U(1)$ factors are anomalous for both the 
5- and the 9-branes and therefore
become massive \cite{Berkooz:1996iz}. In the 
T-dual picture (with six T-dualities along all compact directions), the 
breaking of the gauge group can be understood in terms of D7- and D3-branes 
that are moved away from the origin of the torus $\mathbb{T}^2$.
The main results are formulas (\ref{correct}) and 
(\ref{correct2}), which give the one-loop 
correction to the couplings of non-abelian gauge groups on the 
5-branes, in particular displaying 
the dependence on the open string scalars of both
5- and 9-branes. 

Having repeated the relevant aspects of this model, 
let us now draw some conclusions
for the rho problem described in the introduction. 
To do so, we have to make use of the 
relation between the variables common in the inflationary literature
(e.g.\ in the KKLMMT model) 
introduced in section \ref{inflation} and those common in the 
orientifold literature used in section 
\ref{technicalstuff}.\footnote{Strictly speaking, the KKLMMT model 
has ${\cal N}=1$ before supersymmetry breaking through antibranes 
and in this subsection we are considering our ${\cal N}=2$ orientifold
example. However, the notational dictionary 
works the same way as in the ${\cal N}=1$ examples of the next subsection.
Also, inflation based directly 
on K3$\times \mathbb{T}^2$ compactifications was 
studied in \cite{Dasgupta:2002ew,Hsu:2004hi,Halyo:2003wd}. } 
The volume modulus $\rho$ corresponds to the field $S\, '$ and the
combination of the D3-brane scalars $\phi$ denoting the inflaton 
field corresponds to the Wilson line modulus $A$ on the D9-brane 
which is T-dual to the mobile D3-brane, i.e.\
\be \label{dictionary}
\rho \quad
\longleftrightarrow \quad S\, ' \quad , \qquad \phi \quad 
\longleftrightarrow \quad A\ .
\ee
Other fields present in 
the $\mathbb{T}^4/\mathbb{Z}_2 \times \mathbb{T}^2$ orientifold 
are the modulus $U$, which corresponds to one of the complex structure 
moduli that are supposed to be fixed in the inflationary models 
by fluxes, the modulus $S$, which, in the T-dual picture, 
gives the gauge coupling on the D3-branes, i.e.\ the dilaton, which 
is also supposed to be fixed, and finally the Wilson line moduli 
other than those corresponding to the inflaton. These do not 
have any direct counterpart in the original KKLMMT model, which 
only considered a single mobile D3-brane. In section 
\ref{technicalstuff} we denoted all Wilson line moduli (including the 
one corresponding to the inflaton) by $A_i$, where $i$ enumerated 
the different stacks of branes. In the following, we will use the
notation introduced in section \ref{inflation}, because we want to 
interpret our results in the context of inflationary models in string 
theory. In order to translate the formulas from section 
\ref{technicalstuff}, we have to use the dictionary just outlined,
in particular (\ref{dictionary}). However, we continue 
to use the formulas derived in the T-dual D9/D5-picture, and it is
understood that the D9-branes (resp.\ D5-branes) 
are mapped to D3-branes (resp.\ D7-branes) after 6 T-dualities. Our
formulas equally hold in the T-dual (D3/D7) picture if one maps the 
fields in the usual way (see e.g.\ \cite{Myers:1999ps}).  
At most instances we give our 
formulas including the other Wilson line moduli (corresponding to 
positions of further D3-branes in the T-dual picture that are present 
for consistency in our toy models but not in the KKLMMT model). 
We denote all Wilson line moduli collectively as 
\be \label{phii}
\phi_i \quad \longleftrightarrow \quad A_i\ ,
\ee
where as in (\ref{newa}) the $\phi_i$ are related to the 
D3-brane positions $(a_i)_4$ and $(a_i)_5$ on the third torus 
according to $\phi_i=U (a_i)_4 - (a_i)_5$. Note that
the index $i$ on $\phi_i$ has a completely different meaning now than
the index in (\ref{d3kin}), where the $i$ denoted the internal direction. 
Here, it enumerates the stacks of branes; all $\phi_i$ correspond to 
locations of the branes along the third torus.

The general form of one-loop 
physical gauge couplings in string theory is given by\footnote{Cf.\ 
\cite{Louis:1996ya} and references therein.}
\be \label{physical}
g^{-2}(\mu) = {\rm Re} (f) + \frac{b}{16 \pi^2} \ln 
\left( \frac{{M}^2_{\rm str}}{\mu^2} \right) + \ldots\ ,
\ee
where $b$ is the one-loop beta function coefficient and 
the dots stand for moduli dependent terms that are not real parts of  
holomorphic functions, as opposed to the ${\rm Re} (f)$ term. 
This first term is the 
Wilsonian coupling from integrating out heavy fields, 
whereas the second term and 
the non-holomorphic contributions are 
due to light fields, with masses below the scale $\mu^2$
at which the coupling is probed. Formula (\ref{physical}) is valid
both for all gauge couplings in $\cn =1$ (where $f$ is the gauge 
kinetic function) and for non-abelian couplings in $\cn=2$ supersymmetric 
theories (in which case $f$ is related to the prepotential). 
Abelian gauge couplings in $\cn=2$ theories are a bit more
complicated due to a possible mixing of the abelian gauge fields 
with the graviphoton.

For simplicity, let us now consider the case with 
vanishing Wilson lines on the 5-branes, such that the 
5-brane gauge group is the unbroken $SU(16)$, i.e.\ we take all 
$\vec{a}_a=\vec{0}$. Moreover, in order to make contact to the 
formulas of section \ref{tree}, we assume
that the 9-brane gauge-group is completely broken to its 
abelian subgroup so that all $N_i=1$ and all $\vec{a}_i$ 
are distinct and nonzero.
In this case we can read off the one-loop corrected 
gauge coupling of the 5-brane gauge group from 
the sum of (\ref{gaugec}) (with $\alpha'=1/2$)
and (\ref{correct}), using $\chi\sim \mu^2$, and find
\be \label{1loopgc}
g^{-2}_{(5)} = \frac{1}{\pi \sqrt{2}} \, e^{-\Phi_{10}} 
\sqrt{G} + \frac{1}{8\pi} U_2 \sum_i (a_i)_4^2  
- \frac{1}{16 \pi^2} \sum_i 
\ln \left|\frac{\tht_1(\phi_i,U)}{\sqrt{\eta(U)}}\right|^2
+ \frac{1}{4\pi^2} \ln(8 \pi^3 \mu^2 \sqrt{G} U_2)\ ,
\ee
where $U_2 = {\rm Im} (U)$.
Comparing the first two terms on the 
right hand side with (\ref{ssprime}) (again for $\alpha'=1/2$) 
and (\ref{ss'}) we see that they combine to ${\rm Re}(-i \rho)$ with 
the modified field $\rho$. The third term on the right hand side of 
(\ref{1loopgc}) is an additional one-loop contribution to the gauge coupling 
which is the real part of a holomorphic function in the variables 
$U$ and $\phi_i$, and the last term corresponds to contributions 
from massless modes that are not given by the real part 
of a holomorphic function. From (\ref{1loopgc}) we read off  
\be \label{endresult}
f_{(5)}=-i\rho - \frac{1}{8 \pi^2} \sum_i 
\ln \tht_1(\phi_i,U) + \frac{1}{\pi^2} \ln \eta(U)\ ,
\ee
involving the modified K\"ahler modulus but also 
including an additional dependence on the 9-brane
scalars. This solves the rho problem described in 
the introduction. Of course, we could have done the 
same analysis with 5- and 9-branes exchanged, in which case we would 
have found that the 9-brane gauge kinetic coupling depends holomorphically
on the modified field $S$ of (\ref{ss'}). 


\subsection{$\cn=1$}
\label{intern1}

There are two different $\cn = 1$ models that we considered in 
section \ref{none}, the $\mathbb{T}^6/\mathbb{Z}_6'$ model
\cite{Aldazabal:1998mr,AnBaDu99} and the 
$\mathbb{T}^6/(\mathbb{Z}_2 \times \mathbb{Z}_2)$ model 
\cite{Berkooz:1996dw}. Again, readers already familiar with that section 
can skip the following two paragraphs.\footnote{As for the ${\cal N}=2$ 
case also here we use the notation of section \ref{inflation}.
The dictionary to the variables of section \ref{technicalstuff} is 
basically the same as the one given in the last subsection.} 

The $\mathbb{T}^6/\mathbb{Z}_6'$ model is 
defined in terms of the eigenvalues 
$\exp(2\pi iv)$, $v=(1,-3,2)/6$, of 
its generator $\Theta$. The 
open string sector is similar to the $\cn=2$ model just discussed,
i.e.\ it has 32 D9-branes and 32 D5-branes wrapped around the 
third torus. This leads to a gauge group 
$(U(4)^2 \times U(8))_{{\rm D}9} \times 
(U(4)^2 \times U(8))_{{\rm D}5}$. The moduli space of 
the untwisted moduli is given by 
three copies of $SU(1,1)/U(1)$ for each of the three 
generic K\"ahler moduli of the 
three 2-tori, and one extra copy each for the dilaton and 
the complex structure modulus of the second torus. 
The complex structures of the first 
and the third torus, around which the 5-branes are wrapped,
are no moduli. Rather, they are fixed to some rational 
values.\footnote{For the moduli spaces of the untwisted moduli in $\cn=1$
orientifolds see e.g.\ \cite{Ibanez:1992hc}.}
In this case the dependence of the one-loop correction to the
5-brane gauge couplings on the 9-brane Wilson line moduli is 
given in (\ref{g5n1}) and (\ref{g5n1b}).

The $\mathbb{Z}_2 \times \mathbb{Z}_2$ model is defined by the action of the
three orbifold group elements $\Theta_p$, $p=1,2,3$, on
$\mathbb{T}^6=\mathbb{T}_1^2\times\mathbb{T}_2^2\times\mathbb{T}_3^2$,
where each $\Theta_p$ is a reflection along two 2-tori,
leaving the torus $\mathbb{T}_p^2$ invariant. 
Now the open string sector 
is more complicated. In addition to 
32 D9-branes there are three sets of 
32 D5-branes, wrapped around the three tori that make up the 
$\mathbb{T}^6$. The resulting gauge group is  
symplectic, specifically $Sp(8)_{\rm D9}\times Sp(8)_{{\rm
D}5_1}\times Sp(8)_{{\rm D}5_2}\times Sp(8)_{{\rm
D}5_3}$.\footnote{Note 
footnote \ref{nichsoschlimm}, however.}
Another difference to the $\mathbb{Z}_6'$ case is that 
the complex structure moduli of all three tori are moduli and 
therefore the untwisted moduli space consists of six 
copies of $SU(1,1)/U(1)$ for the geometric moduli and one
copy for the dilaton. 
The dependence of the one-loop correction to one of the
5-brane gauge couplings on the 9-brane Wilson line moduli is 
given in (\ref{g5n1z2}).

We start the discussion of our results with the 
$\mathbb{Z}_6'$ model. Adding (\ref{g5n1}) to (\ref{gaugec}) we derive 
\beqn \label{1loopgcz6}
g^{-2}_{(5)} &=& \frac{1}{3 \pi \sqrt{2}} \, e^{-\Phi_{10}} 
\sqrt{G} + \frac{1}{12\pi} U_2 \sum_i 
n_i \Big((a_i)_4^2 + (a_i)_5^2-(a_i)_4(a_i)_5\Big) \non
&& \mbox{} - \frac{1}{48 \pi^2} \sum_i  n_i \sum_{m=0}^2
\ln|\tht_1(\phi^{\Theta^m}_i,U)|^2
+ \ldots \ ,
\eeqn
where compared to the $\cn=2$ case 
there is an additional factor of $1/3$ in the tree-level contribution 
due to the smaller volume of the orbifolded torus and $n_i$ denotes the
number of 9-branes in the $i$-th stack. This formula is valid 
for all three factors of the gauge group $U(4)^2\times U(8)$ 
on the 5-branes. The dots stand for
further one-loop corrections that do 
not depend on the Wilson line moduli.
These could in principle be extracted from \cite{AnBaDu99} and contain terms 
that would depend on the only complex structure modulus $U'$ of the model, 
in the form 
$\ln|\eta(U')|$, whereas the $U$
appearing in (\ref{1loopgcz6}) is the complex structure of the 
third torus, which is not a modulus, as we already mentioned.

In the $\cn=1$ case we do not know of a derivation of the 
proper K\"ahler coordinates in the presence of open string scalars
from a KK reduction.
The analogy to the $\cn = 2$ case suggests that the 
sum of the first two terms in (\ref{1loopgcz6}) should form the 
imaginary part of the K\"ahler modulus (of the third torus) 
in the $\mathbb{Z}_6'$ model, i.e.
\be \label{modkaehler}
{\rm Im}(\rho) = \frac{1}{3 \pi \sqrt{2}} \, e^{-\Phi_{10}} 
\sqrt{G} + \frac{1}{12\pi} U_2 \sum_i 
n_i \Big((a_i)_4^2 + (a_i)_5^2-(a_i)_4(a_i)_5\Big)\ .
\ee
Geometrically, this just means we propose to define the $\cn=1$ version of 
the corrected coordinate by summing over the three $\Theta$-images of the 
correction that appeared in $\cn=2$, and properly normalized.  
This is supported by the fact that the one-loop correction included 
in (\ref{1loopgcz6}) originates from open strings stretched between 
5-branes and 9-branes that, for large enough Wilson line moduli,
do not have any light fields in their spectrum. Thus they only
contribute to the Wilsonian gauge coupling, i.e.\ their 
contribution to the gauge coupling should be the real part of a
holomorphic function in the proper K\"ahler coordinates. 
In addition, (\ref{1loopgcz6}) contains further dependence
on the open string scalars, one of which is meant to be 
interpreted as the inflaton. Assuming that the Wilson line 
moduli are still given in the form (\ref{newa}), we read off 
the gauge kinetic function for the D5-brane gauge groups
\beqn \label{endresultn1}
f_{(5)} &=& -i\rho - \frac{1}{24 \pi^2} 
\sum_i  n_i \sum_{m=0}^2 \ln \tht_1(\phi^{\Theta^m}_i,U) + \ldots \non
&=& -i\rho - \frac{1}{8 \pi^2} \sum_i  n_i \ln(\phi_i) + 
\co (\phi^6) + \ldots\ ,
\eeqn
where, again, we neglected one-loop corrections that are 
independent of the Wilson line moduli and the second equality 
comes from the fact that the Wilson line moduli 
on the images under $\Theta$ are related 
by multiplication with a phase, i.e.\ $\phi^{\Theta}=e^{2\pi i/3}\phi$.
This implies that $\phi^{2m} + (\phi^{\Theta})^{2m} + (\phi^{2 \Theta})^{2m} = 0$ for 
all integers $m$ that are not multiples of $3$.
A few comments are in order here. 

$i)$ First, we see that 
(\ref{endresultn1}) contains the modified K\"ahler modulus (in fact, it 
was defined so that this would be the case). This is the solution 
to the rho problem that we propose in the $\cn=1$ case. Note 
that there is a slight difference to the model of KKLMMT. In their case 
only one K\"ahler modulus is present, whereas in our case there are 
three untwisted K\"ahler moduli. It is the one measuring the volume 
of the third torus that enters $f_{(5)}$ 
at tree level, so this is is the relevant 
K\"ahler modulus for 
the rho problem in our case. This modulus is mapped via 
T-duality to the volume of the four-cycle (transverse to the third torus) 
around which the 7-branes (the T-duals of the 5-branes) are 
wrapped.\footnote{Note that for this mapping of volumes, it is important 
that there is a factor of $e^{-\Phi_{10}}$ in the definition of $\rho$,
cf.\ (\ref{ssprime}).}

$ii)$ Moreover, the gauge kinetic function (\ref{endresultn1}) 
blows up for $\phi_i\to 0$. The beta function coefficient of the 
$SU(8)$-factor of the 5-brane gauge group without 
Wilson line moduli is $b(SU(8))=-6$ \cite{AnBaDu99} and it can 
only become more negative if bifundamental matter turns massive 
for non-vanishing Wilson lines. 
Therefore the gauge group is asymptotically free, and at low 
energies a non-perturbative 
superpotential due to gaugino condensation is generated.
From (\ref{supergeneral}) we read off  
\be
W_{\rm nonpert} ~\sim~ \exp\Big\{\alpha (i\rho + 
\frac{1}{8 \pi^2} \sum_i  n_i \ln(\phi_i) 
+ \co (\phi^6) + \ldots)\Big\}\ .
\ee
As $\alpha$ is positive for a negative beta function coefficient 
$b$, we see that the 
non-per\-tur\-ba\-tive superpotential vanishes in the limit when,
in the T-dual language,
the 3-branes hit the cycle on which the 7-branes are wrapped, in 
accord with the results found in \cite{Ganor:1996pe}.\footnote{Strictly
speaking, it is no longer valid to integrate out the modes from 59 strings
when determining the gauge kinetic function 
in this limit; one has to introduce an IR-cutoff as in 
(\ref{ghil2}).} 

$iii)$ Furthermore, the superpotential develops an explicit dependence 
on the open string scalars at one-loop level, hence the shift symmetry 
discussed in \cite{Hsu:2003cy,Hsu:2004hi,Firouzjahi:2003zy} 
is violated by the one-loop corrections.

$iv)$ Finally, there is no quadratic term in the open string 
fields in (\ref{endresultn1}). Thus the one-loop corrections to 
the superpotential in the $\mathbb{Z}_6'$ model turn out to be 
incapable of reducing the inflaton mass. However, this is just an 
accident occuring in this model due to the $\mathbb{Z}_3$ symmetry 
of the Wilson line.\footnote{We do expect a mass term to appear 
also in this model if one deforms away from the orbifold limit,
but we will not do so here.}

This last problem is absent in the 
$\mathbb{Z}_2 \times \mathbb{Z}_2$ model to which we turn now. In this case  
the one-loop corrected 
gauge coupling of one of the 5-brane gauge groups is 
\be \label{g5z2}
g^{-2}_{(5)} = \frac{1}{4 \pi \sqrt{2}} \, e^{-\Phi_{10}} 
\sqrt{G} + \frac{1}{4\pi} U_2 \sum_i n_i (a_i)_4^2 
- \frac{1}{4\pi^2} \sum_{i} n_i \ln|\tht_1(\phi_i,U)| + \ldots \ ,
\ee
where, as in the $\mathbb{Z}_6'$ model, the tree-level term contains 
an additional factor (1/4, this time) due to the orbifolding 
of the torus and the dots stand for all the one-loop corrections that do 
not depend on the Wilson line moduli. In analogy to 
\cite{Kaplunovsky:1987rp,AnBaDu99} we expect them to again include
terms of the form $\ln|\eta(U^{(p)})|$ 
for all three complex structure moduli $U^{(p)}$, $p=1,2,3$.
Formula (\ref{g5z2}) 
holds true for each of the three different types 
of 5-branes, and $\sqrt{G}$ and $U\in\{U^{(p)}\}$ denote 
the volume and complex structure 
of the corresponding 
torus around which they are wrapped. To keep the notation 
simple we just focus on one of them, without explicitly indexing the 
coupling or the volume and complex structure of the torus.
Moreover, (\ref{g5z2}) also holds if the gauge group on the stack of 
5-branes is broken to some smaller $Sp$ group by moving some of 
them out of the origin. 

As above, we suggest that the 
K\"ahler modulus for the torus, around which the 5-branes are wrapped, 
is modified at one-loop level to 
\be
{\rm Im}(\rho) = \frac{1}{4 \pi \sqrt{2}} \, e^{-\Phi_{10}} 
\sqrt{G} + \frac{1}{4\pi} U_2 \sum_i n_i (a_i)_4^2
\ee
and the gauge kinetic function is given by
\be \label{gkf5}
f_{(5)} = -i \rho - \frac{1}{4\pi^2} \sum_{i} n_i \ln\tht_1(\phi_i,U) + \ldots\ .
\ee
To calculate the beta function coefficient of the orientifold model 
we need the charged spectrum derived in \cite{Berkooz:1996dw}. At a generic 
point in the moduli space of the $\cn=1$ supersymmetric theory 
considered there, 
all bifundamental matter is massive due to
Wilson lines, and the only massless charged matter resides in 
the vector multiplet and three chiral multiplets transforming 
in the antisymmetric representation of the unbroken $Sp$ group of
the 5-branes under consideration. Since the KKLMMT model furthermore involves 
(at least spontaneous) supersymmetry breaking, one would expect that mass 
terms will be generated 
even for these matter fields, since only 
chiral fermions 
should generically remain massless.\footnote{For possible forms 
of soft breaking terms in (orientifold) 
models with D-branes, see \cite{Camara:2003ku,Kors:2003wf}.}
In any case, if the rank of the symplectic gauge group has been 
broken to a small 
enough value, the beta function is negative even with 
the antisymmetric matter 
remaining massless, and gaugino condensation
occurs at low energies. Substituting (\ref{gkf5}) into the formula 
for the resulting superpotential (\ref{supergeneral}) now gives
\be \label{Wz2z2}
W_{\rm nonpert} ~\sim~ \exp\Big\{\alpha (i\rho + \frac{1}{4 \pi^2} \sum_i  n_i 
\ln\tht_1(\phi_i,U) + \ldots)\Big\}\ .
\ee
When expanded around generic values for the open string scalars,
the potential that follows from this superpotential 
in general possesses both linear and quadratic terms in the $\phi_i$, 
whose coefficients
depend on all the complex structure moduli. 
Thus the inflaton mass correction depends on the values at which the 
complex structure moduli are fixed by the background fluxes. It is 
then plausible that it is possible to fine-tune the fluxes to 
achieve a correction that leads to a value for the mass that is small
enough to allow 
for slow roll inflation, a possibility that was anticipated 
in \cite{Kachru:2003sx}. However, 
to obtain a conclusive answer, the one-loop corrections to the 
K\"ahler potential have to be known as well. 
Since they are not known,
we hope to come back to them in a future publication
\cite{goldandglory}.\footnote{One-loop corrections to the K\"ahler metric 
in the $\mathbb{Z}_6'$ model without Wilson lines were
calculated in \cite{Bain:2000fb}.} 

Our lack of knowledge of the corrections to the 
K\"ahler potential notwithstanding, and ignoring the fact that 
the $\mathbb{Z}_2 \times \mathbb{Z}_2$ orientifold is at best 
a toy model for the actual KKLMMT setup, let us 
conclude by combining our result (\ref{Wz2z2}) with 
the analysis of \cite{Kachru:2003sx},\footnote{See their 
appendix F in particular.} 
in order to get a rough picture 
of fine-tuning the inflaton mass. To do so, we 
focus on a single dynamical D3-brane (i.e.\  
$n=1$ in (\ref{Wz2z2})) whose 
scalar $\phi$ we interpret as the inflaton field. We then want 
to expand the superpotential to quadratic order in $\phi$. 
In the KKLMMT model one considers a 
D3-brane that is well separated from both the D7-branes and the 
anti D3-branes at the tip of the throat. 
In principle, any such value would be a valid
expansion point. In practice, however, it is most convenient 
to perform the explicit expansion of $\tht_1(\phi,U)$ either around $\phi=0$ or
$\phi=1/2$. As we do not want to consider the special point 
$\phi=0$, at which the gauge symmetry gets enhanced and 
new massless states appear, we choose to expand 
(\ref{Wz2z2}) around $\phi=1/2$ for definiteness. 
Shifting $\phi\rightarrow \phi+1/2$, so that $\phi$ now denotes the 
fluctuations around $1/2$,
we can use the relation $\tht_1(\phi+1/2,U)=\tht_2(\phi,U)$ and
(\ref{firstder}), (\ref{secondder}) to expand
\be
\tht_1(\phi+1/2,U) = \tht_2(0,U) \Big(1 - \frac{\pi^2}{6} [E_2(U) 
+ \tht_3^4(0,U) + \tht_4^4(0,U)] \phi^2 +\ldots \Big)\ .
\ee
Substituting this into the formula for the superpotential 
(\ref{Wz2z2}) we obtain
\be \label{superexpand}
W ~\sim~ W_0(U^{(p)}) + C(U^{(p)},\phi_i) e^{\alpha i\rho} 
\Big( 1 - \frac{\alpha}{24} 
[E_2(U) + \tht_3^4(0,U) + \tht_4^4(0,U)] \phi^2 + \ldots \Big)\ ,
\ee
where we reinstated the contribution $W_0$ 
to the superpotential coming from the 3-form fluxes, and
the function $C$ depends on all three 
complex structure moduli $U^{(p)}$ and the Wilson line moduli 
$\phi_i$ other than the inflaton field $\phi$. In principle the additional 
dependence of the non-perturbative superpotential (and possibly also 
of the K\"ahler potential) on the 
complex structure moduli would require re-minimizing 
the potential with respect to them. It is conventional, however,
to assume a separation of scales, such that the complex structure moduli
receive a flux-induced mass term that is much bigger than the scale 
of non-perturbative physics leading to gaugino condensation. 
In this case one can assume that the additional $U^{(p)}$-dependence 
in (\ref{superexpand}) does not alter their values 
at the minimum very much, so that they can be considered constant. 
This is also the philosophy that we follow here.

Comparing (\ref{superexpand}) with 
formula (F.1) of \cite{Kachru:2003sx}\footnote{Note that our $\alpha$
corresponds to their $a$, our $C(U^{(p)},\phi_i)$ to their $A$ 
and there is a relative factor of $i$ 
in our definition of the K\"ahler coordinates as compared to theirs.},
i.e.\ $W=W_0+C e^{\alpha i\rho} (1+\delta \phi^2)$ in our notation,
we read off
\be
\delta = - \frac{\alpha}{24} 
[E_2(U) + \tht_3^4(0,U) + \tht_4^4(0,U)]\ .
\ee
This quantity determines whether the one-loop
correction to the superpotential can help to lower the inflaton mass or not, 
as can be inferred from (F.8) of \cite{Kachru:2003sx},
\be \label{mphi}
{m_{\varphi}^2 \over H^2} \;=\; 2 - {2 |V_{\rm AdS}| \over V_{\rm
dS}}
\,\Delta\ ,
\ee
where $\varphi$ is the canonically normalized inflaton field, 
\be \label{r}
\Delta=\beta - 2\beta^2\  \qquad {\rm with}  \qquad 
\beta = -\frac{\delta}{\alpha}= 
\frac{1}{24} [E_2(U) + \tht_3^4(0,U) + \tht_4^4(0,U)]
\ee
\begin{figure}[h]
\begin{center}
  \resizebox{7cm}{!}{\psfig{figure=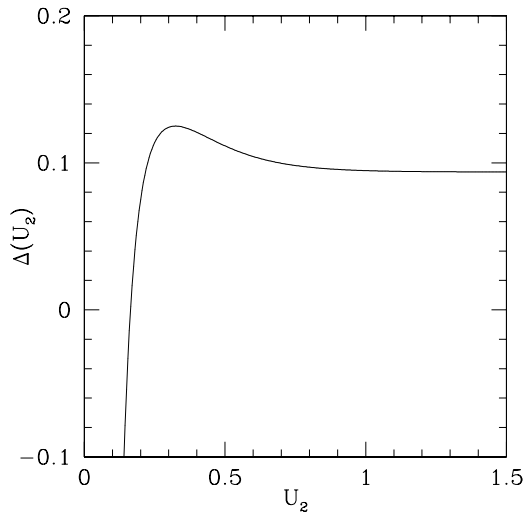,width=6cm}}
\caption{The function $\Delta$ of eq.\ (\ref{r}); 
for positive values the inflaton mass
is lowered by the one-loop corrections to the superpotential.} 
\label{reissdorfkoelsch}
\end{center}
\end{figure}
and $V_{\rm AdS}$ and $V_{\rm dS}$ are explained in 
\cite{Kachru:2003sx}. (However, the
mass formula (\ref{mphi}) does not 
include any contributions 
from one-loop corrections to the K\"ahler potential.)
Obviously, if $\Delta$ 
is positive for some value of $U$, the inflaton mass is lowered.
Note that neither 
the value of the beta function coefficient nor
the function $C(U^{(p)},\phi_i)$ enter into $\Delta$ and so it 
is insensitive to the uncertainties with which these quantities 
are afflicted.  

In fig.\ \ref{reissdorfkoelsch} we plot the dependence
of $\Delta$ on the value of $U$, which, for ease of presentation, 
we assume to be imaginary at its
minimum, i.e.\ we assume that ${\rm Re} (U)=0$ is consistent 
with minimization of 
the flux-induced potential. 
It is clear in the plot that for a large range of 
values for ${\rm Im} (U)$, the function $\Delta(U)$ is positive,
so if the complex structure is 
stabilized in this range, the inflaton mass is lowered by the 
one-loop corrections to the superpotential. The explicit value also 
depends
 on the values of the other complex structure and open string 
moduli, as well as on the beta function coefficient and the one-loop 
corrections to the K\"ahler potential. Moreover, our analysis is not 
able to take into account any possible effects of the warp factor present 
in the actual KKLMMT model. Due to these uncertainties, we refrain from 
giving a numerical correction to the inflaton mass, 
which would require computing the value of 
$|V_{\rm AdS}| / V_{\rm dS}$ 
in (\ref{mphi}) by minimizing the full scalar potential. 

Nevertheless, our conclusion is that the open string one-loop
corrections to the superpotential should, in general, 
provide the added flexibility needed to fine-tune the inflaton 
mass to small values. In the philosophy of 
\cite{Giddings:2001yu,Kachru:2003aw,Kachru:2003sx}
this fine-tuning is achieved by choosing appropriate values for 
the 3-form fluxes, because their values determine the warp factor
(and thus the effective tension of the anti-D3-branes at the tip of 
the throat) and the values at which the complex structure moduli 
are fixed. (As stressed on p.\ 37 of appendix F in \cite{Kachru:2003sx}, 
this fine-tuning is only numerical at the 1\% level.) 
The possibility to lower the inflaton mass in the KKLMMT model via moderate
fine-tuning was already anticipated in \cite{Kachru:2003sx}, 
assuming the superpotential may contain
terms quadratic in the 
inflaton field with moduli-dependent coefficients. Even though 
our calculation is not realistic enough to make this completely quantitative, 
the merit of our result is to show that 
terms quadratic in 
the inflaton field with moduli-dependent coefficients
do indeed appear in the 
superpotential in our explicit string theory model. They are 
induced by open string one-loop (annulus)
corrections to the gauge kinetic function on the D7-branes, 
and this gauge kinetic function
appears in the superpotential after gaugino condensation. We believe
this qualitative result to be generic in models with D3/D7-branes,
not an artifact of the simplifying assumptions in our
explicit calculation, and that it will survive in more 
realistic cases.

\vspace{1cm} 

\begin{center}
{\bf Acknowledgments} 
\end{center}

It is a pleasure to thank Carlo Angelantonj, David Berenstein,
Emilian Dudas, Shamit Kachru, Renata Kallosh,  
Liam McAllister, Joe Polchinski, Massimo Porrati, 
Radu Roiban, Stephan Stieberger, David Tong, and Angel Uranga for 
illuminating insights and helpful advice through discussions 
and email conversations. 
M.~B. was supported by the 
Wenner-Gren Foundations, 
and M.~H. by the German Science Foundation (DFG). 
Moreover, the research of M.~B.\ and M.~H. was supported in part by 
the National Science Foundation under
Grant No. PHY99-07949. 
The work of B.~K.~was supported by 
the German Science Foundation (DFG) and in part by
funds provided by the U.S. Department of Energy (D.O.E.) 
under cooperative research agreement
$\#$DF-FC02-94ER40818.  


\clearpage
\begin{appendix}

\section{One-loop amplitude for $\mathbb{T}^4/\mathbb{Z}_2\times \mathbb{T}^2$}
\label{appampl}

In this appendix we summarize the technical steps to compute 
the relevant one-loop amplitudes. 
The conventional method to incorporate the background gauge 
fields into the amplitude  
is to introduce them in the loop channel by replacing the momentum integration 
in the directions with a magnetic background field by the
degeneracy per unit area of the Landau levels, 
and further shift the modings of the string world sheet fields by 
\beqn \label{eps}
\bfe_i = \frac{1}{\pi} {\rm arctan}( 2\pi\alpha' {\bf F}_i)\ , \quad 
\bfe_a = \frac{1}{\pi} {\rm arctan}( 2\pi\alpha' {\bf F}_a)
\eeqn
according to the sigma model boundary conditions 
\cite{Bachas:bh,Bachas:1996zt}. 
We actually prefer to compute the diagrams 
without the gauge field in the loop channel, and then 
later directly implement its effects on 
the tree channel result. 

The open string amplitudes in (\ref{oneloop}) are defined in the loop channel 
\beqn \label{loopch}
{\cal M} &=& \frac{\sqrt{-g_4}}{(4\pi^2\alpha')^2} \int_0^\infty \frac{dt}{(2t)^3} 
{\rm Tr}_{\rm op}^{\rm NS-R} \left( \frac{\Omega}{2} \frac{ 1+(-1)^F }{2} \frac{1+\Theta}{2} 
  e^{-2\pi t {\cal H}_{\rm op} } \right) 
\\
&=& \sqrt{-g_4} \int_0^\infty \frac{dt}{t} \sum_{k=0,1} 
\left( \sum_i {\cal M}_{i}^{(k)} (-q) + \sum_a {\cal M}_{a}^{(k)} (-q) \right) 
\non
{\cal A} &=& \frac{\sqrt{-g_4}}{(4\pi^2\alpha')^2} 
\int_0^\infty \frac{dt}{(2t)^3} {\rm Tr}_{\rm op}^{\rm NS-R} \left( \frac{1}{2} \frac{ 1+(-1)^F }{2} \frac{1+\Theta}{2} 
  e^{-2\pi t {\cal H}_{\rm op} } \right) 
\non
&=& \sqrt{-g_4} 
\int_0^\infty \frac{dt}{t} \sum_{k=0,1} \left( \sum_{i,j} {\cal A}_{ij}^{(k)} (q) 
 + \sum_{a,b} {\cal A}_{ab}^{(k)} (q) 
 +  \sum_{i,a} \left( {\cal A}_{ia}^{(k)} (q) + {\cal A}_{ai}^{(k)} (q) 
\right) \right),  
\nonumber 
\eeqn
where $k=0,1$ stands for the power of $\Theta$ inserted in the trace. 
The contributions with $k=1$ 
originate from the twisted components of the relevant boundary states, 
and are thus called twisted 
contributions. 
The arguments of theta-functions are abbreviated $q=e^{-2\pi t}$ and 
$\tilde q=e^{-4\pi l}$ (used below), 
i.e.\ 
we leave out the second argument of $\vartheta[{\alpha\atop\beta}](\nu,it)$ 
or $\vartheta[{\alpha\atop\beta}](\nu,2il)$. 
For the 

In the loop channel, the presence of the background gauge field amounts to 
a shift of 
the first argument $\nu$ of the theta functions by $\bfe_i$, 
etc. 
This implies that we can ignore the ${\cal N}=4$ ``subsectors'' of 
the amplitude, since 
\beqn 
{\cal M}_{i}^{(0)} \ , \ {\cal M}_{a}^{(1)} \ , \ 
{\cal A}_{ij}^{(0)} \ , \  {\cal A}_{ab}^{(0)} ~=~ {\cal O}(1) + {\cal O}(\bfe_i^4,\bfe_a^4)  
\eeqn
by (\ref{Riemannzero}). 
The constant tadpole ${\cal O}(1)$ cancels 
after adding up all contributions to the one-loop amplitude.  
So we are interested in the quadratic term in the expansion of 
\beqn
\sum_i {\cal M}_{i}^{(1)} + \sum_a {\cal M}_{a}^{(0)} 
+ \sum_{i,j} {\cal A}_{ij}^{(1)} + \sum_{a,b} {\cal A}_{ab}^{(1)} 
+ \sum_{k=0,1} \sum_{i,a} \left( {\cal A}_{ia}^{(k)} + {\cal A}_{ai}^{(k)} 
\right)\ .
\eeqn
To implement the Wilson lines, we use the notation introduced 
in section \ref{seconeloop}, and write bold ${\bf \vec A}$ 
for the tensor-valued quantities. As well, we extend the 
usual matrices $\gamma$ to this notation and formally write a single trace 
(cf.\ the discussion below (\ref{fa})). 
The results are 
\beqn
{\cal M}_{i}^{(1)} + {\cal M}_{a}^{(0)} &=& 
 - \frac{1}{16 (8 \pi^2 \alpha')^2 t^2} {\rm tr}\left( \gamma_{\Omega\Theta i}^{-1} \gamma_{\Omega\Theta i}^{\rm T}  
  \thba{2{\bf \vec A}_i}{\vec 0}(0,2it G^{-1}\alpha') 
\right. 
\\
&& \left. 
\hspace{1.5cm}  
+ \gamma_{\Omega a}^{-1} \gamma_{\Omega a}^{\rm T} 
   \thba{2{\bf \vec A}_a}{\vec 0}(0,2it G^{-1}\alpha') \right) 
\sum_{\alpha,\beta} \eta_{\alpha\beta} \frac{\vartheta[{\alpha\atop\beta}]^2\vartheta[{\alpha\atop\beta+1/2}]^2}
                                            {\eta^6\frac14 \vartheta[{1/2 \atop 0}]^2}  
\ , 
\non 
{\cal A}_{ij}^{(1)} + {\cal A}_{ab}^{(1)} &=& 
  \frac{1}{16 (8 \pi^2 \alpha')^2 t^2} {\rm tr} \left( \bfg_{\Theta i} \bfg^{-1}_{\Theta j} 
  \thba{{\bf \vec A}_{ij}}{\vec 0}(0,2it G^{-1}\alpha') 
 \right. 
\non
&& \left. 
\hspace{1.5cm} 
+ \bfg_{\Theta a} \bfg^{-1}_{\Theta b} 
  \thba{{\bf \vec A}_{ab}}{\vec 0}(0,2it G^{-1}\alpha') \right) 
\sum_{\alpha,\beta} \eta_{\alpha\beta} \frac{\vartheta[{\alpha\atop\beta}]^2\vartheta[{\alpha\atop\beta+1/2}]^2}
                                            {\eta^6\frac14 \vartheta[{1/2\atop 0}]^2}  
 \ , 
\non
{\cal A}_{ia}^{(0)} + {\cal A}_{ai}^{(0)} &=&  
 \frac{1}{16 (8 \pi^2 \alpha')^2 t^2} {\rm tr}
 \Big( \bfg_{i} \bfg^{-1}_{a} \thba{{\bf \vec A}_{ia}}{\vec 0}(0,2it G^{-1}\alpha') 
\non 
&&\hspace{1.5cm} 
+ \bfg_{a} \bfg^{-1}_{i} 
  \thba{{\bf \vec A}_{ai}}{\vec 0}(0,2it G^{-1}\alpha') \Big)  
\sum_{\alpha,\beta} \eta_{\alpha\beta} \frac{\vartheta[{\alpha\atop\beta}]^2\vartheta[{\alpha+1/2 \atop\beta}]^2}
                                            {\eta^6\vartheta[{0\atop 1/2}]^2}  
 \ , 
\non
{\cal A}_{ia}^{(1)} + {\cal A}_{ai}^{(1)} &=&  
 \frac{1}{16 (8 \pi^2 \alpha')^2 t^2} {\rm tr}
 \Big( \bfg_{\Theta i} \bfg^{-1}_{\Theta a} \thba{{\bf \vec A}_{ia}}{\vec 0}(0,2it G^{-1}\alpha') 
\non
&&  \hspace{1.5cm}        
+ \bfg_{\Theta a} \bfg^{-1}_{\Theta i} 
 \thba{{\bf \vec A}_{ai}}{\vec 0}(0,2it G^{-1}\alpha') \Big)  
\sum_{\alpha,\beta} \eta_{\alpha\beta} \frac{\vartheta[{\alpha\atop\beta}]^2\vartheta[{\alpha+1/2 \atop\beta+1/2}]^2}
                                            {\eta^6\thba{0}{0}^2}  
 \ .
\nonumber
\eeqn
Transforming to the tree-channel by 
$t=1/(2l)$ for the annuli and $t=1/(8l)$ for the M\"obius strip, 
one finds the amplitude in a form 
\beqn \label{directch}
\tilde{\cal M} &=& 
\sqrt{-g_4} \int_0^\infty dl\, \sum_{k=0,1} 
\left( \sum_i \tilde{\cal M}_{i}^{(k)} (-\tilde{q}) + \sum_a \tilde{\cal M}_{a}^{(k)} (-\tilde{q}) \right) \ , 
\\
\tilde{\cal A} &=& 
\sqrt{-g_4} \int_0^\infty dl\,  \sum_{k=0,1} \left( \sum_{i,j} \tilde{\cal A}_{ij}^{(k)} (\tilde{q}) 
 + \sum_{a,b} \tilde{\cal A}_{ab}^{(k)} (\tilde{q}) 
+  \sum_{i,a} \left( \tilde{\cal A}_{ia}^{(k)} (\tilde{q}) 
+ \tilde{\cal A}_{ai}^{(k)} (\tilde{q}) \right) \right)  
\nonumber
\eeqn
with (again focusing on the ${\cal N}=2$ sectors that contribute
to the gauge coupling corrections)  
\beqn \label{treechan}
\tilde{\cal M}_{i}^{(1)} + \tilde{\cal M}_{a}^{(0)} &=& 
 - \frac{4}{(8\pi^2\alpha')^2}\sqrt{G}\alpha'^{-1}\, {\rm tr}\Big( \gamma_{\Omega\Theta i}^{-1} \gamma_{\Omega\Theta i}^{\rm T} 
 \thba{\vec{0}}{\vec{0}}(2{\bf \vec A}_i,4ilG\alpha'^{-1}) 
 \non
&& \hspace{1.5cm}
 + \gamma_{\Omega a}^{-1} \gamma_{\Omega a}^{\rm T} \thba{\vec{0}}{\vec{0}}(2{\bf \vec A}_a,4ilG\alpha'^{-1}) 
 \Big) 
\sum_{\alpha,\beta} \eta_{\alpha\beta} \frac{\vartheta[{\alpha\atop\beta}]^2\vartheta[{\alpha\atop\beta+1/2}]^2}
                                            {\eta^6\vartheta[{1/2 \atop 0}]^2}  
 \ , 
\non
\tilde{\cal A}_{ij}^{(1)} + \tilde{\cal A}_{ab}^{(1)} &=& 
 \frac{1}{4 (8\pi^2\alpha')^2} \sqrt{G}\alpha'^{-1}\, {\rm tr} \Big( \bfg_{\Theta i} 
 \bfg^{-1}_{\Theta j} \thba{\vec{0}}{\vec{0}}({\bf \vec A}_{ij},ilG\alpha'^{-1}) 
\non
&& \hspace{1.5cm}
 + \bfg_{\Theta a} \bfg^{-1}_{\Theta b}  \thba{\vec{0}}{\vec{0}}({\bf \vec A}_{ab} ,ilG\alpha'^{-1}) \Big) 
\sum_{\alpha,\beta} \eta_{\alpha\beta} \frac{\vartheta[{-\beta\atop\alpha}]^2\vartheta[{-1/2-\beta\atop\alpha}]^2}
                                            {\eta^6\vartheta[{0\atop 1/2}]^2}  
 \ , 
\non
\tilde{\cal A}_{ia}^{(0)} + \tilde{\cal A}_{ai}^{(0)} &=&  
 \frac{1}{8 (8\pi^2\alpha')^2} \sqrt{G}\alpha'^{-1}\, {\rm tr} \left( \bfg_{i} \bfg^{-1}_a 
 \thba{\vec{0}}{\vec{0}}({\bf \vec A}_{ia},ilG\alpha'^{-1}) \right)  
\non
&& \hspace{1.5cm} \times  
\sum_{\alpha,\beta} \eta_{\alpha\beta} \frac{\vartheta[{-\beta\atop\alpha}]^2\vartheta[{-\beta\atop\alpha+1/2}]^2}
                                            {\eta^6\vartheta[{1/2 \atop 0}]^2}  
 \ , 
\non
\tilde{\cal A}_{ia}^{(1)} + \tilde{\cal A}_{ai}^{(1)} &=&  
 \frac{1}{8 (8\pi^2\alpha')^2} \sqrt{G}\alpha'^{-1}\, {\rm tr}\left( \bfg_{\Theta i} \bfg^{-1}_{\Theta a}  
 \thba{\vec{0}}{\vec{0}}({\bf \vec A}_{ia},ilG\alpha'^{-1}) \right) 
\non
&& \hspace{1.5cm} \times  
 \sum_{\alpha,\beta} \eta_{\alpha\beta} \frac{\vartheta[{-\beta\atop\alpha}]^2\vartheta[{-1/2-\beta\atop1/2+\alpha}]^2}
                                            {\eta^6\thba{0}{0}^2}  
 \ .
\eeqn
The Klein bottle is normalized such that the untwisted tadpole 
cancellation (which involves also contributions 
from the ${\cal N}=4$ sectors that we did not write out in (\ref{treechan})) 
is achieved with 
\beqn 
&&
\sum_i {\rm tr} (\gamma_{i}) = \sum_a {\rm tr} (\gamma_{a}) = 
\sum_i {\rm tr} ( \gamma_{\Omega i}^{-1} \gamma_{\Omega i}^{\rm T} ) = 
\sum_a {\rm tr} ( \gamma_{\Omega\Theta a}^{-1} \gamma_{\Omega\Theta a}^{\rm T} ) = 32 \ , 
\eeqn
which is solved by 
\beqn \label{gamma1}
\gamma_{i} = 
 {\rm diag}(\underbrace{0,\, ...\, ,0}_{p_i\ {\rm entries}},{\bf 1}_{2N_{i}},0,\, ...\, ,0) \ , 
\quad 
\gamma_{a} = 
 {\rm diag}(\underbrace{0,\, ...\, ,0}_{p_a\ {\rm entries}},{\bf 1}_{2N_{a}},0,\, ...\, ,0) \ , 
\eeqn
and 
\beqn \label{gammaomega}
&&
\gamma_{\Omega i}^{\rm T}=\gamma_{\Omega i}^{-1}\ , \quad 
\gamma_{\Omega i}^{-1} \gamma_{\Omega i}^{\rm T}=
{\rm diag}( \underbrace{0,\, ...\, ,0}_{p_i\ {\rm entries}},{\bf 1}_{2N_i},0,\, ...\, ,0)\ , 
\non
&&
\gamma_{\Omega \Theta a}^{\rm T}=\gamma_{\Omega\Theta  a}^{-1}\ , \quad 
\gamma_{\Omega \Theta a}^{-1} \gamma_{\Omega \Theta a}^{\rm T}={
\rm diag}( \underbrace{0,\, ...\, ,0}_{p_i\ {\rm entries}},{\bf 1}_{2N_a},0,\, ...\, ,0)\ ,
\eeqn
with the same $p_i$ and $p_a$ as in (\ref{wi}) and (\ref{qi}). The twisted contribution vanishes for 
\beqn
\sum_i {\rm tr} (\gamma_{\Theta i}) = \sum_a {\rm tr} (\gamma_{\Theta a}) = 0 \ .
\eeqn 
The solution to the latter condition can be achieved by 
\cite{AnBaDu99}
\beqn \label{gamma2}
\gamma_{\Theta i} &=& 
{\rm diag}(\underbrace{0,\, ...\, ,0}_{p_i\ {\rm entries}},i{\bf 1}_{N_{i}},-i{\bf 1}_{N_{i}},0,\, ...\, ,0) \ , 
\non 
\gamma_{\Theta a} &=& 
{\rm diag}(\underbrace{0,\, ...\, ,0}_{p_a\ {\rm entries}},i{\bf 1}_{N_{a}},-i{\bf 1}_{N_{a}},0,\, ...\, ,0) \ . 
\eeqn
The notation assumes that all $32\times 32$ CP matrices are subdivided into the $(2N_i) \times (2N_i)$ blocks 
(and similarly for $a$) referring to the factors of the gauge group (\ref{gauge}), such that the matrices in 
(\ref{gamma1}) and (\ref{gamma2}) exactly act on the blocks $i$ and $a$. 
We are using the conventions of \cite{Aldazabal:1998mr}, 
such that $\gamma_{\Theta i}^2=\gamma_{\Theta a}^2=-{\bf 1}_{32}$, and later similarly for the 
$\mathbb{Z}_6'$ model $\gamma_{\Theta 9}^6=\gamma_{\Theta 5}^6=-{\bf 1}_{32}$. In this basis, the operation of $\Omega$ 
on the CP labels is off-diagonal, given by \cite{Aldazabal:1998mr} 
\beqn 
\gamma_{\Omega 9} = {\oplus}_{i}\left( 
\begin{array}{cc}
{\bf 0}_{N_i} & {\bf 1}_{N_i} \\ 
{\bf 1}_{N_i} & {\bf 0}_{N_i} 
\end{array}
\right) \ , \quad 
\gamma_{\Omega 5} = \oplus_{a} \left( 
\begin{array}{cc}
{\bf 0}_{N_a} & i{\bf 1}_{N_a} \\ 
-i{\bf 1}_{N_a} & {\bf 0}_{N_a} 
\end{array}
\right) \ . 
\eeqn
To incorporate the background gauge fields, denoted ${\bf F}$ as in section \ref{seconeloop}, 
we make the replacement (cf.\ the discussion in \cite{Blumenhagen:2000wh})
\beqn 
&& {\rm tr} \left( \gamma \thba{\vec{0}}{\vec{0}}({\bf \vec A},ilG\alpha'^{-1})  \right) 
 \frac{\vartheta[{\alpha\atop\beta}](0)}{\eta^3} ~\longrightarrow~
{\rm tr} \left( \gamma (-2\sin(\pi\bfe))  \, 
  \thba{\vec{0}}{\vec{0}}({\bf \vec A},ilG\alpha'^{-1}) \frac{\vartheta[{\alpha\atop\beta}]({\bf \bfe}) } 
 {\vartheta[{1/2 \atop 1/2}]({\bf \bfe})} \right)\ .
\nonumber 
\eeqn
Effectively, we have added phase factors for the world sheet oscillators along 
the space-time directions, where the magnetic field is pointing, 
and the phase prefactor in the 
numerator cancels against that of the denominator. 
Expanding the prefactor $-2\sin(\pi\bfe)$ in 
${\cal F}$ to first order gives back the semiclassical 
result $-4\pi\alpha'{\bf F}$, 
 in accord with point particles in a background magnetic field. 
Using the identity (\ref{Thetader}),
one can expand the oscillator sums in $\bfe$ and 
up to ${\cal O}({\bf F}^4)$ we find
\beqn \label{total} 
\tilde{\cal M}_{i}^{(1)} + \tilde{\cal M}_{a}^{(0)} &=& 
-\frac{8}{(8\pi^2\alpha')^2} \sqrt{G}\alpha'^{-1}\,  {\rm tr}\left( \gamma_{\Omega\Theta i}^{-1} \gamma_{\Omega\Theta i}^{\rm T} 
  \thba{\vec{0}}{\vec{0}}(2{\bf \vec A}_i,4ilG\alpha'^{-1}) (2\pi\alpha'{\bf F}_i)^2 
\right. \non
&& \hspace{1.5cm} \left. 
 + \gamma_{\Omega a}^{-1} \gamma_{\Omega a}^{\rm T} \thba{\vec{0}}{\vec{0}}(2{\bf \vec A}_a,4ilG\alpha'^{-1}) 
(2\pi\alpha'{\bf F}_a)^2 \right) 
 \ , 
\\ 
\tilde{\cal A}_{ij}^{(1)} + \tilde{\cal A}_{ab}^{(1)} &=& 
\frac{1}{2 (8\pi^2\alpha')^2} \sqrt{G}\alpha'^{-1}\, 
{\rm tr} \left( \bfg_{\Theta i} \bfg_{\Theta j}^{-1}  
\thba{\vec{0}}{\vec{0}}({\bf \vec A}_{ij} ,ilG\alpha'^{-1}) (2\pi\alpha'{\bf F}_{ij})^2  
\right. \non
&& \hspace{1.5cm} \left. 
 + \bfg_{\Theta a} \bfg_{\Theta b}^{-1} \thba{\vec{0}}{\vec{0}}({\bf \vec A}_{ab} ,ilG\alpha'^{-1}) (2\pi\alpha'{\bf F}_{ab})^2 \right) 
 \ , 
\non
\tilde{\cal A}_{ia}^{(0)} + \tilde{\cal A}_{ai}^{(0)} &=&   
\frac{1}{4 (8\pi^2\alpha')^2} \sqrt{G}\alpha'^{-1}\, {\rm tr}\left(  \bfg_{i}\bfg_{a}^{-1}  
 \thba{\vec{0}}{\vec{0}}({\bf \vec A}_{ia} ,ilG\alpha'^{-1})  (2\pi\alpha'{\bf F}_{ia})^2   \right) 
 \ , 
\non
\tilde{\cal A}_{ia}^{(1)} + \tilde{\cal A}_{ai}^{(1)} &=&  
\frac{1}{4 (8\pi^2\alpha')^2} \sqrt{G}\alpha'^{-1}\, {\rm tr}\left( \bfg_{\Theta i}\bfg_{\Theta a}^{-1}  
 \thba{\vec{0}}{\vec{0}}({\bf \vec A}_{ia} ,ilG\alpha'^{-1}) (2\pi\alpha'{\bf F}_{ia})^2   
  \right) 
 \ .
\nonumber  
\eeqn
For $\alpha'=1/2$ and vanishing Wilson lines 
we recover the formulas given in (3.18) of 
\cite{AnBaDu99} if 
in the final result we take into account the different factors which were 
factored out in (2.1) and (2.7) of \cite{AnBaDu99}, and we correct their 
formula (3.11) by adding an additional factor $1/2$ in the exponent, which 
leads to an additional factor of $2$ on the r.h.s.\ of their (3.18).


\section{One-loop amplitude for $\mathbb{T}^6/\mathbb{Z}_6'$}
\label{z6}

Here we collect some formulas which are relevant for our discussion 
of the $\mathbb{Z}_6'$ orientifold 
in section \ref{none}. The one-loop amplitude without Wilson lines 
can be directly copied from \cite{AnBaDu99}, 
taking into account the remarks at the end of the previous section, 
and adapting to our notation. 
We include them here for completeness, and to make our statement concrete
that the part important for the discussion of the rho problem 
is just identical to the $\mathbb{Z}_2$ result, 
up to an overall numerical factor $1/3$. 

We use an analogous normalization as for the $\mathbb{Z}_2$ model 
and absorb all 
relative factors into the integrands of the amplitudes, writing 
\beqn \label{loopch6}
{\cal M} &=& \frac{\sqrt{-g_4}}{(4\pi^2\alpha')^2} \int_0^\infty \frac{dt}{(2t)^3} 
{\rm Tr}_{\rm op}^{\rm NS-R} \left( \frac{\Omega}{2} \frac{ 1+(-1)^F }{2} \frac{1+\Theta+ \, \cdots\, +\Theta^5}{6} 
  e^{-2\pi t {\cal H}_{\rm op} } \right) 
\\
&=& \sqrt{-g_4} \int_0^\infty \frac{dt}{t} \sum_{k=0}^5 
\left( \sum_i {\cal M}_{i}^{(k)} (-q) + \sum_a {\cal M}_{a}^{(k)} (-q) \right) 
\non
{\cal A} &=& \frac{\sqrt{-g_4}}{(4\pi^2\alpha')^2} 
\int_0^\infty \frac{dt}{(2t)^3} {\rm Tr}_{\rm op}^{\rm NS-R} \left( \frac{1}{2} \frac{ 1+(-1)^F }{2} 
 \frac{1+\Theta+ \, \cdots\, +\Theta^5}{6} 
  e^{-2\pi t {\cal H}_{\rm op} } \right) 
\non
&=& \sqrt{-g_4} 
\int_0^\infty \frac{dt}{t} \sum_{k=0}^5 \left( \sum_{i,j} {\cal A}_{ij}^{(k)} (q) 
 + \sum_{a,b} {\cal A}_{ab}^{(k)} (q) 
 +  \sum_{i,a} \left( {\cal A}_{ia}^{(k)} (q) + {\cal A}_{ai}^{(k)} (q) 
\right) \right)\ .  
\nonumber 
\eeqn
The only amplitudes that depend on the D9-brane Wilson 
lines on the third 2-torus and on the background 
field strength on the D5-branes are given (to order $\co ({\bf F}^2)$) by 
\beqn \hspace{-1cm} 
\tilde{\cal A}_{ia}^{(0)} + \tilde{\cal A}_{ai}^{(0)} &=&   
\frac{1}{12(8\pi^2\alpha')^2} \sqrt{G}\alpha'^{-1}\, {\rm tr}\left(  \bfg_{i}\bfg_{a}^{-1}  
 \thba{\vec{0}}{\vec{0}}({\bf \vec A}_{ia} ,ilG\alpha'^{-1})  (2\pi\alpha'{\bf F}_{ia})^2   \right) 
 \ , 
\non\hspace{-1cm} 
\tilde{\cal A}_{ia}^{(3)} + \tilde{\cal A}_{ai}^{(3)} &=&  
\frac{1}{12(8\pi^2\alpha')^2} \sqrt{G}\alpha'^{-1}\, {\rm tr}\left( \bfg_{\Theta i}^3\bfg_{\Theta a}^{-3}  
 \thba{\vec{0}}{\vec{0}}({\bf \vec A}_{ia} ,ilG\alpha'^{-1}) (2\pi\alpha'{\bf F}_{ia})^2   
  \right) \label{keq03}\ .
\eeqn
In the $\mathbb{Z}_6'$ model there will also emerge a dependence 
of the correction to the gauge couplings 
on the moduli of the second 2-torus, whose metric we denote $G_2$. 
This is evident from the classification of all amplitudes in 
table \ref{amptable}.

\begin{table}
\vspace{-5mm}
\[
\begin{array}{|c|c|c|}\hline
\rule{0mm}{4.5mm}
\mbox{SUSY} & \mbox{Untwisted $\mathbb{T}^2$} & \mbox{Amplitudes} \\ \hline\hline
\rule{0mm}{4.5mm}\cn=1 & - & \ca_{ij}^{(k=1,5)},\ca_{ia}^{(k=1,2,4,5)}
,\cm_{i}^{(k=1,5)}  \\ \hline 
\rule{0mm}{4.5mm} \cn=2 &  \mbox{2nd} & \ca_{ij}^{(k=2,4)},
\cm_{i}^{(k=2,4)} \\ \hline
\rule{0mm}{4.5mm}\cn=2 &  \mbox{3rd} & \ca_{ij}^{(k=3)},\ca_{ia}^{(k=0,3)},
\cm_{i}^{(k=3)} \\ \hline 
\end{array}
\]
\vspace{-8mm}
\caption{Amplitude summary}
\label{amptable}
\end{table}
Explicitly, the other contributions are given by 
(suppressing the $\cn=4$ sector again, 
and leaving out terms that can be restored trivially be using 
the symmetry between 9- and 5-branes) 
\beqn 
\tilde{\cal M}_{i}^{(3)} &=& 
-\frac{8}{3(8\pi^2\alpha')^2} \sqrt{G}\alpha'^{-1}\,  {\rm tr}\left( \gamma_{\Omega\Theta i}^{-3} 
  (\gamma_{\Omega\Theta i}^{\rm T})^3
  \thba{\vec{0}}{\vec{0}}(\vec 0,4ilG\alpha'^{-1}) (2\pi\alpha'{\bf F}_i)^2 \right) 
 \ , 
\non 
\tilde{\cal A}_{ij}^{(3)} &=& 
\frac{1}{6(8\pi^2\alpha')^2} \sqrt{G}\alpha'^{-1}\, 
{\rm tr} \left( \bfg_{\Theta i}^3 \bfg_{\Theta j}^{-3} 
 \thba{\vec{0}}{\vec{0}}(\vec 0 ,ilG\alpha'^{-1}) (2\pi\alpha'{\bf F}_{ij})^2 \right) 
 \ , 
\non
\tilde \cm_{i}^{(k=1,5)}
&=& {2 \over 3\pi(8\pi\alpha')^2  } \tr
 \left( (\gamma_{\Omega\Theta i}^{\rm T})^k 
 \gamma_{\Omega\Theta i}^{-k}  
 (2\pi\alpha' {\bf F}_{i} )^2  \right)
 \prod_{i=1}^3 \sin (\pi kv_i)
 \sum_{i=1}^3 
 {\tht'\bw{1/2}{1/2+kv_i}(0)
 \over \thbw{1/2}{1/2+kv_i}(0)} \ , 
\non
\tilde \cm_{i}^{(k=2,4)}
&=& {8 \over 3(8\pi\alpha')^2 } \sqrt{G_2}\alpha'^{-1} 
 \tr \left( (\gamma_{\Omega\Theta i}^{\rm T})^k 
 \gamma_{\Omega \Theta i}^{-k}  
  \thba{{\vec 0}}{\,\vec 0}(\vec 0,4il G_2\alpha'^{-1})
  (2\pi\alpha' {\bf F}_{i} )^2  \right) \non 
&& \qquad\qquad\qquad\qquad 
\times  \sin (\pi kv_1)\sin (\pi kv_3)\ , 
\non
\tilde \ca_{ij}^{(k=1,5)}
&=& -{1 \over 3\pi (8\pi\alpha')^2 } 
\tr
 \left( \bfg_{\Theta i}^k
 \bfg_{\Theta j}^{-k}  
 (2\pi\alpha' {\bf F}_{ij} )^2  \right)
 \prod_{i=1}^3 \sin (\pi kv_i)
 \sum_{i=1}^3 
 {\tht'\bw{1/2}{1/2+kv_i}(0)
 \over \thbw{1/2}{1/2+kv_i}(0)} \ , 
\non
\tilde \ca_{ij}^{(k=2,4)} 
&=& -{1 \over 6 (8\pi\alpha')^2 } \sqrt{G_2}\alpha'^{-1} 
 \tr \left(\bfg_{\Theta i}^k \bfg_{\Theta j}^{-k}  
  \thba{{\vec 0}}{\vec 0}(\vec 0,il G_2\alpha'^{-1})
 (2\pi\alpha' {\bf F}_{ij} )^2 
 \right)\non 
&& \qquad\qquad\qquad\qquad 
\times  
 \sin (\pi kv_1)\sin (\pi kv_3)\ , 
\non
\tilde \ca_{ia}^{(k=1,2,4,5)} 
&=& -{2 \over 3\pi (8\pi\alpha')^2 } \tr
 \left( \bfg_{\Theta i}^k 
 \bfg_{\Theta a}^{-k}  
 (2\pi\alpha' {\bf F}_{ia} )^2 \right)
 \sin (\pi kv_3) \non [-1mm]
&& \qquad\qquad\qquad\qquad \times
 \left({\tht'\bw{1/2}{1/2-kv_3}(0)
 \over \thbw{1/2}{1/2-kv_3}(0)}
 +\sum_{i=1}^2 
 {\tht'\bw{0}{1/2-kv_i}(0)
 \over \thbw{0}{1/2-kv_i}(0)}\right) \ . 
\eeqn
We have given these amplitudes only for vanishing Wilson 
lines. Turning on Wilson lines would lead to non-vanishing first 
arguments $2{\bf \vec A}_i$ and ${\bf \vec A}_{ij}$ 
in the theta functions appearing in $\tilde{\cal M}_{i}^{(3)}$ and 
$\tilde{\cal A}_{ij}^{(3)}$, respectively. In the other amplitudes
(twisted by $\Theta^k$ with $k \notin \{0,3\}$),
turning on Wilson lines would require summing over fixed points of 
$\Theta^k$ on the third torus and inserting powers of the matrices 
$\gamma_{W9}$ (\ref{w6}) (and the analogs $\gamma_{W5}$ for D5-branes), 
appropriate to the fixed point, into the traces, cf. 
\cite{Aldazabal:1998mr,Cvetic:2000st}. 
Obviously, the combined set of $k=0,3$ amplitudes is
formally identical to the results (\ref{total}) for the $\cn=2$ model, 
up to the overall normalization factor $1/6$ instead of $1/2$, 
and the necessary replacements of the appropriate matrices $\gamma$. 
The rest will contribute terms independent of the Wilson line 
moduli, for example a term $\sim \ln|\eta(U')|$, where 
$U'$ is the only complex structure modulus of the model as explained
in section \ref{inter}. These contributions could in principle 
be extracted from \cite{AnBaDu99}.


\section{Formulas}
\label{thetas}

Here we collect some formulas about elliptic functions
that are all available in various corners of the literature, 
but usually in different notation.
We make an effort to consistently
follow the conventions of 
the textbook by Polchinski \cite{Polchinski:rr}.

The eta and theta functions we use are 
\beqn 
\eta(\tau) &=& q^{1/24} \prod_{n=1}^\infty \left( 1-q^n \right) \ , \non
\tht\ba{\vec \alpha}{\vec\beta}(\vec \nu,G) &=& \sum_{\vec n\in \mathbb{Z}^N} 
 e^{i\pi(\vec n+\vec \alpha)^{\rm T} G (\vec n+\vec \alpha)} 
e^{2\pi i(\vec \nu+\vec \beta)^{\rm T}(\vec n+\vec\alpha)}  \ , 
\label{thetamatrix}
\eeqn
where $G$ is an $N\times N$ matrix with  ${\rm Im}(G)>0$, and $q=e^{2\pi i \tau}$.
The case $N=1$ is the usual set of genus one theta functions.
For $N=1$ and half-integer characteristics we use the notation
\beqn
\tht \ba{0}{0}(\nu,\tau) = \tht_3(\nu,\tau)\ , \quad 
\tht \ba{1/2}{0}(\nu,\tau) = \tht_2(\nu,\tau)\ , \non
\tht \ba{0}{1/2}(\nu,\tau) = \tht_4(\nu,\tau)\ , \quad
\tht \ba{1/2}{1/2}(\nu,\tau) = -\tht_1(\nu,\tau)\ .
\label{Z2theta}
\eeqn
Comparing to another good source for theta identities,
the lecture notes by Kiritsis \cite{Kiri97}, 
we have
$\thba{\alpha}{\beta}=\tht^{\rm K}\bb{-2\alpha}{-2\beta}$,
where $\tht^{\rm K}$ is that of Kiritsis.
A word of warning:
\[
\tht^{\rm K}_1(\nu,\tau)\equiv \tht^{\rm K}\bb{1}{1}(\nu,\tau)
= \thba{-1/2}{-1/2}(\nu,\tau)=-\thba{1/2}{1/2}(\nu,\tau) \equiv
\tht_1(\nu,\tau)\ . 
\]
The modular transformation property of (\ref{thetamatrix})
that will be relevant to us is 
\beqn
\tht\ba{\vec \alpha}{\vec 0}(0,it G^{-1}) &=& \sqrt{G} \, t^{-N/2}\,  
 \tht\ba{\vec 0}{\vec 0 }(\vec \alpha,it^{-1} G) \ , 
\eeqn
where the $G$ under the square root 
denotes the determinant of the matrix $G$ in the argument. 
The modular transformations for $N=1$ read 
\beqn
\thba{\alpha}{\beta}(\nu,\tau)
&=& e^{\pi i \alpha(\alpha+1)}
\thba{\alpha}{\beta-\alpha-1/2}(\nu,\tau+1) \label{T} \ , \non
\thba{\alpha}{\beta}(\nu,\tau)
&=& (-i \tau)^{-1/2} e^{2\pi i \alpha \beta - \pi i \nu^2/\tau}
\thba{-\beta}{\alpha}(\nu/\tau,-1/\tau) \label{S} \ . 
\eeqn
For the M\"obius strip the following sequence of modular transformations is
useful: 
\[
\tau ~\rightarrow~ -{1 \over \tau} ~\rightarrow~ -{1 \over \tau}
+2 ~\rightarrow~ -{1 \over -{1 \over \tau}+2} \ ,
\]
giving
\beqn
\thba{\alpha}{\beta}(\nu,\tau)
&=& (1-2\tau)^{-1/2} 
e^{-2\pi i \beta} e^{-\pi i\nu^2/(\tau-1/2)}
\thba{\alpha+2\beta}{\beta}\Big(\frac{\nu}{1-2\tau},
\frac{\tau}{1-2\tau}\Big) \; .
\eeqn

%
%
%
%

For $\nu$-derivatives we use the  notation 
$\vartheta'[{\alpha\atop\beta}](0,\tau)=
\partial_\nu\vartheta[{\alpha\atop\beta}](\nu,\tau)|_{\nu=0}$.
For the four special theta functions (\ref{Z2theta}),
we have
\beqn \label{firstder}
\tht'_2 (0,\tau) = \tht'_3(0,\tau) = \tht'_4(0,\tau) = 0 \ , 
\quad  \tht'_1(0,\tau) = 2\pi \eta(\tau)^3 \ .
\eeqn
In section \ref{seconeloop}, we also make use of the third $\nu$-derivative 
(cf.\ (F.14) in \cite{Kiri97})
\be \label{thirdder}
\tht'''_1(0,\tau) = - \pi^2 \tht'_1(0,\tau) 
E_2(\tau)\ ,
\ee
where $E_2(\tau)$ is the holomorphic second Eisenstein series,
\be \label{Eisen2}
E_2(\tau) = 1  - 24 \sum_{n=1}^{\infty}{n q^n \over 1 -q^n} \; .
\ee
Moreover, in section \ref{inter}, we use the second $\nu$-derivative 
(cf.\ (A.25) in \cite{Kiri97})
\be \label{secondder}
\tht''_2(0,\tau) = - \frac{\pi^2}{3} \tht_2(0,\tau) \Big(E_2(\tau) 
+ \tht_3^4(0,\tau) + \tht_4^4(0,\tau) \Big)\ .
\ee
From the basic quartic Riemann identity
\beqn
{1 \over 2}\sum_{\alpha,\beta} \eta_{\alpha\beta}
\prod_{i=1}^4 \thba{\alpha}{\beta}(g_i,\tau)&=&  
-\prod_{i=1}^4 \thba{1/2}{1/2}(g_i',\tau)
\eeqn
a number of useful identities follow.
Here 
\beqn \label{gs}
g_1' &=& {1 \over 2}(g_1+g_2+g_3+g_4) \; , \qquad
g_2' = {1 \over 2}(g_1+g_2-g_3-g_4) \; , \non
g_3' &=& {1 \over 2}(g_1-g_2+g_3-g_4) \; , \qquad
g_4' = {1 \over 2}(g_1-g_2-g_3+g_4) 
\eeqn
are those of \cite{Polchinski:rr}. (The identity holds
for other sign combinations as well.)
For instance, setting $g_1=g_2=g_3=0$, $g_4=\nu$ and expanding in $\nu$,
one has
\beqn \label{Riemannzero}
\sum_{\alpha,\beta} \eta_{\alpha\beta} \thbap{\alpha}{\beta}(0) 
\thba{\alpha}{\beta}^3(0) = 0 \ . 
\eeqn
It will be useful to have the following slightly more general
theta identity, which allows for shifts not
only in the $\nu$ argument (or equivalently in the $\beta$ characteristic)
but also in the $\alpha$ characteristic.
It can be proven from the standard
one using periodicity properties (see e.g.\ \cite{Kiri97}).
For $\sum_i g_i=0$, $\sum_i h_i=0$, the most useful form is to
include an additional spin-structure independent denominator: 
\beqn
\sum_{\alpha,\beta} \eta_{\alpha \beta} {\tht''\bw{\alpha}{\beta}(0) \over
\tht_1'(0)}
{\prod_{i=1}^{3}\thbw{\alpha+h_i}{\beta+g_i}(0)
\over \prod_{i=1}^{3}\thbw{1/2+h_i}{1/2+g_i}(0)}
= -\sum_{i=1}^3 
{\tht'\bw{1/2+h_i}{1/2+g_i}(0)
\over \thbw{1/2+h_i}{1/2+g_i}(0)} \ , 
\eeqn
where we set the $g_1$ of (\ref{gs}) to zero and relabeled
the other $g_i \rightarrow g_{i-1}, i=2,3,4$.
We now turn to two useful special cases.\\
{\it Special case 1:} $h_1=0$, $h_2=1/2$, $h_3=-1/2$:
\beqn
&&\hspace{-2cm}\sum_{\alpha,\beta} \eta_{\alpha \beta} {\tht''\bw{\alpha}{\beta}(0) \over
\tht_1'(0)}
{\thbw{\alpha}{\beta+g_1}(0) 
\thbw{\alpha+ 1/2}{\beta+g_2}(0)\thbw{\alpha- 1/2}{\beta+g_3}(0)
\over \thbw{1/2}{1/2+g_1}(0) 
\thbw{0}{1/2+g_2}(0)\thbw{0}{1/2+g_3}(0)}
\non
&& \hspace{3cm} = -\left(
{\tht'\bw{1/2}{1/2+g_1}(0) \over \thbw{1/2}{1/2+g_1}(0)}+
\sum_{i=2}^3 
{\tht'\bw{0}{1/2+g_i}(0)
\over \thbw{0}{1/2+g_i}(0)} \right) \; .
\eeqn
In applying this identity, it is useful to note that
all theta functions have periodicity 1 in the upper characteristic. 
\ \\
{\it Special case 2:}
Let us in addition to the previous assumptions assume
$g_1=0$ (untwisted first two-torus). Then the last two terms on the
right hand side cancel, as they must since $\tht_1(0)$ must cancel
out of the denominators for the expression to remain regular.
When $h_1=g_1=0$, we can denote $h_2=-h_3=:h$, and similarly
$g_2=-g_3=:g$.
This is the familiar case that the sum collapses to a number:
\beqn \label{Thetader}
\sum_{\alpha,\beta}  \eta_{\alpha \beta} {\tht''\bw{\alpha}{\beta}(0)
\thbw{\alpha}{\beta}(0) 
\over\eta^6}
{\thbw{\alpha+ h}{\beta+g}(0)\thbw{\alpha-h}{\beta-g}(0)
\over
\thbw{1/2+ h}{1/2+g}(0)\thbw{1/2- h}{1/2-g}(0)}
= -4\pi^2 \; .
\eeqn

\end{appendix}


\end{document}